\documentclass[aps,pra,reprint,twocolumn,nofootinbib,showpacs,showkeys,address,longbibliography]{revtex4-2}
\usepackage{amsmath,amssymb,amstext}
\usepackage[usenames,dvipsnames]{color}
\usepackage{graphicx}
\usepackage{bm,bbold,braket}
\usepackage{natbib}
\usepackage{txfonts, comment,stmaryrd}
\usepackage{dcolumn}
\usepackage{color}
\usepackage{xcolor}
\usepackage{soul}
\usepackage[colorlinks,bookmarks=false,citecolor=blue,linkcolor=red,urlcolor=blue]{hyperref}

\begin{document}

\preprint{APS/123-QED}
\title{Properties of Krylov state complexity in qubit dynamics}

\author{Siddharth Seetharaman}

\author{Chetanya Singh}%

\author{Rejish Nath}
\affiliation{%
 Department of Physics, Indian Institute of Science Education and Research, Pune 411008, India
}%

\date{\today}

\begin{abstract}
We analyze the properties of Krylov state complexity in qubit dynamics, considering a single qubit and a qubit pair. A geometrical picture of the Krylov complexity is discussed for the single-qubit case, whereas it becomes non-trivial for the two-qubit case. Considering the particular case of interacting Rydberg two-level atoms, we show that the Krylov basis obtained using an effective Hamiltonian minimizes the time-averaged spread complexity compared to that which is obtained from the original Hamiltonian. We further generalize the latter property to an arbitrary Hamiltonian in which the entire Hilbert space comprises of two subspaces provided a weak coupling between them. \end{abstract}

\maketitle


\section{\label{sec:level1}Introduction} 

The concept of complexity has found various implications in physics. For instance, in computation, it can be the number of resources one requires, like that of bits/qubits, operations, or the depth of the circuits, etc. \cite{Complexity-Watrous2009}. For implementing a multi-qubit unitary operation, it can be the minimum number of predefined elementary unitary operations \cite{osborne_Hamiltonian_complexity_2012, aaronson2016_complexity_quantumstates_transformations}. It is also interpreted as the minimum geodesic distance between the identity and a targeted unitary operator in an operator space with an appropriately chosen metric, as in the case of  Nielsen complexity \cite{Nielson2006, Nielson-Dowling-2008, Nielson-Quantum-Computing-Journal-2006, Nielsen_Optimal_control_Geometry_QuantumComputing_PRA2006}. Similarly, one can define the complexity of a quantum state as the minimum number of unitary operations from a predefined set of operators required to synthesize the given target state from a reference state \cite{aaronson2016_complexity_quantumstates_transformations, Nielsen_Chuang_Book_2010, Complexity_Black_Hole_Review_Chapman2022}. The latter also led to studies for finding the minimal geodesics under various metrics for the manifold of unitary operators that act on the Hilbert space. \cite{Quantum_Complexity_Time_Evolution_Chaos_VijayBala, Complexity_Growth_VijayBala2021, Complexity_Measures_Magan_JHEP2021, Preskill_Models_of_Quantum_Complexity_Growth_PRXQuantum2021, Smooth_Quantum_Complexity_Sondhi_JHEP2021, ABrown_Quantum_Complexity_LowerBound_Nature2023, Black_Holes_Complexity_Chaos_JMagan_JHEP2018, Geometry_of_Quantum_Complexity_Nielsen_Auzzi_PRD2021}. The concept of complexity also finds applications in black holes, where correspondence between the growth of an interior volume of a black hole and growth of the complexity of its dual quantum state on the boundary is conjectured, termed as the holographic complexity \cite{susskind2014_computational_complexity_blackhole, Susskind_Complexity_ShockWaveGeom_PhysRevD2014, susskind2014_switchback_complexity, Alishahiha_Holographic_Complexity_PRD2015, Susskind_Holographic_Complexity_PRL2016, susskind_complexity_action_blackholes_PRD2016, Holographic_fluc_and_principle_of_minimal_complexity_Chemissany_JHEP2016, Sussking_Quantum_Complexity_Neg_Curvature_PRD2017, bouland2019-computational-pseudorandomness-wormhole-growth, Complexity_Black_Hole_Review_Chapman2022, Holographic_Duality_Review_Chen2022}. Despite the developments, the above complexity measures suffer from ambiguities, for instance, in the choice of metric for the case of the Nielsen complexity, in the choice of the set of elementary unitary operations for achieving a target unitary operation from the identity operator or achieving a target quantum state from an initial state. 

Subsequently, a new measure of complexity was put forward for states \cite{Caputa2022-State-Complexity-Original}, motivated by a similar measure for operators \cite{Universal-Growth-Hypothesis-2019}, called the Krylov complexity \cite{Krylov-Review2024, sanchezgarrido_thesis_2024krylovcomplexity}. It is quantified as the depth or spread of an initial state in its Krylov basis over time evolution. The advantages of the Krylov complexity are twofold. First, given an initial state and a Hamiltonian, the complexity is unambiguously defined using the Krylov basis obtained via the Lanczos algorithm \cite{Lanczos1950, Vishwanath_Mueller_Book_Recursion_Method}. Second, the Krylov basis has been shown to minimize the \textit{spread} of an initial state among all choices of ordered bases \cite{Caputa2022-State-Complexity-Original}. The Krylov complexity further gained attention in studying quantum chaos and integrability \cite{Hashimoto2023-Billiards, Xian2023-Universal-chaotic-dynamics, scialchi2023_integrability_chaos_krylov_arxiv, vijay_bala_dec2023_quantumchaos_integrability_arxiv, Bhattacharjee-PXP-PRB2022, SNandy-PXP-Numerical-IOP2024, Krylov_Complexity_Saddle-dominated_Scrambling_Chaos_Pedraza_JHEP2024, Krylov_Chaos_Mixed_Field_Ising_Model_2024, nizami2024_Krylov_chaos_periodic_driving_arxiv, Nishida_Krylov_Billiard_MostlyOperator_PRD2024, Krylov_triangular_billiards_balasubramanian_arxiv2024, krylov_fractality_ergodic_localized_regimes_transitions_generic_pnandy_arxiv2024, krylov_complexity_order_parameter_chaotic_integrable_baggioli_hyun_sik_jeong_arxiv2024} and found applications in quantum phase transitions \cite{Caputa2022-Topological-Phases-of-Matter, Caputa2023-Kitaev, Krylov_Quenched_LMG_QPT_Afrasiar_JStat2023, Krylov_state_free_fermion_theories_QPT_Gautam_EurPhysJ_2024, Krylov-DynamicalPhaseTransition-LMGModel-PRB2024, Lanczos-Krylov-PowerLawRandomBandedModel-2024}, black holes \cite{Krylov_Complexity_Black_Hole_States_Wolfgang_Muck_PRD2024}, neutrino oscillations \cite{Krylov_Neutrino_Oscillations_Dixit2024_Arxiv}, modular Hamiltonians \cite{Caputa_Modular_Hamiltonian_PRD2024}, quenched systems \cite{Krylov_Quenched_LMG_QPT_Afrasiar_JStat2023, Krylov_Comp_Quenched_Interacting_Systems_Tapobrata_PRB2024, Krylov_Quench_TwoPoint_Measurement_Schemes_Tapobrata_PRB2024}, non-Hermitian systems \cite{Krylov_NonHermitian_Appendix_PNandy_JHEP2023, Krylov-NonHermitianSkinEffect-IPR-2024}, periodically-driven systems \cite{Krylov_Driven_Systems_Nizami_PRE2023, nizami2024_Krylov_chaos_periodic_driving_arxiv}, and others \cite{Thermalization_KrylovBasis_Alishahiha2024arxiv, Krylov_Nonunitary_Zeno_Effect_Erdmenger_JHEP2024, zhou_Krylov_2ModeBEC_Arxiv2024, krylov_syk_sparse_holographic_non-holographic_jha_arxiv2024}. Recently, there have been efforts to establish a connection between Krylov complexity and other complexity measures. For instance, the time average of the Krylov complexity is shown to be connected to an upper bound for the Nielsen complexity of the corresponding unitary operator, using a metric that is determined by the Krylov basis \cite{Krylov-Nielsen-PRL2024}. A unified framework for describing the operator and state complexity was discussed in \cite{Density-matrix-SciPost2023}, which has generalized this measure to mixed states and reduced density matrices \cite{Density-matrix-SciPost2023, Caputa-Density-Matrix2024}. There are efforts to interpret the Krylov complexity geometrically \cite{Krylov_state_comp_dilaton_geometrical_vanZyl_PRD2023, Aguilar_Gutierrez_Krylov_Not_Distance_PRD2024}, and a correspondence between the Krylov complexity and the wormhole length is found \cite{Wormhole-length-JHEP2023}. 

In this work, we study the Krylov complexity of a single-qubit and two-qubit systems in search of fundamental insights into the Krylov state complexity  in qubit dynamics. While it has been shown in \cite{Aguilar_Gutierrez_Krylov_Not_Distance_PRD2024} that the Krylov complexity or its square root cannot act as a measure of distance between states in general, we find that the square root of the complexity does, indeed, relate to the distance between time-evolved states for a single qubit. However, once we consider two non-interacting qubits, which could have been described independently, we find that such a geometrical interpretation is no longer possible precisely because the overall Krylov complexity is more than just a sum of the individual complexities. We further extend our analysis to interacting two-level Rydberg atoms \cite{Gallagher_1994, Quantum-Info-Rydberg-Review-Saffman2010, Rydberg_Review_Rejish_2024}, which exhibit correlated dynamics under Rydberg blockade  \cite{Browaeys-Rydberg-Review-NatPhys2020, Lukin_51Qbits_Nature2017, Lukin_256Qbits_Nature2021}. In the case of a pair of Rydberg atoms, the doubly excited states are inhibited in the Rydberg blockade dynamics and an effective two-level picture emerges. In that case, one expects the dynamics of Krylov complexity to be that of a single two-level system, i.e., it periodically oscillates between zero and one. However, we observe that the Krylov basis obtained using the original Hamiltonian exhibits a complexity having an amplitude larger than one. Interestingly, we find an ordered basis, obtained as the Krylov basis of an effective Hamiltonian, that minimizes the time-averaged spread complexity compared to the original Krylov basis  and in which the amplitude is one, as expected. We further generalize this aspect for an arbitrary Hamiltonian for which the entire Hilbert space is decomposed into two subspaces with a weak coupling between them.

The paper is organized as follows: in section \ref{section:Krylov-state-complexity}, we review the Krylov state complexity and some of its properties. In section \ref{sec:qu1}, we study the single qubit and provide a geometrical interpretation of the Krylov complexity. We extend the study to two qubits in section \ref{sec:two_qubits}, considering both uncoupled and coupled qubits. The particular case of a pair of Rydberg atoms is discussed in Sec.~\ref{iqs}. We generalize the minimization of the time-averaged spread complexity in different bases and discuss their relevance in Sec.~\ref{sec:minimization_of_complexity}. Finally,  we summarize our results in Sec.~\ref{sec:summary}. 


\section{Krylov State Complexity}\label{section:Krylov-state-complexity}

For a closed system described by a time-independent Hamiltonian, $\hat H$, the time-evolution of an initial state $|\psi_0\rangle$ is 
\begin{equation}
    |\psi (t) \rangle = e^{-i\hat{H}t} |\psi_0 \rangle = \sum\limits_{n=0}^\infty \dfrac{(-it)^{n}}{n!} |\psi_n \rangle,
\end{equation}
where $|\psi_n \rangle = \hat{H}^n |\psi_0\rangle$ and we have set $\hbar = 1$. Employing the Gram-Schmidt process to the set of states $\{\psi_0, \hat H\psi_0, \hat H^2\psi_0, ...\}$ generates an ordered, orthonormal basis, $\mathcal K=\{|K_0\rangle, |K_1\rangle, |K_2\rangle, ..., |K_n\rangle \}$ for the part of the Hilbert space explored under time-evolution. The new basis $\mathcal K$ is called the Krylov basis. The zeroth element in the Krylov basis is the initial state itself, i.e., $|K_0\rangle=|\psi_0\rangle$. The other Krylov basis states ($n \geq 1$) are constructed recursively via the Lanczos algorithm \cite{Lanczos1950, Vishwanath_Mueller_Book_Recursion_Method} from the initial state and the Hamiltonian as follows:
\begin{equation}
    |K_{n+1} \rangle =b_{n+1}^{-1}|A_{n+1} \rangle
\end{equation}
where
\begin{equation}
    |A_{n+1} \rangle = \big(\hat{H} - a_{n} \big) |K_{n} \rangle - b_{n} |K_{n-1} \rangle
\end{equation}
with $a_n=\langle K_n|\hat{H}|K_n\rangle$ and $b_n= \langle A_n|A_n\rangle^{1/2}$ are the Lanczos coefficients. Note that $b_0=0$. By construction, the dimension of the Krylov space can be smaller than or equal to that of the full Hilbert space. 

Given an initial state, we can define a cost function on any ordered basis, $\mathcal{B} = \{|B_0\rangle, |B_1\rangle, \hdots, |B_{D-1} \rangle \}$, of the Hilbert space as
\begin{equation}
\label{costB}
    \mathcal{C}_{\mathcal{B}} (t) = \sum\limits_{n=0}^{D-1} c_n \left| \langle \psi (t)|B_n\rangle \right|^2 = \sum\limits_{n=0}^{D-1} c_n P_{\mathcal{B}} (n,t)
\end{equation}
where $c_n$ is a monotonically increasing function of $n$ with $c_n \geq 0$, $P_\mathcal{B} (n,t) = \left| \langle \psi (t)|B_n\rangle \right|^2$, and $D$ is the dimension of the Hilbert space. For $c_n = n$, the cost function \( \mathcal{C}_{\mathcal{B}} (t)\) measures the average spread of the initial state in the basis \(\mathcal{B}\) and is dubbed the \textit{spread complexity} \cite{Caputa2022-State-Complexity-Original}. The spread complexity is minimized in the Krylov basis in the vicinity of $t=0$ and is termed as the Krylov complexity, given by \cite{Caputa2022-State-Complexity-Original}, 
\begin{eqnarray}\label{eq:krylov_complexity_definition}
    \mathcal{C}_{\mathcal{K}} (t) = \sum\limits_{n = 0} n |\langle \psi (t)|K_n\rangle|^2.
    \label{kc}
\end{eqnarray}
By construction, $\mathcal{C}_{\mathcal{K}} (t=0)=0$ and in general, $\mathcal{C}_{\mathcal{K}} (t)\geq 0$. 

The Krylov complexity can be viewed as a measure of the complexity of the dynamics of an initial state under a Hamiltonian evolution. The two states related by time-evolution have the same Lanczos coefficients and the Krylov complexity. To show that, consider two different initial states, $|\psi_0\rangle=|K_0\rangle$ and $|\psi_0^\prime \rangle = e^{-i\hat{H}t} |\psi_0 \rangle$, connected by a unitary time evolution. We can see that $a_0^\prime = \langle \psi_0^\prime|\hat{H}|\psi_0^\prime \rangle = \langle K_0|e^{i\hat{H}t} \hat H e^{-i\hat{H}t}|K_0\rangle = a_0$ and $b_1^\prime = \langle A_1^\prime | A_1^\prime \rangle^{1/2} = b_1$ since $|A_1^\prime \rangle = e^{-i\hat{H}t} (\hat{H} - a_0) |K_0\rangle = e^{-i\hat{H}t} |A_1 \rangle$. Similarly, we can show that $a_1^\prime = a_1$, $b_2^\prime = b_2$, $|K_1^\prime \rangle = e^{-i\hat{H}t} |K_1\rangle$, and thereby, $|A_2^\prime \rangle = e^{-i\hat{H}t} |A_2\rangle$. Following the same, we get $|A_{n+1}^\prime\rangle = e^{-i\hat{H}t}|A_{n+1}\rangle$ assuming that $|A_j^\prime \rangle = e^{-i\hat{H}t} |A_j\rangle$, $a_j^\prime = a_j$ and $b_j^\prime = b_j$ for all $j \leq n$. Thus, $|\psi_0^\prime \rangle$ has the same set of Lanczos coefficients as $|\psi_0\rangle$ and hence, the Krylov complexity \cite{Aguilar_Gutierrez_Krylov_Not_Distance_PRD2024}. 

In the Krylov basis, the Hamiltonian satisfies,
\begin{equation}\label{Eq:Tridiagonal-Ham-in-Krylov-Basis}
    \hat{H} |K_n\rangle = a_n |K_n\rangle + b_n |K_{n-1} \rangle + b_{n+1} |K_{n+1} \rangle,
\end{equation}
and forms a tridiagonal matrix, which resembles the Hamiltonian of a tight-binding model on a chain with a site-dependent hopping amplitude $b_n$ between $n$ and $n-1$ sites, while $a_n$ acts as a local energy offset or a chemical potential. Now, the Krylov complexity $\mathcal{C}_{\mathcal{K}} (t)$ can be interpreted as the expectation value of the position on this lattice, where the origin is at the first site [$n=0$ in Eq.~(\ref{kc})]. In other words, the Krylov complexity measures how far the state has evolved away from the initial site in the above tight-binding model.

\section{Single qubit}
\label{sec:qu1}

In this section, we discuss the properties of the Krylov complexity in the dynamics of a single qubit. Let the eigenstates of the Hamiltonian be $|\pm\rangle$ with eigenenergies $\pm \omega/2$, so that the Hamiltonian in the energy eigenbasis takes the form $\hat{H} = (\omega/2) \hat{\sigma}_z$. The initial state $|\psi_0\rangle = \cos (\theta_0/2) |+\rangle + \sin (\theta_0/2) e^{i\phi} |-\rangle$ forms the first Krylov basis state $|K_0\rangle$ in Bloch sphere notation, with $0 \leq \theta_0 \leq \pi$ and $0 \leq \phi < 2\pi$. The second one is orthogonal to the initial state and is $|K_1\rangle = \sin (\theta_0/2) |+\rangle - \cos (\theta_0/2) e^{i\phi} |-\rangle$. $|K_1\rangle$ is defined up to a global phase factor. The Hamiltonian of a qubit in the Krylov basis can be written as (after neglecting constants), 
\begin{equation}
    \hat{H}=\frac{(a_0-a_1)}{2}\hat \sigma_z+b_1\hat\sigma_x,
\end{equation}
where $a_0= (\omega \cos \theta_0)/2$, $a_1=-a_0$, $b_1 = (\omega \sin \theta_0)/2$ and $\hat \sigma_{x, z}$ are the Pauli spin-1/2 matrices. 

The time-evolved state can be written in the Krylov basis as $|\psi(t)\rangle=c_0(t)|K_0\rangle+c_1(t)|K_1\rangle$, where $c_{0,1}$ is the probability amplitude of finding the qubit in the corresponding Krylov basis state. The Krylov complexity of a qubit for the initial state $|\psi_0\rangle$ is obtained as \cite{Aguilar_Gutierrez_Krylov_Not_Distance_PRD2024, Caputa-Density-Matrix2024}
\begin{equation}\label{eq:single-qubit-Krylov-complexity} 
    \mathcal{C}_{\mathcal{K}} (t) = |c_1(t)|^2 = \sin^2 \theta_0 \sin^2 \left( \dfrac{\omega t}{2} \right),
\end{equation}
 satisfying $0 \leq \mathcal{C}_{\mathcal{K}} (t) \leq 1$ with its maximum value of $\sin^2 \theta_0$ and a time-averaged value of $\left(\sin^2 \theta_0 \right)/2$. The maximum value of one is attained when the initial state is an equal superposition of energy eigenstates, i.e., when $\theta_0 = \pi/2$. In terms of Lanczos coefficients, the Krylov complexity becomes,
\begin{equation}\label{eq:two-level-result-using-Lanczos}
    \mathcal{C}_{\mathcal{K}} (t) = \dfrac{4b_1^2}{4b_1^2 + (a_0 - a_1)^2} \sin^2 \left( \dfrac{t\sqrt{4b_1^2 + (a_0 - a_1)^2}}{2} \right),
\end{equation}
which is symmetric in interchange between $a_0$ and $a_1$. Thus, an initial state and the corresponding orthogonal state of a qubit exhibit the same complexity dynamics. If the initial state is stationary or if the two levels are degenerate, $\mathcal{C}_{\mathcal{K}} (t)$ remains zero, and those cases are disregarded from further discussion. Once the energy eigenvalues of the Hamiltonian are kept fixed, the initial state determines the amplitude of the oscillation of $\mathcal{C}_{\mathcal{K}} (t)$, i.e., how far the state is evolved away from the initial state. 

\subsection{Shannon entropy and inverse participation ratio}

\begin{figure}
    \centering
\includegraphics[width=1\linewidth]{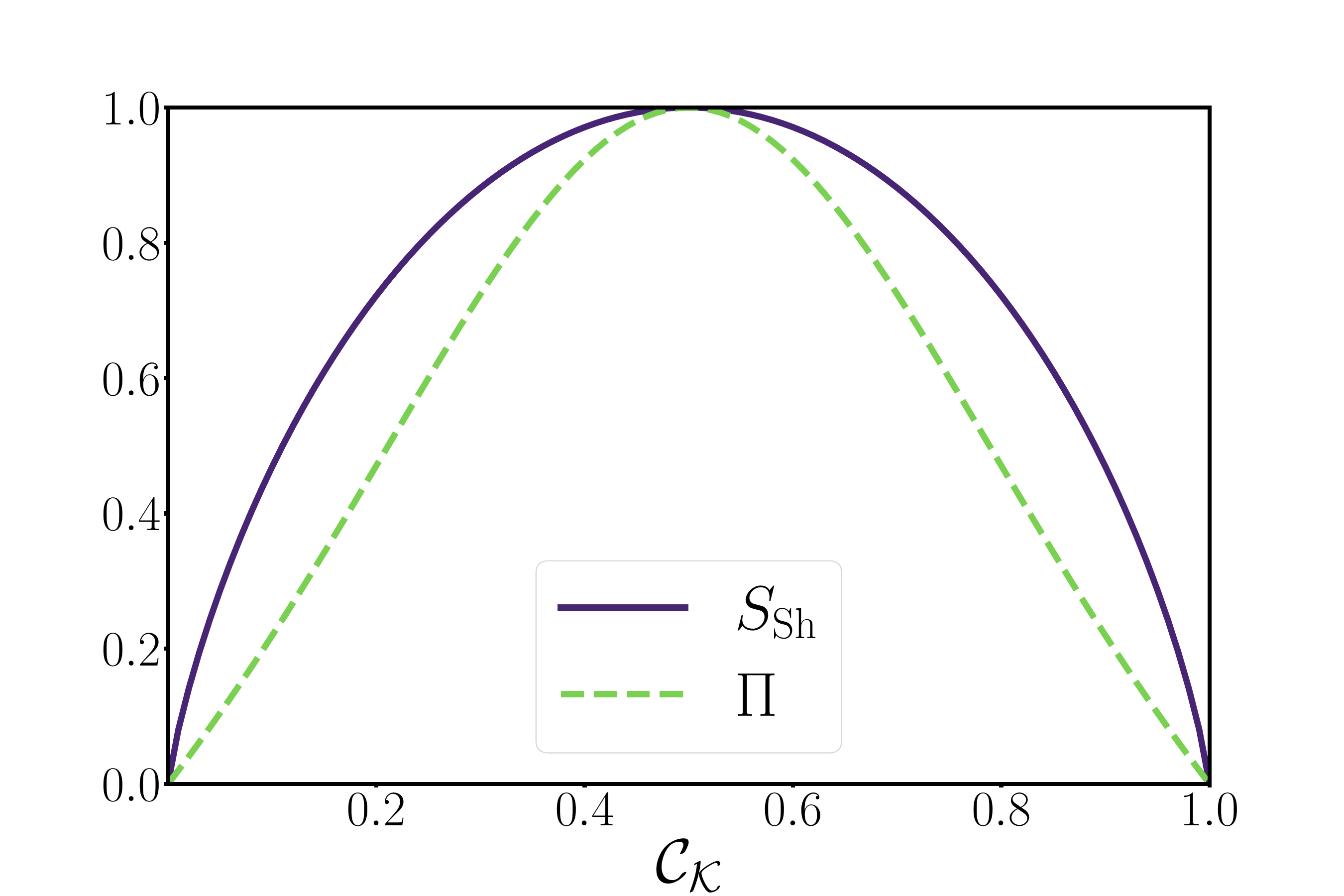}
    \caption{Shannon entropy ($S_{\text{Sh}}$) and the inverse participation ratio ($\Pi$) in the Krylov basis as a function of the Krylov complexity $\mathcal{C}_{\mathcal{K}}$.}   \label{fig:shannon_ipr_as_a_function_of_krylov_complexity}
\end{figure}

Now we look at the Shannon entropy $S_{\rm Sh}(t)$ and the inverse participation ratio $\Pi(t)$ computed using the Krylov basis \cite{Caputa2022-State-Complexity-Original, K-Shannon-Entropy-OperatorComplexity-JHEP2019} and provide the relation with the Krylov complexity. The Shannon entropy determines how a given quantum state is spread over the Hilbert space spanned by the Krylov basis \cite{K-Shannon-Entropy-OperatorComplexity-JHEP2019}, whereas $ \Pi(t)$ is typically used to analyze the localization/delocalization properties of quantum dynamics \cite{PhysRevA.102.022816}. We check if these two quantities exhibit any simple relation with the Krylov complexity. The Shannon entropy and the inverse participation ratio are defined as,
\begin{eqnarray}
S_{{\rm Sh}}= -\sum_n P_n\log_2 P_n \\
    \Pi=\frac{1}{\sum_n P_n^2}-1,
\end{eqnarray}
where $P_n(t)=|\langle\psi(t)|K_n\rangle|^2$. In terms of Krylov complexity, we get,
\begin{eqnarray}
\label{sha}
 S_{\text{Sh}} &=&-\mathcal{C}_{\mathcal{K}} \log_2 \mathcal{C}_{\mathcal{K}} - (1 - \mathcal{C}_{\mathcal{K}}) \log_2 (1 - \mathcal{C}_{\mathcal{K}}) \\
    \Pi&=&\dfrac{1}{2\mathcal{C}_{\mathcal{K}}^2 - 2 \mathcal{C}_{\mathcal{K}} + 1}-1.
    \label{ipr}
\end{eqnarray}
 The dependence of $S_{\text{Sh}}$ and $\Pi$ on $\mathcal{C}_K$ is shown in Fig.~\ref{fig:shannon_ipr_as_a_function_of_krylov_complexity}. When $\mathcal{C}_K=0$ or 1, the system is in a Krylov basis state, and both $S_{\text{Sh}}$ and $\Pi$ vanish, indicating the localization in the Krylov space. When $\mathcal{C}_{\mathcal{K}}=0.5$, they attain the maximum value of unity, indicating the delocalization. Above relations indicate a non-trivial functional dependence of $S_{\rm Sh}(t)$ or $\Pi(t)$ on $\mathcal{C}_{\mathcal{K}}$.

\subsection{Geometrical interpretation}\label{geo-single-qbit}

Here, we establish a geometrical interpretation of the Krylov state complexity for a single qubit. The Krylov complexity itself does not act as a measure of distance between states \cite{Aguilar_Gutierrez_Krylov_Not_Distance_PRD2024}. In particular, it is shown that $\mathcal{C}_{\mathcal{K}}$ and some functions of the Krylov complexity do not satisfy the triangle inequality, i.e. $\mathcal{C}_{\mathcal{K}} (t_3 - t_1) \lneq \mathcal{C}_{\mathcal{K}} (t_3 - t_2) + \mathcal{C}_{\mathcal{K}} (t_2 - t_1)$. Now, we show that in the case of a single qubit, the \textit{square root} of the Krylov complexity does act as a measure of distance between the states connected by unitary time-evolution and satisfies the triangle inequality.

\begin{figure}
    \centering
    \includegraphics[width=1\linewidth]{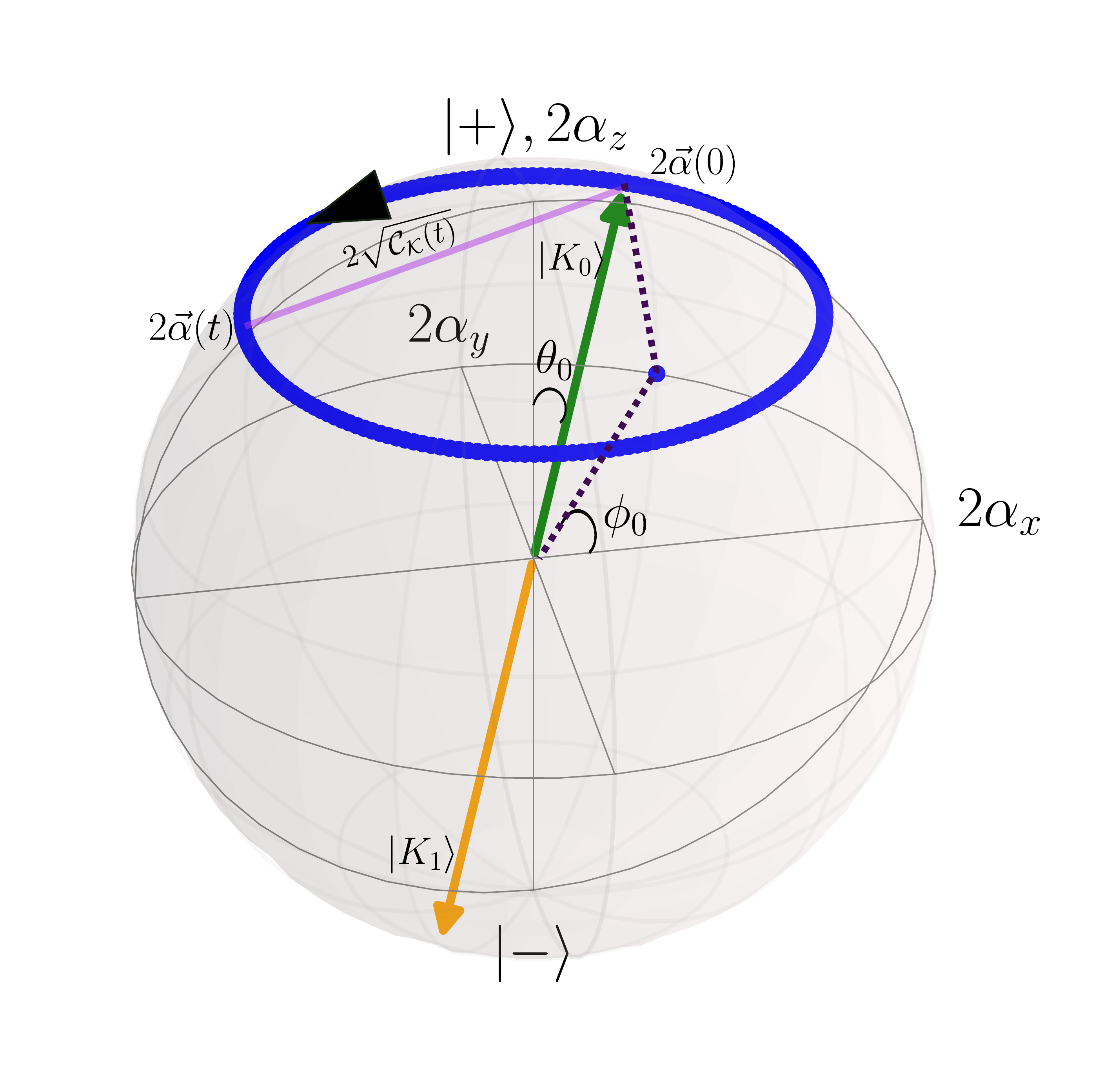}
    \caption{(Color online) Krylov basis states visualized on the Bloch sphere with $|K_0\rangle = \cos (\theta_0/2) |+\rangle + \sin (\theta_0/2) e^{i\phi_0} |-\rangle$ (green) and $|K_1\rangle = \sin (\theta_0/2) |+\rangle - \cos (\theta_0/2) e^{i\phi_0}$ (yellow). For a given $(\theta_0, \phi_0)$, the Bloch sphere coordinates of $|K_1\rangle$ are given by $(\pi - \theta_0, \pi + \phi_0)$. The states on the Bloch sphere map to points on the surface of a sphere of half the radius, spanned by the components of $\vec{\alpha}$. The trajectory of the quantum state forms a circle with $\theta$ determined by the initial state. The Euclidean distance between a time-evolved state at time, $t$ and the initial state is equal to the square root of the Krylov complexity at that time.}
    \label{fig:geom-single-qubit}
\end{figure}

The state vector $|\psi\rangle$ is mapped to a point, $\vec{\alpha}_{|\psi\rangle}$, in a three-dimensional parameter space via the density matrix written in the energy eigenbasis as
\begin{equation}\label{eq:define-alpha-vector}
    \hat \rho (t) = |\psi (t)\rangle \langle \psi (t)| = \dfrac{\mathcal{I}}{2} + \vec{\alpha}_{|\psi\rangle} \cdot \hat {\mathbf{\sigma}},
\end{equation}
where $\mathcal{I}$ is the identity matrix and $\hat{\mathbf{\sigma}}=(\hat \sigma_x, \hat \sigma_y, \hat \sigma_z)^T$, the Pauli spin-1/2 matrices. Henceforth, we drop the subscript $|\psi\rangle$ in $\vec{\alpha}_{|\psi\rangle}$ for convenience. For an initial state $|\psi_0\rangle = \cos (\theta_0/2) |+\rangle + \sin (\theta_0/2) e^{i\phi_0} |-\rangle$ under the unitary evolution governed by the Hamiltonian, $\hat{H} = (\omega/2) \hat{\sigma}_z$, the components of the vector $\vec{\alpha} (t)$ are,
\begin{align}
    \alpha_x &= \dfrac{\sin \theta_0}{2} \cos (\phi_0 + \omega t),\\
    \alpha_y &= \dfrac{\sin \theta_0}{2} \sin (\phi_0 + \omega t), \\
    \alpha_z &= \dfrac{\cos \theta_0}{2}.
\end{align}
 Thus, the time-evolution of an initial state $|\psi_0\rangle$ represents a circular motion in the $\alpha_x-\alpha_y$ plane, while keeping $\theta(t)=\theta_0$, a constant, as shown in Fig. \ref{fig:geom-single-qubit}. Now, we can define the distance between two states as the Euclidean distance between the corresponding points in the $\vec{\alpha}$- space and is obtained as,
\begin{equation}
    |\vec{\alpha} (t_2) - \vec{\alpha} (t_1)|^2 = \sin^2 \theta_0 \sin^2 \left( \dfrac{\omega (t_2 - t_1)}{2} \right) = \mathcal{C}_{\mathcal{K}} (t_2 - t_1),
\end{equation}
where the Krylov complexity is independent of $\phi_0$. Strikingly, the square root of the Krylov complexity can be identified as the magnitude of displacement of the state vector from the initial state.

{\em Triangle inequality}. We now prove that this does satisfy the triangle inequality. In the following, we set $t_{ij} = t_i - t_j$ for brevity. The proof of the triangle inequality then follows from:

\begin{align}
        \sqrt{\mathcal{C}_{\mathcal{K}} (t_{31})} &= \sqrt{\mathcal{C}_{\mathcal{K}} (t_{32} + t_{21})} \nonumber \\
        &= \sin \theta_0 \left|\sin \left(\dfrac{\omega (t_{32} +  t_{21})}{2}\right)\right| \nonumber \\
        &= \sin \theta_0 \left|\sin \left(\dfrac{\omega t_{32}}{2}\right) \cos \left(\dfrac{\omega t_{21}}{2}\right) + \sin \left(\dfrac{\omega t_{21}}{2}\right) \cos \left(\dfrac{\omega t_{32}}{2}\right) \right| \nonumber \\
        &\leq \sin \theta_0 \left( \left|\sin \Bigg(\dfrac{\omega t_{32}}{2}\right) \right| \left| \cos \left(\dfrac{\omega t_{21}}{2}\right) \right| \nonumber \\
        &\hspace{1.5cm} + \left|\sin \left(\dfrac{\omega t_{21}}{2}\right) \right| \left| \cos \left(\dfrac{\omega t_{32}}{2}\right) \right|  \Bigg) \nonumber \\
        &\leq \sin \theta_0 \left( \left|\sin \left(\dfrac{\omega t_{32}}{2}\right) \right| + \left|\sin \left(\dfrac{\omega t_{21}}{2}\right) \right| \right) \nonumber \\
        &= \sqrt{\mathcal{C}_{\mathcal{K}} (t_{32})} + \sqrt{\mathcal{C}_{\mathcal{K}} (t_{21})}.
\end{align}
 Note that this is still consistent with the results shown in \cite{Aguilar_Gutierrez_Krylov_Not_Distance_PRD2024}, where it is argued that $\sqrt{\mathcal{C}_{\mathcal{K}}}$ violates the triangle inequality for three states separated by small time intervals. This is shown in \cite{Aguilar_Gutierrez_Krylov_Not_Distance_PRD2024} by Taylor expanding $\sqrt{\mathcal{C}_{\mathcal{K}} (t)}$ as,

\begin{equation}
    \sqrt{\mathcal{C}_{\mathcal{K}} (t)} \approx b_1 t - \dfrac{\left[ \left(a_0 - a_1\right)^2 + 2 \left(2b_1^2 - b_2^2 \right) \right]}{24} t^3 + \mathcal{O} (t^5).
\end{equation}
Depending on the initial state and its Lanczos coefficients, the coefficient of $t^3$ may be positive, in which case $\sqrt{\mathcal{C}_{\mathcal{K}} (t_1+t_2)} \nleq \sqrt{\mathcal{C}_{\mathcal{K}} (t_1)} + \sqrt{\mathcal{C}_{\mathcal{K}} (t_2)}$, i.e., it is not sub-additive. However, this coefficient is negative for all initial states for the single qubit as $b_2 = 0$. This necessitates looking at higher-order terms in the series expansion, or the function itself, which we have explicitly shown above does satisfy the triangle inequality. 

Finally, we remark that while our mapping to the $\alpha$-parameter space appears to be dependent on our initial choice of basis (the energy eigenbasis), a change in basis only results in mapping to different points in the parameter space and does not affect the displacement, as we show below. We consider a unitary transformation, $|\psi^\prime \rangle = \mathcal{U} |\psi\rangle$, under which the density matrix transforms as

\begin{equation}
\begin{split}
    \rho^\prime (t) = \mathcal{U} \rho (t) \mathcal{U}^\dagger &=\dfrac{1}{2} \cdot \mathcal{I} + \vec{\alpha}^\prime (t) \cdot \hat{\sigma}
\end{split} 
\end{equation}
Now we show that $d{\vec{\alpha}}^{\prime 2} (t)=d\vec{\alpha}^2(t)=\mathcal{C}_{\mathcal{K}} (t)$. We consider the unitary transformation in its general form,
\begin{equation}
    \mathcal{U} = (\cos a) \cdot \mathcal{I} + i (\sin a) (\vec{n} \cdot \hat{\sigma})
\end{equation}
where $\vec{n} \cdot \vec{n} = 1$ and 2$a$ is an angle of rotation about an axis along $\vec n$. By a straightforward, albeit slightly tedious calculation using the properties of Pauli matrices, it can be shown that $\mathcal{U} (\vec{\alpha} \cdot \hat{\sigma}) \mathcal{U}^\dagger = \vec{\alpha}^\prime \cdot \hat{\sigma}$ where
\begin{equation}
    \vec{\alpha}^\prime = (\cos 2a) \Vec{\alpha} - (\sin 2a) (\Vec{n} \times \Vec{\alpha}) + 2 \left(\sin^2 a \right) (\Vec{n} \cdot \Vec{\alpha}) \Vec{n}
\end{equation}
For brevity, we set $\alpha (t_1=0) = \alpha_1$ and $\alpha (t_2 = t) = \alpha_2$ and $\Vec{\alpha}_{21} = \Vec{\alpha}_2 - \Vec{\alpha}_1$, so that $(d\Vec{\alpha})^2 (t) = \alpha_{21}^2$. Then, we have 
\begin{align}
        (d\vec{\alpha}^{\prime})^2 (t) &= \big[ (\cos 2a) \vec{\alpha}_{21} - (\sin 2a) (\vec{n} \times \vec{\alpha}_{21}) + 2 (\sin^2 a) (\vec{n} \cdot \vec{\alpha}_{21}) \vec{n} \big]^2 \nonumber \\
        &= (\cos^2 2a) \alpha_{21}^2 + (\sin^2 2a) \left[\alpha_{21}^2 - \left( \vec{n} \cdot \vec{\alpha}_{21} \right)^2 \right] \nonumber \\
        & + 4 \sin^4 a \left( \vec{n} \cdot \vec{\alpha}_{21} \right)^2 + 4 \sin^2 a \cos 2a \left( \vec{n} \cdot \vec{\alpha}_{21} \right)^2 \nonumber \\ 
        &=(d\vec{\alpha})^2 (t),
\end{align}
and hence, the geometric interpretation is basis-independent. 

We note that our geometrical interpretation of $\mathcal{C}_{\mathcal{K}}$ shares similarities with that of the operator Krylov complexity, discussed for systems exhibiting SU(2) symmetry \cite{building_krylov_complexity_from_circuit_complexity_chenwei_PRR2024}. In the latter case, a Krylov basis comprising $2l+1$ states in the operator space is mapped to the eigenstates of $\hat{J}_z$ with total angular momentum $j=l$. These states are then geometrically represented on a 2-sphere of diameter $2l$, with $|K_0\rangle$ and $|K_{2l}\rangle$ occupying two diametrically opposite poles of the sphere and, in particular, the initial state on the south pole. Any state on the sphere is projected onto a vertical axis whose bottom end corresponds to the initial state. Thus, the Krylov operator complexity, at a time $t$, is obtained as the projected height of the time-evolved operator state on the vertical axis. It is then easy to verify that for a single qubit, by rotating the sphere in Fig.~\ref{fig:geom-single-qubit} by $\pi - \theta_0$ about the $\alpha_y$-axis so that the two Krylov basis states lie at the vertical poles of the sphere, projecting the time-evolved state onto the vertical axis gives $(\vec{\alpha} (t) - \vec{\alpha} (t=0)) \cdot (\vec{\alpha}_{|K_1\rangle} - \vec{\alpha} (t=0)) = \mathcal{C}_{\mathcal{K}} (t)$, where $\vec{\alpha}_{|K_1\rangle}$ is the point corresponding to $|K_1\rangle$ (the north pole of the sphere after rotation). That is, the Krylov complexity is obtained after projecting it onto the axis connecting $|K_0\rangle$ and $|K_1\rangle$. 

Note that this geometric interpretation of $\mathcal{C}_{\mathcal{K}} (t)$ is still consistent with \cite{Aguilar_Gutierrez_Krylov_Not_Distance_PRD2024}, where it has been epxlicitly shown that the single qubit Krylov complexity violates the triangle inequality. Considering the initial state along with two time-evolved states at times $t_1$ and $t_2$, $\mathcal{C}_{\mathcal{K}} (t_1)$ and $\mathcal{C}_{\mathcal{K}} (t_2)$ give the projected height onto the vertical axis (as defined above); however, $\mathcal{C}_\mathcal{K} (t_2-t_1)$ gives the projection onto a different axis that passes through the Krylov states of $|\psi (t_1)\rangle$. Thus, $\mathcal{C}_{\mathcal{K}} (t_1)$, $\mathcal{C}_{\mathcal{K}} (t_2)$ and $\mathcal{C}_{\mathcal{K}} (t_2-t_1)$ do not need to satisfy the triangle inequality.

\subsection{A two level atom}
\label{tla}
At this point, we consider a two-level atom in which the ground state $|g\rangle$ is coupled to the excited state $|e\rangle$ by a light field of Rabi frequency $\Omega$ and with a detuning $\Delta$, described by the Hamiltonian,
\begin{equation}\label{hams}
    \hat{H} = - \Delta \hat\sigma_{ee} + \dfrac{\Omega}{2} \hat\sigma_x,
\end{equation}
where the operator $\hat \sigma_{ab}=|a\rangle \langle b|$ with $a, b\in \{g, e\}$ and $\hat\sigma_x=\hat\sigma_{eg}+\hat\sigma_{ge}$. The energy eigenvalues of $\hat{H}$ are $E_\pm = \pm \sqrt{\Delta^2 + \Omega^2}/2$. In terms of the energy eigenstates $|\pm\rangle$, the bare states are,
    $|e\rangle = \sin (\theta/2) |+\rangle + \cos (\theta/2) |-\rangle$ and $|g\rangle = \cos (\theta/2) |+\rangle - \sin (\theta/2) |-\rangle$,
where $\cos (\theta/2) = [(\sqrt{\Delta^2 + \Omega^2} + \Delta)/(2\sqrt{\Delta^2 + \Omega^2})]^{1/2}$. If we take either $|g\rangle$ or $|e\rangle$ as the initial state, the Krylov complexity is the same and is
\begin{equation}
 \mathcal{C}_{\mathcal{K}}(t) = \dfrac{\Omega^2}{\Delta^2 + \Omega^2} \sin^2 \left( \dfrac{\sqrt{\Delta^2 + \Omega^2}t}{2}\right),
\end{equation}
which characterizes the well known Rabi oscillations. Using $\mathcal{C}_{\mathcal{K}}(t)$, we can compute the Shannon entropy and the inverse participation ratio using Eqs.~(\ref{sha}) and (\ref{ipr}) as well.

\section{Two Qubits}\label{sec:two_qubits}
Now, we extend the calculations to two qubits. In particular, we consider both uncoupled and coupled qubits. For the latter, we take the case of a pair of interacting two-level Rydberg atoms.

\subsection{Non-interacting qubits}\label{sec:two_ni_qubits}
\begin{figure}
    \centering
    \includegraphics[width=1\linewidth]{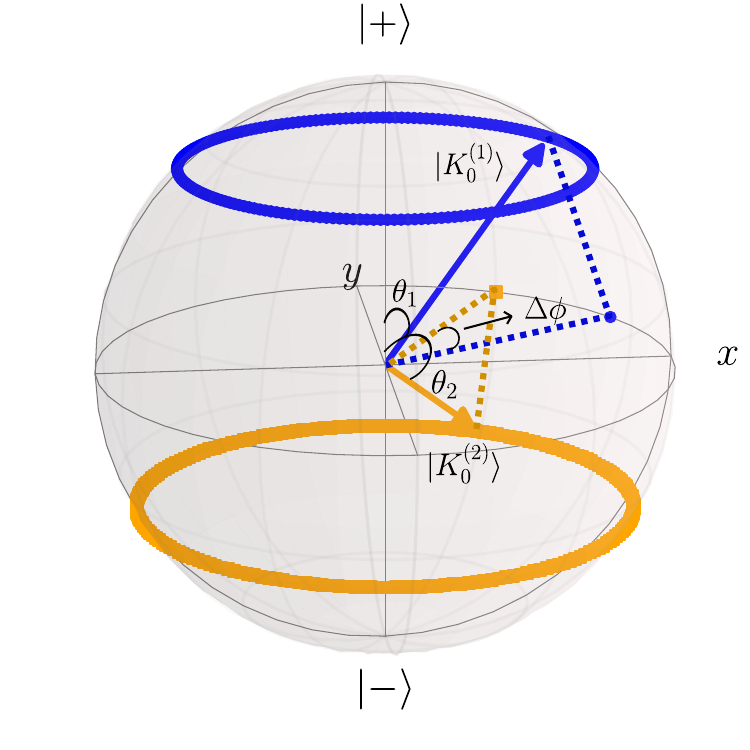}
    \caption{The trajectories of two initial single-qubit states, $|K_0^{(1)} \rangle$ and $|K_0^{(2)} \rangle$, visualized on the Bloch sphere, with $|K_0^{(i)} \rangle = \cos (\theta_i/2) |+\rangle_i + \sin (\theta_i/2) e^{i\phi_i} |-\rangle_i$, with $\Delta \phi = \phi_2 - \phi_1$. For $\omega_1$ = $\omega_2$, $\mathcal{F} \rightarrow 0$ when $\theta_1 = \theta_2$, i.e. both the single-qubit states lie on the same disc of constant $\theta$.}
    \label{fig:geom-two-qubit}
\end{figure}

In the following, we discuss the Krylov complexity of a pair of non-interacting qubits governed by the Hamiltonian, $\hat{H}=\hat{H}_1 + \hat{H}_2$, where $\hat{H}_1$ $\left(\hat{H}_2\right)$ is the Hamiltonian of the first (second) qubit. Let the energy eigenstates of $H_i$ be $|\pm\rangle_i$ and eigenvalues $E_{\pm}^{(i)}=\omega_i/2$. Considering a general initial product state for the two qubits: $(\cos (\theta_1/2) |+\rangle_1 + \sin (\theta_1/2) e^{i\phi_1} |-\rangle_1) \otimes (\cos (\theta_2/2) |+\rangle_2 + \sin (\theta_2/2) e^{i\phi_2} |-\rangle_2)$, we obtain the Krylov complexity as,
\begin{equation}\label{eq:non-interacting-qubits-gen-krylov}
    \mathcal{C}_{\mathcal{K}} (t) = \mathcal{C}_{\mathcal{K}}^{(1)} (t) + \mathcal{C}_{\mathcal{K}}^{(2)} (t) + \mathcal F\big(t\big)
\end{equation}
where $\mathcal{C}_{\mathcal{K}}^{(i)} (t)$ is the Krylov complexity of the individual qubit given in Eq.~(\ref{eq:single-qubit-Krylov-complexity}), and  

\begin{widetext}
    \begin{equation}
        \begin{split}
            \mathcal{F} (t) &= \dfrac{(\omega_1 \cos \theta_1 - \omega_2 \cos \theta_2)^2}{\omega_1^2 \sin^2 \theta_1 + \omega_2^2 \sin^2 \theta_2} \mathcal{C}_{\mathcal{K}}^{(1)} (t) \mathcal{C}_{\mathcal{K}}^{(2)} (t) \\
            &\hspace{0.5cm} + \left[ \dfrac{\sin^2 \theta_1 \sin^2 \theta_2 [(\omega_1 \cos \theta_1 - \omega_2 \cos \theta_2)^2 + 2 (\omega_1^2 \sin^2 \theta_1 + \omega_2^2 \sin^2 \theta_2)]}{(\omega_1^2 + \omega_2^2 - 2\omega_1 \omega_2 \cos \theta_1 \cos \theta_2)(\omega_1^2 \sin^2 \theta_1 + \omega_2^2 \sin^2 \theta_2)} \right]\left[\dfrac{\omega_-}{2} \sin \left(\dfrac{\omega_+t}{2} \right) - \dfrac{\omega_+}{2} \sin \left(\dfrac{\omega_-t}{2} \right) \right]^2,
        \end{split}
    \end{equation}
\end{widetext}
where $\omega_{\pm}=\omega_1\pm\omega_2$. See Fig.~\ref{fig:geom-two-qubit} for the Bloch sphere representation of the product state of two independent qubits as they evolve independently on the Bloch sphere with two circular trajectories. The Krylov basis states and Lanczos coefficients are calculated in Appendix \ref{appendix:two-non-interacting-qubits}.

\begin{figure}
    \centering
\includegraphics[width=0.95\linewidth]{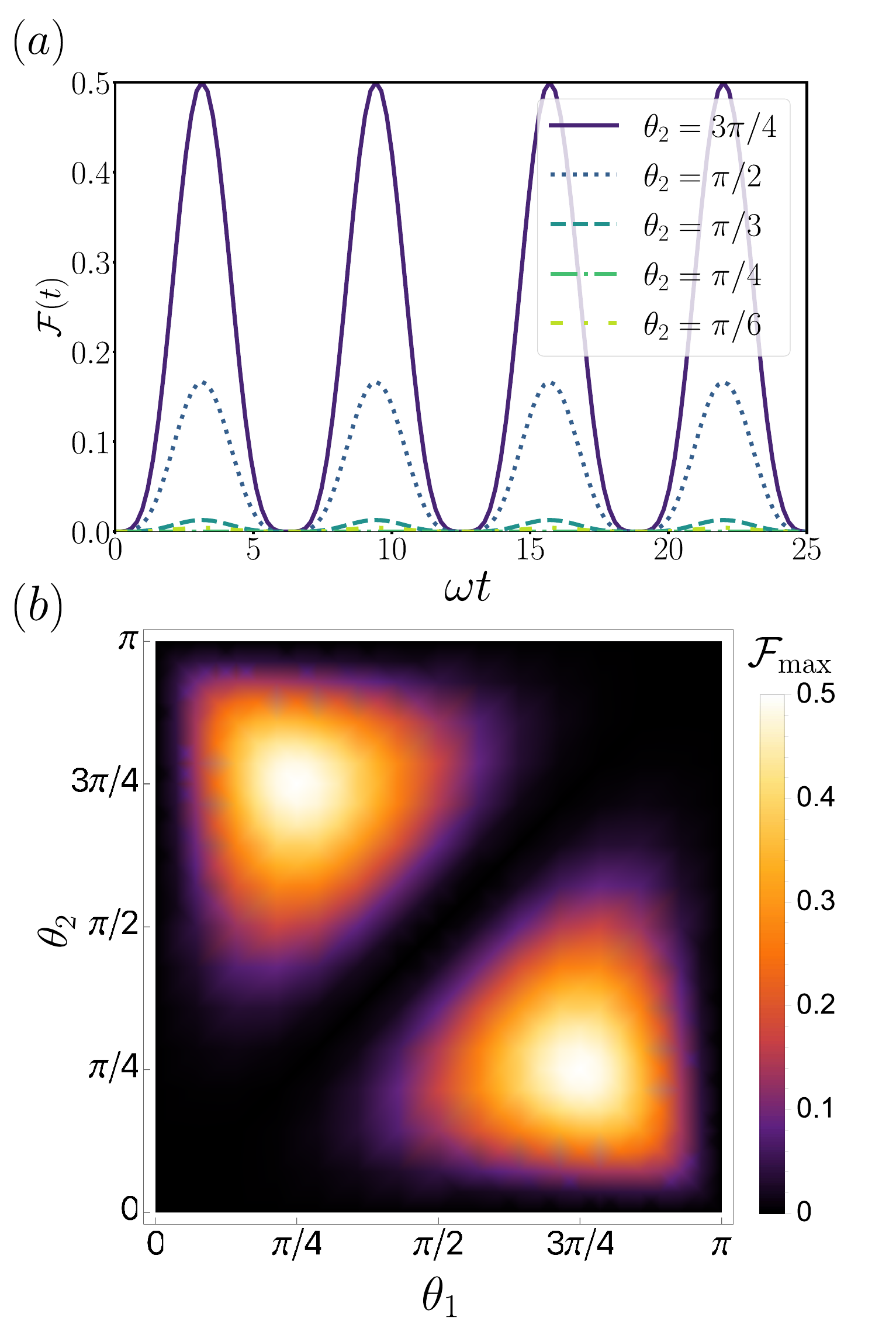}
    \caption{The behaviour of $\mathcal{F} (t)$ when $\omega_1 = \omega_2 = \omega$. (a) $\mathcal{F} (t)$ vs $\omega t$ for  different values of $\theta_2$, with $\theta_1 = \pi/4$. (b) The amplitude of $\mathcal{F} (t)$ in $\theta_1-\theta_2$ plane.}
\label{fig:F_max_theta1_theta2_iden}
\end{figure}

\begin{figure}
    \centering
    \includegraphics[width=1\linewidth]{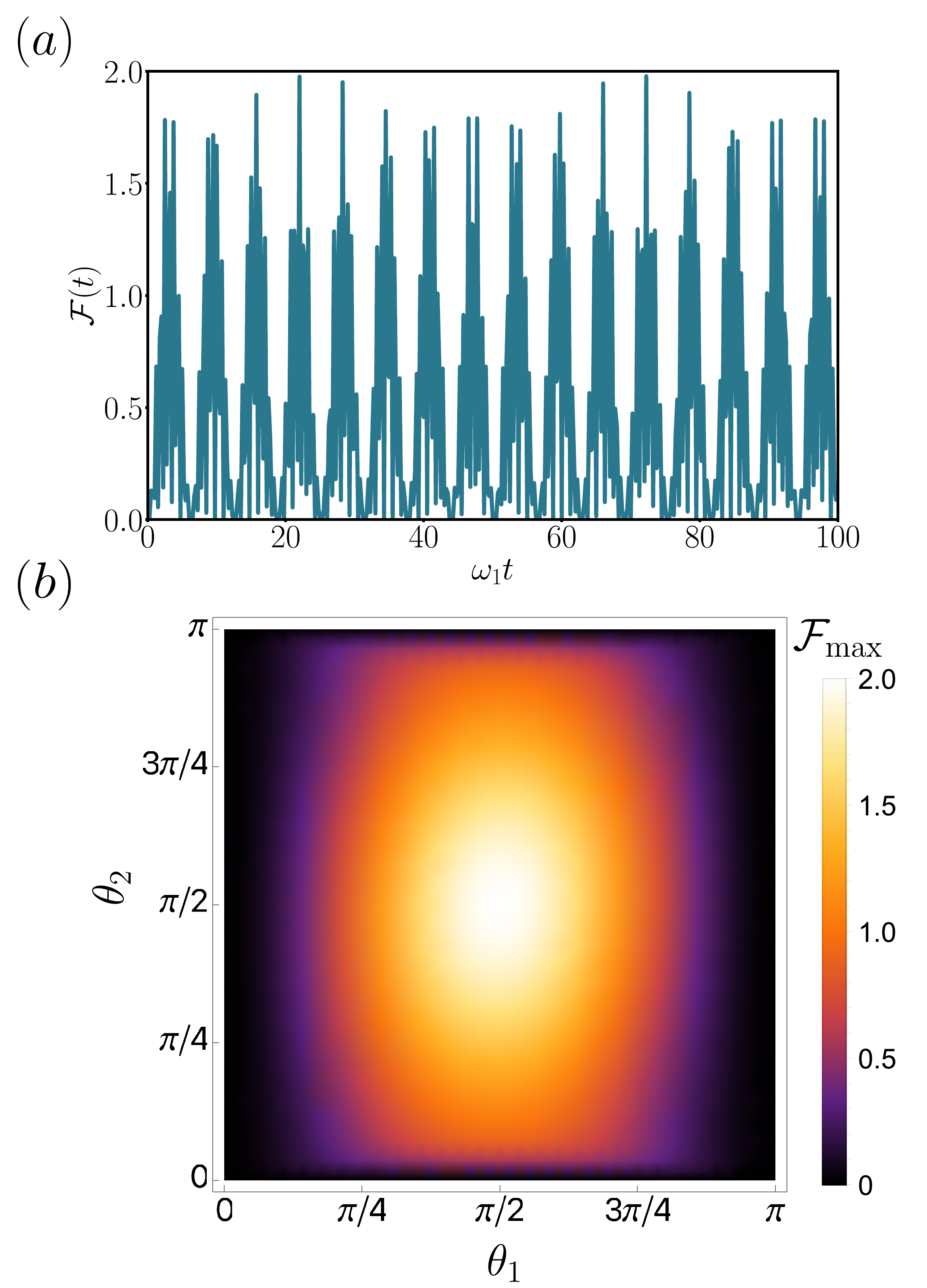}
    \caption{The behaviour of $\mathcal{F} (t)$ for $\omega_2 = 10\omega_1$. (a) $\mathcal{F} (t)$ vs $\omega_1 t$ for $\theta_1 = \theta_2 = \pi/2$. (b) The amplitude of $\mathcal{F} (t)$ in $\theta_1-\theta_2$ plane.}
\label{fig:F_gen_behaviour_w1_notequal_w2}
\end{figure}

As we find, the Krylov complexity is the sum of the Krylov complexity of each qubit and an additional term, $\mathcal F$, which is positive-valued at any instant $t$. If $\omega_1 = \omega_2$, the function $\mathcal F$ simplifies to
\begin{equation}\label{eq:krylov-comp-id-Ham-1}
    \mathcal{F} (t)= \Bigg[\dfrac{(\cos \theta_1 - \cos \theta_2)^2}{\sin^2 \theta_1 + \sin^2 \theta_2} \Bigg] \mathcal{C}_{\mathcal{K}}^{(1)} (t) \mathcal{C}_{\mathcal{K}}^{(2)} (t),
\end{equation}
which exhibits periodic oscillations and vanishes when $\theta_1=\theta_2$, as shown in Fig.~\ref{fig:F_max_theta1_theta2_iden}(a). The amplitude or the maximum of $\mathcal{F} (t)$ as a function of $\theta_1$ and $\theta_2$ for $\omega_1=\omega_2=\omega$ is shown in Fig.~\ref{fig:F_max_theta1_theta2_iden}(b), and is independent of $\omega$. The maximum amplitude is 0.5 when $\theta_1=\pi/4$ and $\theta_2=3\pi/4$ or vice versa. For $\omega_1\neq\omega_2$, $\mathcal{F} (t)$ is still periodic but exhibits more complex behaviour as shown in Fig.~\ref{fig:F_gen_behaviour_w1_notequal_w2}(a) and the maximum of the peak value of $\mathcal{F} (t)$ occurs when $\theta_1=\theta_2=\pi/2$.

For vanishing $\mathcal{F} (t)$, the geometric interpretation for the single qubit can be straightforwardly extended to the case of two non-interacting qubits. As the Krylov complexity, in this case, is simply the sum of the complexities of the individual qubits, the total complexity can be viewed simply as a restatement of the Pythagoras theorem. The individual complexities can represent squared distances in orthogonal directions, and the total complexity can represent the hypotenuse squared. However, when $\mathcal{F} (t)\neq 0$, the square root of the Krylov complexity generally does not satisfy the triangle inequality. Likewise, for $\omega_1 = \omega_2$ and identical initial states for both the qubits [i.e. for $\mathcal{F}(t)=0$], the geometric interpretation found in \cite{building_krylov_complexity_from_circuit_complexity_chenwei_PRR2024} can similarly be applied here taking $l=1$. However, such an interpretation does not seem possible in general when $\mathcal{F} (t) \neq 0$.

{\em Two-level atoms}. For a pair of two non-interacting two-level atoms coupled by light fields, the Hamiltonian is
\begin{equation}
    \hat{H} = \sum\limits_{i=1}^2 \Big[ - \Delta_i \hat{\sigma}^i_{ee} + \dfrac{\Omega_i}{2} \hat{\sigma}_x^i \Big].
\end{equation}
The two atom bare states are $|gg\rangle$, $|ge\rangle$, $|eg\rangle$ and $|ee\rangle$. For a global driving, i.e., when $\Delta_1 = \Delta_2 = \Delta$ and $\Omega_1 = \Omega_2 = \Omega$, only the symmetric state $|+\rangle=(|ge\rangle + |eg\rangle)/\sqrt{2}$ is relevant to the dynamics and the antisymmetric state $|-\rangle=(|ge\rangle - |eg\rangle)/\sqrt{2}$ can be disregarded. In that case, we obtain the Krylov complexity $\mathcal{C}_{\mathcal{K}}^{\alpha} (t)$ for an initial state $|\alpha\rangle$ as, 
\begin{align}
    \mathcal{C}_{\mathcal{K}}^{gg} (t) = \dfrac{2\Omega^2}{\Delta^2 + \Omega^2} &\sin^2 \left(\dfrac{\sqrt{\Delta^2 + \Omega^2}t}{2} \right), \label{eq:rydberg-noninteracting-ggrr} \\
    \mathcal{C}_{\mathcal{K}}^{ge} (t) = \dfrac{2\Omega^2}{\Delta^2 + \Omega^2} &\sin^2 \left( \dfrac{\sqrt{\Delta^2 + \Omega^2} t}{2} \right) \nonumber \\  &+ \dfrac{2\Delta^2 \Omega^2}{\big(\Delta^2 + \Omega^2\big)^2} \sin^4 \left( \dfrac{\sqrt{\Delta^2+\Omega^2}t}{2} \right),
\end{align}    
$\mathcal{C}_{\mathcal{K}}^{ee} (t)=\mathcal{C}_{\mathcal{K}}^{gg}(t)$ and $\mathcal{C}_{\mathcal{K}}^{eg} (t)=\mathcal{C}_{\mathcal{K}}^{ge}(t)$. For the symmetric state, we get 
\begin{eqnarray}
    \mathcal{C}_{\mathcal{K}}^+ (t) = \dfrac{\Omega^2}{\Delta^2 + \Omega^2} \sin^2 \left(\sqrt{\Delta^2 + \Omega^2}t \right) + \nonumber \\ \dfrac{8\Omega^2 \Delta^2}{(\Delta^2 + \Omega^2)^2} \sin^4 \left(\dfrac{\sqrt{\Delta^2 + \Omega^2} t}{2} \right),
\end{eqnarray}
which cannot be expressed in terms of $\mathcal{C}_{\mathcal{K}}^{eg} (t)$ and $\mathcal{C}_{\mathcal{K}}^{ge} (t)$. Although in a single qubit, the Krylov complexity is identical for initial bare states $|g\rangle$ and $|e\rangle$, in a pair of qubits, despite being entirely uncoupled, $\mathcal{C}_{\mathcal{K}}$ is found to depend on the initial state, i.e. whether the two qubits are initially in the same single-qubit states or not. Such a dependence can be attributed to the function $\mathcal F(t)$ in Eq.~(\ref{eq:krylov-comp-id-Ham-1}). When $\Delta=0$, all states are degenerate, $\mathcal F(t)$ vanishes and the complexity simplifies to $\mathcal{C}_{\mathcal{K}}^{gg} = \mathcal{C}_{\mathcal{K}}^{ge} = 2 \sin^2 (\Omega t/2)$ and  $\mathcal{C}_{\mathcal{K}}^+ = \sin^2 (\Omega t)$.

\subsection{Interacting qubits: a Rydberg atom pair}
\label{iqs}
At this point, we consider the coupling between the qubits, and in particular, we study the case of a pair of interacting two-level Rydberg atoms. Henceforth, we take $\Delta_i=0$. When both qubits are in the Rydberg state $|e\rangle$, they interact repulsively, and the governing Hamiltonian in the interaction picture is,
\begin{equation}
    \hat{H} = \sum\limits_{i=1}^2  \dfrac{\Omega_i}{2} \hat{\sigma}_x^i + V_0 \hat{\sigma}^1_{ee} \hat{\sigma}^2_{ee}
\end{equation}
where $V_0 = C_6/R^6$ is the interaction strength, and $R$ is the distance between the two Rydberg atoms.    


\subsubsection{Rydberg blockade}
\label{RB}
\begin{figure}
    \centering
    \includegraphics[width=1\linewidth]{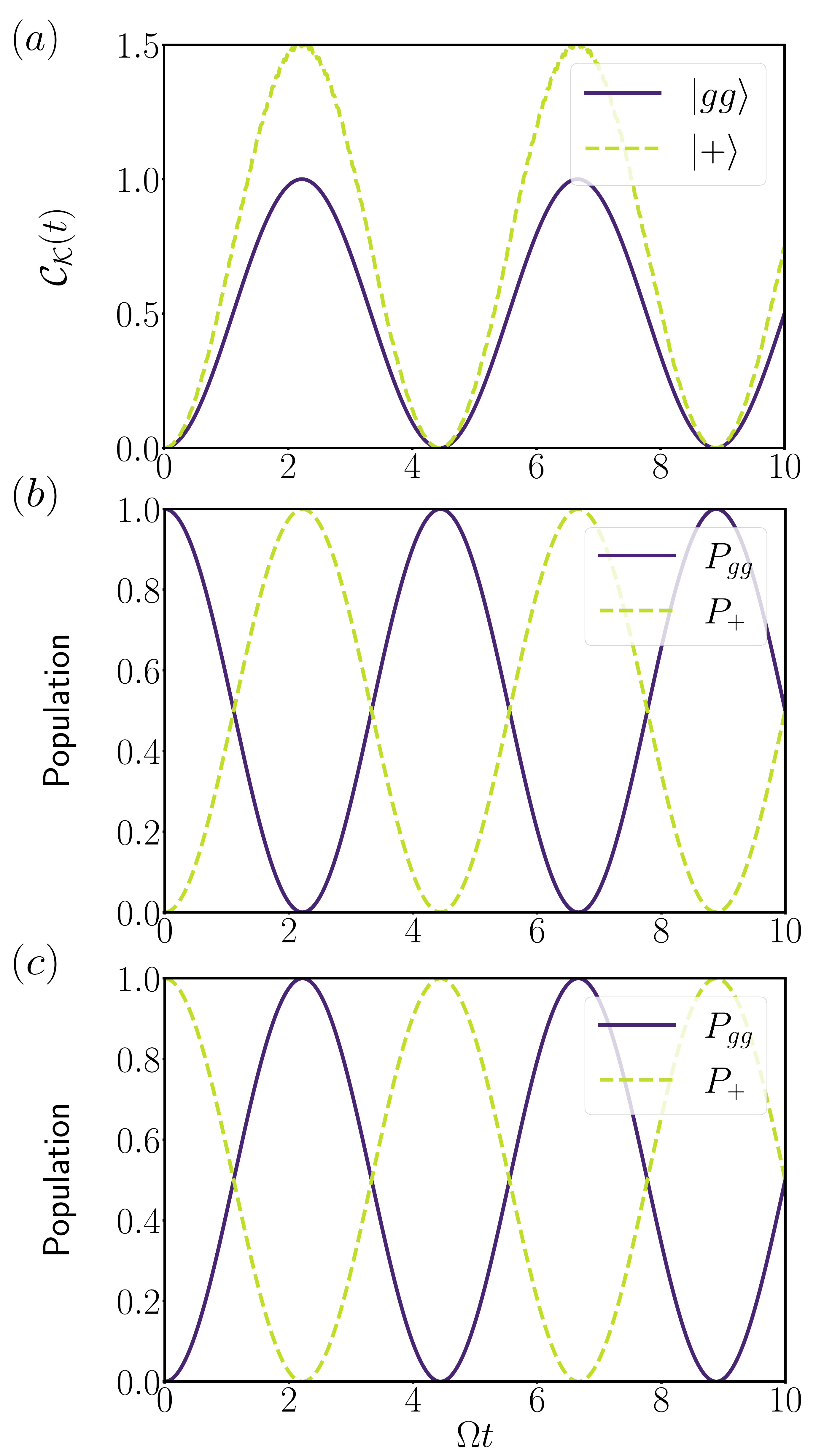}
    \caption{(a) Krylov complexity as a function of time for different initial states and $V_0 = 100 \Omega$. (b) Population in the states $|gg\rangle$ and $|+\rangle$ as a function of time for initial state $|gg\rangle$. (c) Population in the states $|gg\rangle$ and $|+\rangle$ as a function of time for initial state $|+\rangle$.}
    \label{fig:blockade-V100}
\end{figure}

For $V_0\gg\Omega$, the system exhibits coherent Rabi oscillations between the states $|gg\rangle$ and $|+\rangle$ with an enhanced Rabi frequency of $\sqrt{2}\Omega$, resulting in an effective two-level atom, or the ``superatom" \cite{Rydberg-Blockade-Jaksch-PRL2000, Rydberg-Blockade-Lukin-PRL2001, Coherent-Collective-Excitation-Rydberg-Blockade-PRL2007, Rydberg-Blockade-Exp-NatPhys2009, Collective-Excitation-NatPhys2009, Entanglement-Rydberg-Blockade-PRL2010, Quantum-Info-Rydberg-Review-Saffman2010, Many-Body-Rabi-Oscillations-Exp-NatPhys2012, Rydberg_Review_Rejish_2024}. The doubly excited state $|ee\rangle$ is completely inhibited in the dynamics and is termed the Rydberg blockade. We would, therefore, expect the Krylov state complexity of the two initial states, $|gg\rangle$ and $|+\rangle$, to be identical and resemble that of a two-level system, with the complexity peaking at one and exhibiting oscillations at the enhanced Rabi frequency. However, as we show in Fig.~\ref{fig:blockade-V100}(a), the dynamics of $\mathcal{C}_{\mathcal{K}} (t)$ is different for these two initial states, and in particular, the amplitude of oscillation of $\mathcal{C}_{\mathcal{K}} (t)$ for initial state $|+\rangle$ exceeds one, although the population dynamics is effectively identical for both initial states as shown in Figs.~\ref{fig:blockade-V100}(b) and \ref{fig:blockade-V100}(c). As shown below, they differ since the Krylov basis differs for the two initial states. 

For the initial state $|gg\rangle$, the Krylov basis, apart from $|gg\rangle$, consists of,
\begin{align}
\label{kg1}
    |K_1\rangle^{gg} &= \dfrac{1}{\sqrt{2}} \big(|ge\rangle + |eg\rangle\big) = |+\rangle \\
    \label{kg2}
    |K_2\rangle^{gg} &= |ee\rangle \\
    |K_3\rangle^{gg} &= 0,
\end{align}
with Lanczos coefficients $a_0 = a_1 = 0$, $a_2 = V_0$, and $b_1 = b_2 = \Omega/\sqrt{2}$. Note that only the first two Krylov basis states ($|K_0\rangle^{gg}$ and $|K_1\rangle^{gg}$) are relevant in the blockade regime. In that case, we can use Eq. (\ref{eq:two-level-result-using-Lanczos}) and obtain the Krylov complexity as $\mathcal{C}_{\mathcal{K}} (t) = \sin^2 \left( \dfrac{\Omega^\prime t}{2} \right)$ with $\Omega^\prime = \sqrt{2} \Omega$, the enhanced Rabi frequency. 

In contrast, for the initial state $|+\rangle$, the Krylov basis comprises of $|K_0\rangle^{+}=|+\rangle$ and,
\begin{align}
    |K_1\rangle^{+} &= \dfrac{1}{\sqrt{2}} \big(|gg\rangle + |ee\rangle \big) \label{eq:K1_plus_blockade} \\
    |K_2\rangle^{+} &= \dfrac{1}{\sqrt{2}} \big(-|gg\rangle + |ee\rangle \big) \\
    |K_3\rangle^{+} &= 0,
\end{align}
 with Lanczos coefficients $a_0 = 0$, $a_1 = a_2 = V_0/2$, $b_1 = \Omega$, and $b_2 = V_0/2$. Here, the first three Krylov basis states are needed to describe the blockade dynamics, indicating that $\mathcal{C}_{\mathcal{K}}(t)$ can go beyond the value of one, as shown by the dashed line in Fig.~\ref{fig:blockade-V100}(a). 

Considering the similar population dynamics observed for the initial states $|gg\rangle$ and $|+\rangle$,  one would expect identical Lanczos coefficients and Krylov complexity, as discussed in section \ref{section:Krylov-state-complexity}. Using an ordered basis $\mathcal B$ consisting of  $\{|+\rangle, |gg\rangle,|ee\rangle\}$, the spread complexity ($\mathcal{C}_{\mathcal{B}}^{+}$) dynamics starting from $|+\rangle$ is found to be  effectively identical to the Krylov complexity obtained for the initial state $|gg\rangle$, i.e., it oscillates between zero and one. It implies that for the initial state $|+\rangle$, another ordered basis exists in which the amplitude of the spread complexity is one and the average spread is minimized over long times, but not the Krylov basis constructed using the original Hamiltonian. As we show later, the newly ordered basis $\mathcal B$ which minimizes the average spread complexity over long times, can be obtained as the Krylov basis resulting from a zeroth order effective Hamiltonian (along with $|ee\rangle$ to complete the basis). 
 
The Krylov complexity attains an even larger amplitude ($\sim 3$) under Rydberg blockade [see the dashed line in Fig.~\ref{fig:krylov-popu-ge-V100}(a)] if the initial state is $|eg\rangle$ or $|ge\rangle$. As we see below, all four Krylov states become relevant to characterize the dynamics, resulting in an amplitude of three. The population dynamics shown in Fig.~\ref{fig:krylov-popu-ge-V100}(b) is characterized by the periodic oscillations between $|ge\rangle$ and $|eg\rangle$ via $|gg\rangle$. This implies that we can choose an ordered basis $\{|ge\rangle, |gg\rangle, |eg\rangle, |ee\rangle\}$ such that the spread complexity amplitude is two [see the solid line in Fig.~\ref{fig:krylov-popu-ge-V100}(a)], making it less than that of the Krylov complexity. For the initial state $|K_0\rangle = |ge\rangle$, we get
\begin{align}
    |K_1\rangle^{ge} &= \dfrac{1}{\sqrt{2}} \big(|gg\rangle + |ee\rangle \big) \label{eq:K1_ge_blockade} \\
    |K_2\rangle^{ge} &= \dfrac{1}{\sqrt{2} \Omega_V} \big(-V_0 |gg\rangle + 2\Omega |eg\rangle + V_0 |ee\rangle \big) \\
    |K_3 \rangle^{ge} &= \dfrac{1}{\Omega_V} \big(-\Omega |gg\rangle - V_0 |eg\rangle + \Omega |ee\rangle,
\end{align}
where $\Omega_V = \sqrt{2\Omega^2 + V_0^2}$. The corresponding Lanczos coefficients are $a_0 = 0$, $a_1 = V_0/2$, $a_2 = V_0^3/(2\Omega_V^2)$, $b_1 = \Omega/\sqrt{2}$, $b_2 = \Omega_V/2$ and $b_3 = V_0^2\Omega/(\sqrt{2}\Omega_V^2)$. All four Krylov basis states have significant projections to $|gg\rangle$, $|eg\rangle$, or $|ge\rangle$, making them all relevant for the blockade dynamics and resulting in a larger Krylov complexity. Instead, using the Krylov basis of an effective Hamiltonian that describes the dynamics in the truncated space minimizes the complexity amplitude. For instance, in the perfect blockade scenario, where the population in $|ee\rangle$ is zero, we can arrive at an effective Hamiltonian (zeroth order) for $V_0\gg\Omega$ \cite{Rydberg-Biased-Freezing-IOP2019}, 
 \begin{equation}
   \hat{H}_{\rm eff} = \dfrac{\Omega}{2}\left(\hat{\sigma}_{gg}^1\hat{\sigma}_{x}^2+\hat{\sigma}_{x}^1\hat{\sigma}_{gg}^2\right), 
 \end{equation}
describing the dynamics in the remaining Hilbert space. For the initial state $|ge\rangle$, using the effective Hamiltonian, we obtain the Krylov basis as $\{|ge\rangle, |gg\rangle, |eg\rangle\}$, and to complete the basis, we add $|ee\rangle$ to them. The Krylov complexity is obtained as $\mathcal{C}_{\mathcal{B}}^{ge}(t) = 2 \sin^2 \left(\Omega t/2\sqrt{2} \right)$, which now has an amplitude of two, instead of three. Note that in the absence of blockade ($V_0\ll \Omega$), irrespective of initial bare states, the complexity exceeds one.

\begin{figure}
    \centering
    \includegraphics[width=1\linewidth]{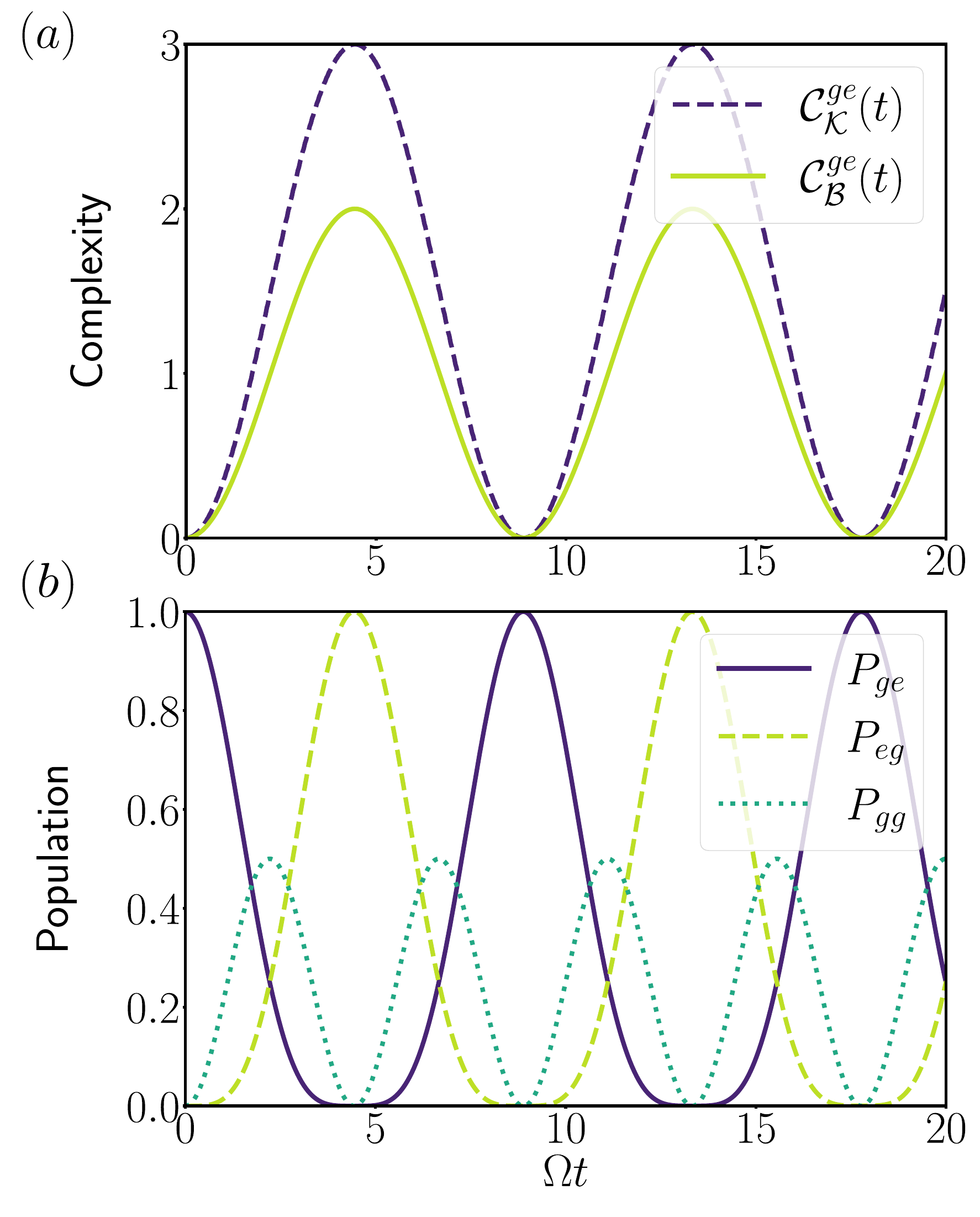}
    \caption{State complexity and population dynamics of initial state $|ge\rangle$ with $V_0 = 100 \Omega$. (a) State complexity calculated using the Krylov basis ($\mathcal{C}_{\mathcal{K}}^{ge}$, dashed line) and the ordered basis, $\mathcal{B} = \{|ge\rangle, |gg\rangle, |eg\rangle\}$,  obtained as the Krylov basis of $|ge\rangle$ using the effective Hamiltonian,  denoted as $\mathcal{C}_{\mathcal{B}}^{ge}$ (solid line). (b) Population dynamics in $|ge\rangle$,  $|gg\rangle$, and  $|eg\rangle$.}
    \label{fig:krylov-popu-ge-V100}
\end{figure}

\begin{figure}
    \centering
    \includegraphics[width=1\linewidth]{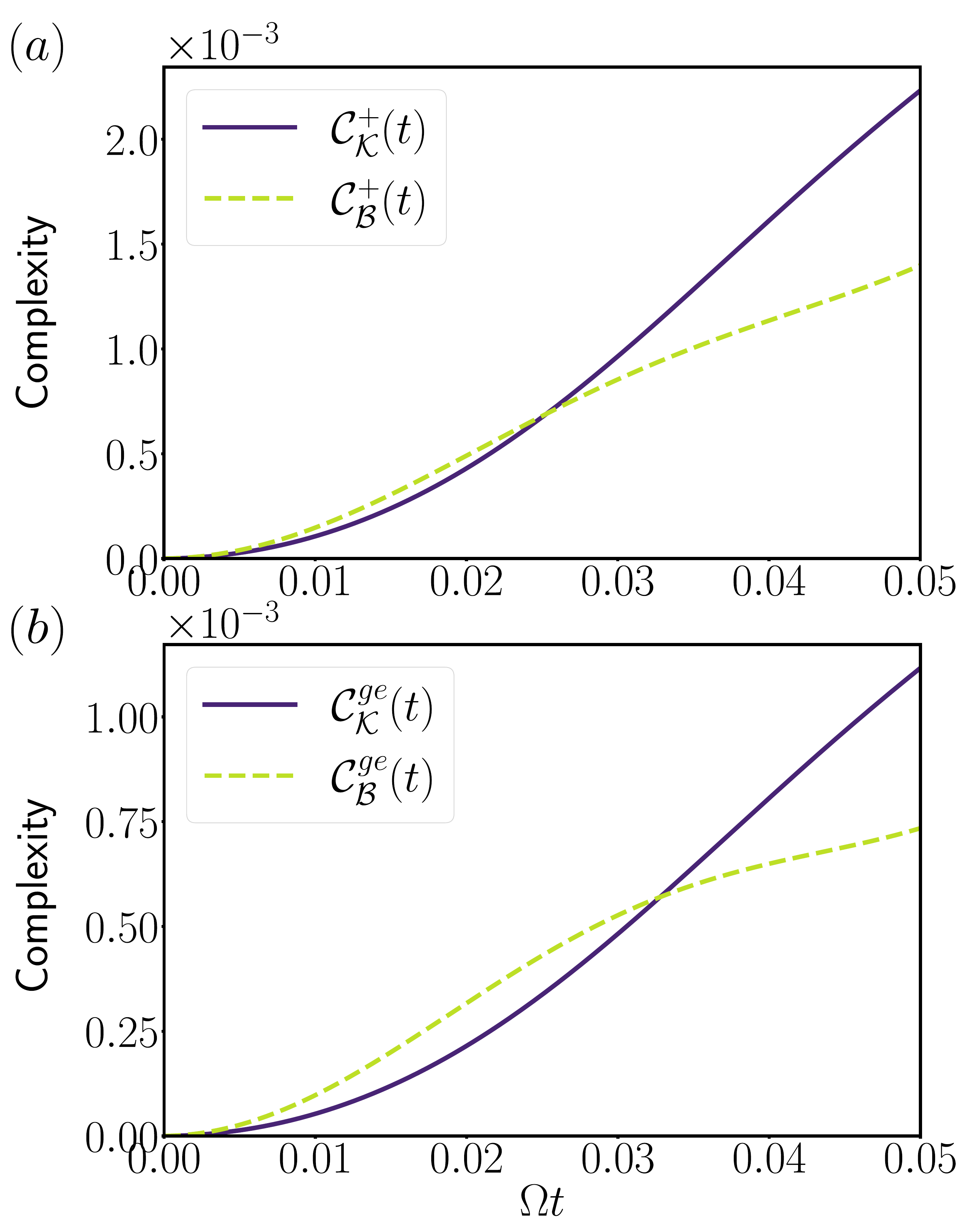}
    \caption{Evolution of the spread complexity in the bases $\mathcal{K}$ and $\mathcal{B}$ at early times for $V_0 = 100 \Omega$ and the initial states (a) $|+\rangle$, and (b) $|ge\rangle$.}
    \label{fig:krylov-comp-V100-early-times}
\end{figure}

\begin{figure}
    \centering
    \includegraphics[width=1\linewidth]{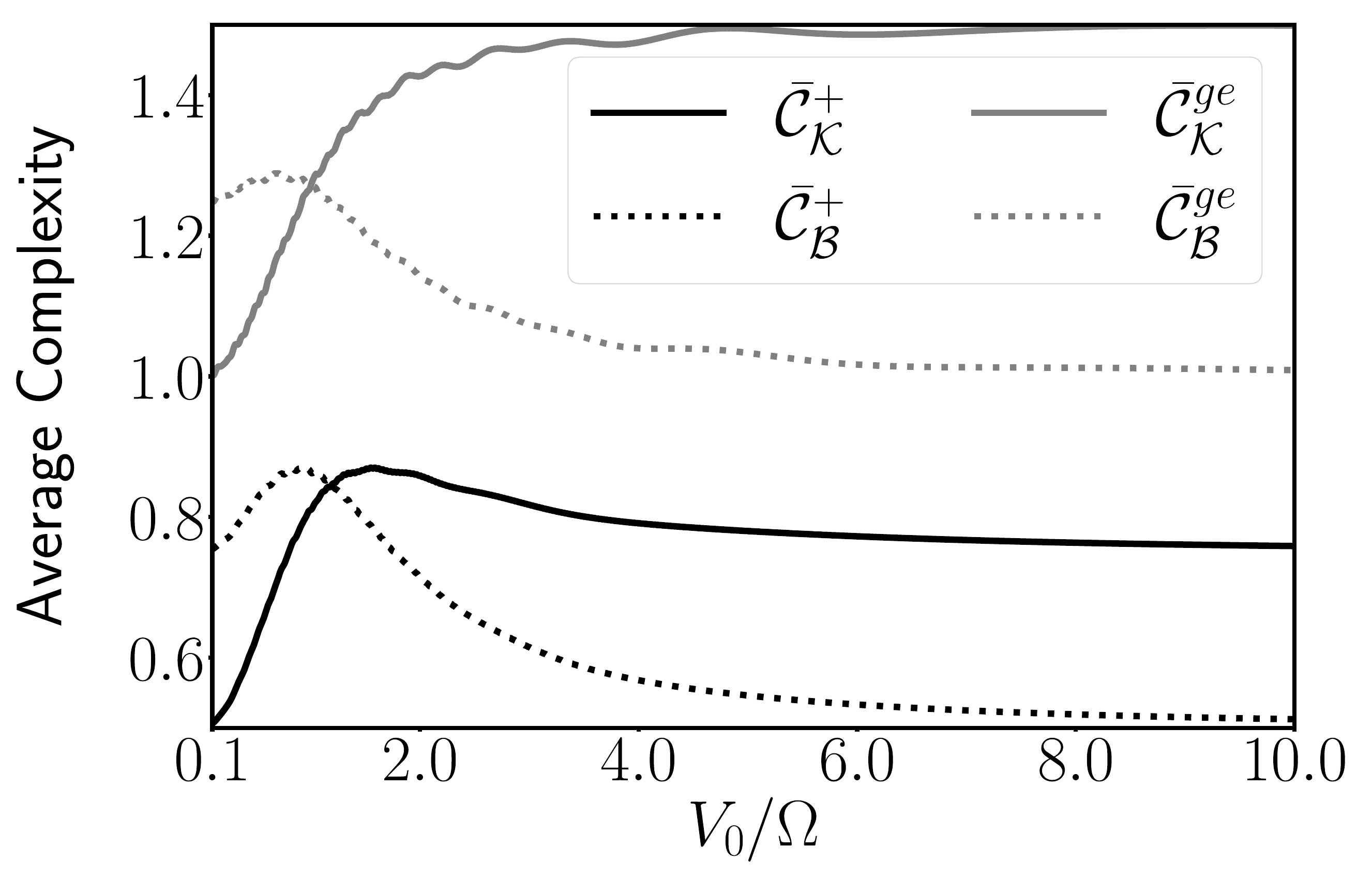}
    \caption{Time-averaged spread complexities in bases  $\mathcal{K}$ and $\mathcal{B}$ over a time of $\Omega t=240$ for initial states $|+\rangle$ and $|ge\rangle$ as a function of the interaction strength, $V_0$. They cross approximately at $V_0=\Omega$, which is roughly the onset of Rydberg blockade.}
    \label{fig:krylov-avg-ge-Vmax10}
\end{figure}

These results are not, however, in contradiction to \cite{Caputa2022-State-Complexity-Original}, in which the minimization of the spread complexity in the Krylov basis among all possible choices of ordered bases is established in the vicinity of $t=0$. It is also the same for the cases discussed above, as we show in Fig.~\ref{fig:krylov-comp-V100-early-times}, where we compare the complexity dynamics obtained using both bases in the vicinity of $t=0$ for the initial states $|+\rangle$ [Fig.~\ref{fig:krylov-comp-V100-early-times}(a)] and $|ge\rangle$ [Fig.~\ref{fig:krylov-comp-V100-early-times}(b)].
To understand this, let's take the case of the initial state $|+\rangle$. Initially, the population gets transferred to both $|gg\rangle$ and $|ee\rangle$ and, in particular, to $|K_1^+\rangle$, requiring only the first two states in the original Krylov basis to describe the dynamics completely. However, the same dynamics necessitate the first three states in the new basis $\mathcal B$. At the later stage,  $|K_2^+\rangle$ also becomes equally significant as the population in $|ee\rangle$ remains significantly low, making $\mathcal{C}_{\mathcal{K}}^{+}>\mathcal{C}_{\mathcal{B}}^{+}$. Note that the Krylov complexity is always minimal in the absence of blockade, i.e., for small $V_0$. In Fig.~\ref{fig:krylov-avg-ge-Vmax10}, we show the time-averaged spread complexity calculated in the bases $\mathcal K$ and $\mathcal B$ as a function of $V_0$. As expected from the blockade criteria, when $V_0\gtrsim\Omega$ for both the initial states, the average spread complexity is minimal in the basis $\mathcal B$.

The above results indicate that if the exact dynamics is effectively restricted to a part of the full Hilbert space, the spread complexity is generally not minimal in the Krylov basis obtained via the original Hamiltonian and does not always reflect the true reduction in dimensionality of the Hilbert space explored in the dynamics. However, it may not always be possible to find such alternate bases, as we shall see in Sec.~\ref{sec:RB-freezing}. In Sec.~\ref{sec:minimization_of_complexity}, we discuss more generally when we can expect to find such an alternate basis that can minimize the amplitude and average of the spread complexity. 

\subsubsection{Rydberg-biased freezing}\label{sec:RB-freezing}

\begin{figure}
    \centering
    \includegraphics[width=1\linewidth]{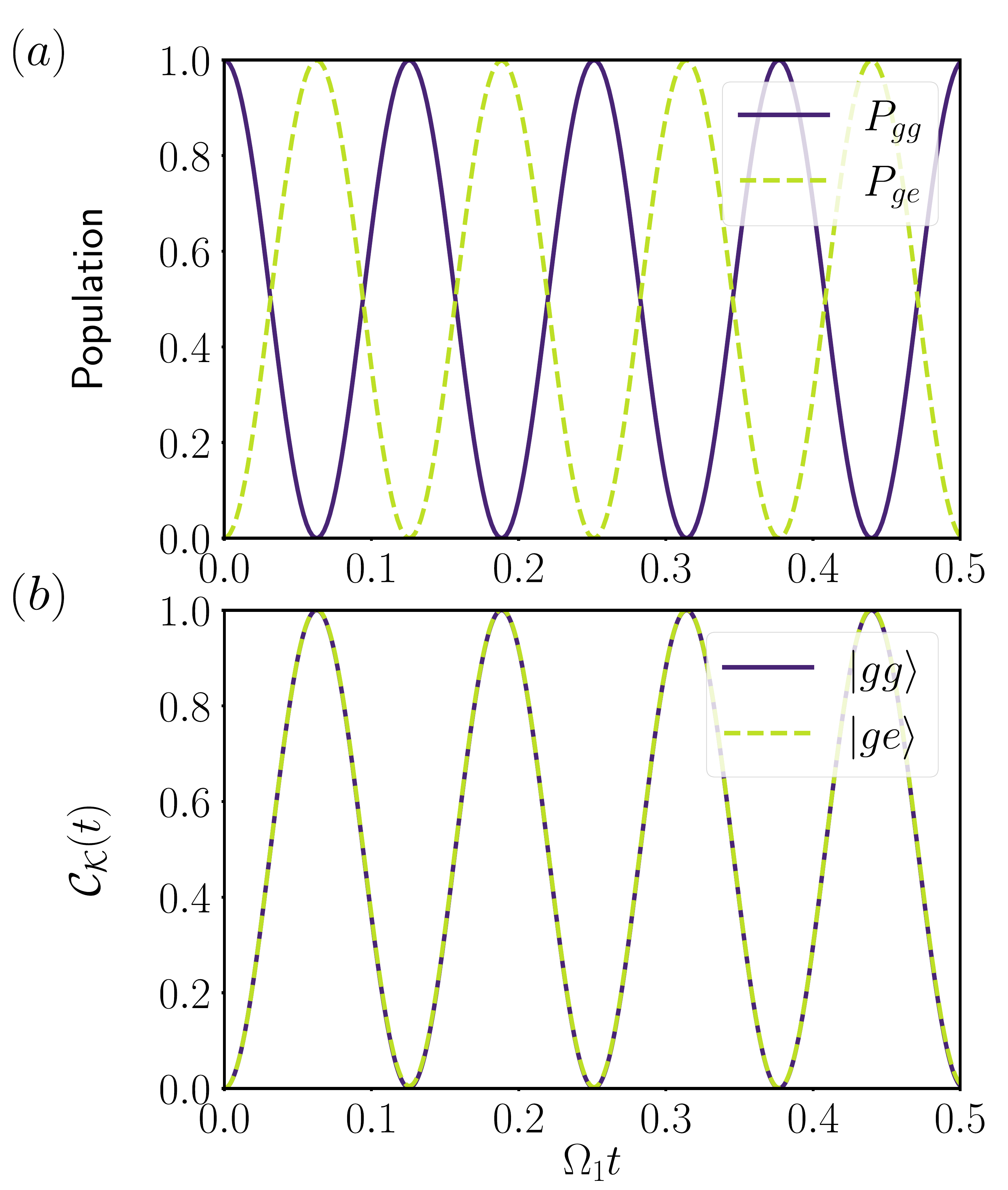}
    \caption{(a) Population in the states $|gg\rangle$ and $|ge\rangle$ as a function of time for initial state $|gg\rangle$, $V_0 = 100 \Omega_1$, $\Omega_2 = 50 \Omega_1$. (b) Time evolution of Krylov complexity for different initial states for the same parameters.}
    \label{fig:Rydberg-biased-freezing-prob-densities-V100-omega2_25}
\end{figure}

Now, we examine the dynamics of the Krylov state complexity under the phenomenon of Rydberg-biased freezing, which arises from a combined effect of Rybderg blockade and an offset in the Rabi couplings of the two Rydberg atoms \cite{Rydberg-Biased-Freezing-IOP2019, Rydberg-Biased-Freezing-Expt-PRR2021}. When $\Omega_2$ and $V_0$ are much greater than $\Omega_1$, starting from $|gg\rangle$, the first atom is essentially frozen in the ground state ($|g\rangle$) while the second atom undergoes Rabi oscillations between $|g\rangle$ and $|e\rangle$. In other words, Rabi oscillations occur between $|gg\rangle$ and $|ge\rangle$ as shown in Fig. \ref{fig:Rydberg-biased-freezing-prob-densities-V100-omega2_25}(a), and the $|eg\rangle$ and $|ee\rangle$ states do not contribute to the dynamics. 

For the initial state $|K_0\rangle^{gg} = |gg\rangle$, the Krylov basis states are,
\begin{align}
    |K_1\rangle^{gg} &= \dfrac{1}{\bar \Omega} \left(\Omega_2 |ge\rangle + \Omega_1 |eg\rangle \right) \label{eq:biased-freezing-basis-gg-1} \\
    |K_2\rangle^{gg} &= |ee\rangle \label{eq:biased-freezing-basis-gg-2} \\
    |K_3\rangle^{gg} &= \dfrac{1}{\bar \Omega} \big( -\Omega_1 |ge\rangle + \Omega_2 |eg\rangle \big) \label{eq:biased-freezing-basis-gg-3}.
\end{align}
where $\bar \Omega = \sqrt{\Omega_1^2 + \Omega_2^2}$. When $\Omega_1=\Omega_2$, this Krylov basis coincides with Eqs.~(\ref{kg1}) and (\ref{kg2}), i.e. the Krylov basis obtained for the blockade case. Even though three Krylov basis states ($|K_{0}\rangle^{gg}, |K_{1} \rangle^{gg},$ and $|K_{3}\rangle^{gg}$) look relevant to the dynamics, in the biased freezing limit  $\{V_0,\ \Omega_2\}\gg\Omega_1$, and therefore $|K_2\rangle^{gg}$ and $|K_3\rangle^{gg}\approx |eg\rangle$ can be disregarded. The corresponding Lanczos coefficients are $a_0 = a_1 = a_3 = 0$, $a_2 = V_0$, $b_1 = \bar \Omega/2$, $b_2 = (\Omega_1 \Omega_2)/\bar \Omega \approx \Omega_1$, and $b_3 = \left(\Omega_2^2 - \Omega_1^2\right)/\bar \Omega \approx \Omega_2$, and the Krylov complexity is obtained as $\mathcal{C}_{\mathcal{K}} (t) \approx \sin^2 (\Omega_2 t/2)$, which oscillates between zero and one as expected, in agreement with the numerical results shown in Fig. \ref{fig:Rydberg-biased-freezing-prob-densities-V100-omega2_25}(b).

If the initial state is $|ge\rangle$, then the Krylov basis states are given by 
\begin{align}
    |K_1\rangle^{ge} &= \dfrac{1}{\bar \Omega} \big(\Omega_2 |gg\rangle + \Omega_1 |ee\rangle \big) \label{eq:biased-freezing-ge-1} \\ 
    |K_2 \rangle^{ge} &= \dfrac{1}{\Omega_V \bar \Omega} \big(-V_0 \Omega_1 |gg\rangle + \tilde{\Omega}^2 |eg\rangle + V_0 \Omega_2 |ee\rangle \big) \label{eq:biased-freezing-ge-2} \\ 
    |K_3\rangle^{ge} &= \dfrac{1}{\Omega_V} ( -\Omega_1 |gg\rangle - V_0 |eg\rangle +  \Omega_2 |ee\rangle ) \label{eq:biased-freezing-ge-3}
\end{align}
where $\Omega_V = \sqrt{\Omega_1^2 + \Omega_2^2 + V_0^2}$ here. The Lanczos coefficients are obtained as $a_0 = 0$, $a_1 = V_0 \Omega_1^2/\bar{\Omega}^2$, $a_2 = V_0(\Omega_2^2 \Omega_V^2 - \Omega_1^2 \bar{\Omega}^2)/(\Omega_V^2 \bar{\Omega}^2)$, $a_3 = V_0\Omega_1^2/\Omega_V^2$, $b_1 = \bar{\Omega}/2$, $b_2 = \Omega_1 \Omega_2 \Omega_V/\bar{\Omega}^2$, and $b_3 = \sqrt{\dfrac{\bar{\Omega}^2}{2} + \dfrac{\Omega_1^2 \Omega_2^2 \Omega_V^2}{\bar{\Omega}^4} - \dfrac{\Omega_1^2 \bar{\Omega}^2(\Omega_2^2 + V_0^2)}{\Omega_V^4} - \dfrac{\Omega_1 \Omega_2 \Omega_V}{\bar{\Omega}}}$.
In the biased freezing limit, the Krylov basis states become $|K_1\rangle^{ge} \rightarrow |gg\rangle$, and $|K_2\rangle^{ge}$ and $|K_3\rangle^{ge}$ are mutually orthogonal superposed states of $|eg\rangle$ and $|ee\rangle$. Therefore, the Krylov complexity oscillates between zero and one and is identical to what we obtained for initial state $|gg\rangle$. Thus, the complexity behaved as expected for the freezing case and is minimal when using the original Krylov basis, even though the dynamics is effectively restricted to a smaller Hilbert space, in contrast to the Rydberg blockade case discussed above. 

\section{Discussion: Minimization of the time-averaged State Complexity}\label{sec:minimization_of_complexity}

It is proved that among the available ordered bases, the cost function in Eq.~(\ref{costB}) is minimized in the Krylov basis obtained from a given initial state and the original Hamiltonian, such that for $0 \leq t \leq \tau$ for some $\tau > 0$, $\mathcal{C}_{\mathcal{K}} (t) \leq \mathcal{C}_{\mathcal{B}} (t)$ \cite{Caputa2022-State-Complexity-Original}, where $\mathcal{B}$ is any ordered basis other than the Krylov basis. However, in Sec.~\ref{RB}, we have shown that $\tau$ is significantly short in the dynamics governed by the Rydberg blockade. Furthermore, the spread complexity is generally minimized when considering the Krylov basis based on the effective Hamiltonian, and is further illustrated through the time-averaged complexity.

Here, we discuss this in a more general context. We consider the Hamiltonian, $\hat{H} = \hat{H}_A + \hat{H}_B + \hat{H}_{AB}$ with 
\begin{align}
    \hat{H}_A &= \sum\limits_{i,j = 1}^{N_A} c_{ij}^{A} |a_i\rangle \langle a_j| \\
    \hat{H}_B &= \sum\limits_{i,j = 1}^{N_B} c_{ij}^{B} |b_i\rangle \langle b_j| \\
    \hat{H}_{AB} &= \sum\limits_{i=1}^{N_A} \sum\limits_{j=1}^{N_B} \left( d_{ij} |a_i\rangle \langle b_j| + \text{ h.c. } \right)
\end{align}
where $A$ and $B$ are orthogonal subspaces of the Hilbert space, spanned by $\{|a_i\rangle, \ 0 \leq i \leq N_A \}$ and $\{|b_j\rangle, \ 0 \leq j \leq N_B \}$ respectively. The coefficients $c_{ij}^{X}$ ($X = A, B$)  represent the matrix elements of the Hamiltonian in the subspace $X$, while $d_{ij}$  provide the coupling between the two subspaces. We assume that the subspaces $A$ and $B$ are energetically well-separated with only a weak coupling between them, i.e.,
\begin{equation}\label{eq:perturbation_limit}
    |d_{ij}| \ll \text{min}(|\{\mathcal{E}_B\} - \{\mathcal{E}_A\}|).
\end{equation}
If the dynamics of an initial state is entirely contained within the $A$-subspace, it can be well described by an effective Hamiltonian on $A$, which at the zeroth order is $\hat H_A$. Higher-order effective Hamiltonians can be computed by incorporating the effects of $\hat{H}_{AB}$ (see \cite{Schrieffer-Wolff-Other-Effective-Bravyi2011} and references therein).

Let $\mathcal{K}$ and $\mathcal{K}_A$ be the Krylov basis obtained from a given initial state $|\psi_0\rangle$, but using the Hamiltonians $\hat H$ and $\hat H_A$, respectively. To make the $\mathcal{K}_A$ basis complete, we need to add the remaining states from the $B$-subspace. We assume that the first $M$ states in $\mathcal{K}$ lie completely in $A$, in which case they coincide with the first $M$ states of $\mathcal{K}_A$.  This is because $\hat{H}_{AB} |K_j\rangle = 0$, where $j \leq M-2$. We denote the basis states in $\mathcal{K}_A$ as $|K_{A,j}\rangle$ and the corresponding Lanczos coefficients as $a_{A,j}$ and $b_{A,j}$. We further assume that $M$ is smaller than the size of $\mathcal{K}_A$; the case when $M$ is equal to the size of $\mathcal{K}_A$ will be discussed shortly. Then following the Lanczos algorithm, the $(M+1)$th state, $|K_{M}\rangle$, is given by
\begin{equation}
    b_M |K_M\rangle=b_{A,M}|K_{A,M}\rangle+\hat{H}_{AB} |K_{A,M-1}\rangle
\end{equation}
where we have used $\hat{H}_B |K_{A,M-1}\rangle=0$, $\hat{H}_A |K_{A,M-1}\rangle=a_{A,M-1}|K_{A,M-1}\rangle+b_{A,M}|K_{A,M}\rangle+b_{A,M-1}|K_{A,M-2}\rangle$, which is obtained using Eq.~(\ref{Eq:Tridiagonal-Ham-in-Krylov-Basis}), and $|K_{A,j}\rangle = |K_{j}\rangle$ for $0 \leq j \leq M-1$. The latter implies that we have identical Lanczos coefficients for $j \leq M-1$. Given that $b_{A,M} \propto c_{ij}^A$ and $\hat{H}_{AB} |K_{A,M-1}\rangle \propto d_{ij} |b\rangle$, where $|b\rangle$ is some normalized state in subspace $B$, the contribution from $B$ in $|K_M\rangle \propto \mathcal{O} \left( d_{ij}/\sqrt{\left(c_{ij}^A\right)^2 + \left(d_{ij}\right)^2} \right)$. Note that the initial state determines which of the $d_{ij}$ (if any) could contribute. This mixing of $A$ and $B$-subspaces in $\mathcal K$ basis can result in a higher Krylov complexity, over a time-period for which those states become relevant in describing the exact dynamics. Such a mixing can be avoided in the Krylov basis constructed using the effective Hamiltonian, thereby resulting in a smaller time-averaged complexity for $\mathcal K_A$.

If $M$ is equal to the dimension of $\mathcal{K}_A$, then $|K_M\rangle$, if it exists, lies entirely in $B$. Any remaining states in $\mathcal{K}$, however, may not necessarily lie completely in $B$ but may have projections onto orthogonal states in $A$ that are not coupled to the other states in $\mathcal{K}_A$ by $\hat{H}_A$. In such cases, the basis we had constructed using $\mathcal{K}_A$ and states from $B$ would not be complete. Rather, this basis must be constructed using a higher-order effective Hamiltonian that takes all orders of virtual processes to subspace $B$ and back (mediated by $\hat{H}_{AB}$). However, the amplitude and average of the spread complexity will once again be lower in this constructed basis rather than the Krylov basis, where the presence of $|K_{M}\rangle$ (that lies in $B$) ahead of Krylov states with contributions from $A$ makes the average complexity larger. This, however, does not arise for the initial states considered in our case of two interacting Rydberg qubits. 
 
Now, let us revisit the case of the Rydberg blockade along the lines of the discussion above. The two subspaces can be identified as $A\in\{|gg\rangle, |ge\rangle, |eg\rangle \}$ and $B \in \{|ee\rangle\}$, and accordingly, the Hamiltonian can be split as,
\begin{align}
    \hat{H}_{A} &= \dfrac{\Omega}{2}\left(\hat{\sigma}_{gg}^1\hat{\sigma}_{x}^2+\hat{\sigma}_{x}^1\hat{\sigma}_{gg}^2\right), \\
    \hat{H}_{B} &= V_0 \hat{\sigma}_{ee}^1 \hat{\sigma}_{ee}^2, \\
    \hat{H}_{AB} &= \dfrac{\Omega}{2}\left(\hat{\sigma}_{ee}^1\hat{\sigma}_{x}^2+\hat{\sigma}_{x}^1\hat{\sigma}_{ee}^2\right).
\end{align}
The energy separation between the subspaces $A$ and $B$ is about $V_0 \gg \Omega$, where $\Omega$ provides the coupling between them, hence, satisfying Eq. (\ref{eq:perturbation_limit}). Further, we see that $c_{i,j}^{A} \sim \Omega$ and $d_{ij} \sim \Omega$, thus equally mixing the state $|ee\rangle$ with the subspace A. The latter makes the average $\mathcal{C}_\mathcal{K} >\mathcal{C}_{\mathcal{K}_A}$ for initial states $|+\rangle$ and $|ge\rangle$, where $\mathcal{K}_A$ is given by the corresponding basis $\mathcal{B}$ discussed in Sec.~\ref{RB}. This is in high contrast with the case of Rydberg-biased freezing, where we have seen that the Krylov basis obtained from the original Hamiltonian minimizes the complexity at all times. With $A \in \{|gg\rangle, |ge\rangle\}$ and $B \in \{|eg\rangle, |ee\rangle \}$, this is evident since we get $d_{ij} \propto \Omega_1$ and $c_{ij}^A \propto \Omega_2$, and the contribution from $B$ in the Krylov basis vanishes in the limit $\Omega_2 \gg \Omega_1$. 

In the case of Rydberg blockade, as we have also shown, the Krylov basis minimizes the spread complexity at the very initial stage of the dynamics, which is consistent with the results in \cite{Caputa2022-State-Complexity-Original}. By treating the coupling between the subspaces $A$ and $B$ as a perturbation, we can obtain order-of-magnitude estimates of the time up to which the original Krylov basis minimizes the complexity, and the maximum difference between the spread complexities measured in the two bases during this time. The unperturbed energy eigenstates of the Hamiltonian $\hat H_A$+$\hat H_B$ are given by $|\tilde{\pm} \rangle^{(0)} = (1/\sqrt{2}) (|gg\rangle \pm |+\rangle)$, and $|ee\rangle^{(0)}=|ee\rangle$, with energies $E_{\tilde{\pm}}^{(0)} = \pm \Omega/\sqrt{2}$, and $E_{ee}^{(0)} = V_0$ respectively. The first order corrections to the eigenstates are, $|\tilde{\pm} \rangle^{(1)}=\mp (\Omega/2V_0) |ee\rangle$ and $|ee\rangle^{(1)}=(\Omega/\sqrt{2} V_0)|+\rangle$. For initial states $|+\rangle$ and $|ge\rangle$, under blockade, we find the populations in $|ee\rangle$ as $P_{ee}^+ (t) = (\Omega^2/2V_0^2) [1 + \cos^2 (\Omega t/\sqrt{2}) - 2 \cos (\Omega t/\sqrt{2}) \cos (V_0 t)]$ and $P_{ee}^{ge} (t) = (\Omega^2/4V_0^2) [1 + \cos^2 (\Omega t/\sqrt{2}) - 2 \cos (\Omega t/\sqrt{2}) \cos (V_0 t)]$, respectively. Subsequently, the spread complexity in bases $\mathcal{K}$ and $\mathcal{K}_A$ are obtained for the initial state $|+\rangle$ as, 
\begin{align}
    \mathcal{C}_{\mathcal{K}}^{+} (t) &= \dfrac{3}{2} \sin^2 \left( \dfrac{\Omega t}{\sqrt{2}} \right) - \dfrac{\Omega}{\sqrt{2} V_0} \sin \left( \dfrac{\Omega t}{\sqrt{2}} \right) \sin (V_0 t) \nonumber \\
    &+ \dfrac{3\Omega^2}{4V_0^2} \left[ 1 + \cos^2 \left(\dfrac{\Omega t}{\sqrt{2}} \right) - 2 \cos (V_0 t) \cos \left( \dfrac{\Omega t}{\sqrt{2}} \right) \right] \\
    \mathcal{C}_{\mathcal{K}_A}^{+} (t) &= \sin^2 \left(\dfrac{\Omega t}{\sqrt{2}} \right) \nonumber \\
    &+ \dfrac{\Omega^2}{V_0^2} \left[ 1 + \cos^2 \left( \dfrac{\Omega t}{\sqrt{2}} \right) - 2 \cos (V_0 t) \cos \left( \dfrac{\Omega t}{\sqrt{2}} \right) \right]
\end{align}
It is easy to verify from the equations above that when $V_0 \gg \Omega$, for $t < \tau \sim \mathcal{O} (\Omega/V_0)$, $\mathcal{C}_{\mathcal{K}}^{+}(t) < \mathcal{C}_{\mathcal{K}_A}^{+}(t)$. The maximum of $\mathcal{C}_{\mathcal{K}}^{+}-\mathcal{C}_{\mathcal{K}_A}^{+}$ attained during this period is of the order of $\Omega^2/V_0^2$, which can be negligible for large interaction strengths. At later times, we have $\mathcal{C}_{\mathcal{K}}^{+}(t) > \mathcal{C}_{\mathcal{K}_A}^{+}(t)$, and $\mathcal{C}_{\mathcal{K}_A}^{+} -\mathcal{C}_{\mathcal{K}}^{+}$ can become significantly large, say a value of 0.5 in this case. As the dynamics is periodic, this behavior periodically repeats. We obtain similar results for initial state $|ge\rangle$ as well.

\section{Summary and Outlook}\label{sec:summary}

We analyzed the Krylov state complexity in the quantum dynamics of a single qubit and a pair of qubits. In the single qubit case, we demonstrated that the square root of the Krylov complexity measures the distance between time-evolved states by explicitly constructing an associated parameter space on which the states in the Hilbert space are mapped. In the case of two non-interacting qubits, we found that the total Krylov complexity was not simply the sum of the individual complexities of the two isolated qubits as one may have expected but consisted of an additional term that only vanishes uniformly for certain initial states. We further noted that this extra term in the complexity breaks the subadditivity of the square root of the complexity, rendering it impossible to treat the square root of the Krylov complexity as a measure of distance between states in general for this system.

We further considered the case of a pair of Rydberg-atom qubits, particularly in the blockade regime, where the two atoms simultaneously occupying the excited Rydberg state are inhibited due to strong interactions. Hence, there is a redundancy in the Hilbert space as far as the dynamics is concerned. As a result, depending on the initial state, the spread complexity in the Krylov basis may have a larger amplitude and time-averaged value compared to a Krylov basis that may be obtained from an effective Hamiltonian on the lower-dimensional subspace in which the dynamics effectively takes place. We also noted that two initial states that exhibit near-identical dynamics may have significantly different complexities in the Krylov basis, while the complexity in the Krylov basis obtained from the aforementioned effective Hamiltonian yielded similar values for these states. We generalize these concepts by considering a Hilbert space that can be partitioned into two subspaces with a weak coupling between them.

As an outlook, it would be interesting to extend the above analysis to more than a pair of qubits, including many-body models and other non-trivial initial states. Such studies could bring deeper insights into the Krylov complexity and uncover additional complexity measures that are relevant to actual experimental setups.


\begin{acknowledgments}

We wish to acknowledge J. Bharathi Kannan for useful discussions. We also thank the anonymous referee for identifying an error in our interpretation of minimizing the spread complexity in the alternate Krylov bases, which allowed us to reformulate and significantly improve our results. S.S. acknowledges funding support from the Junior Research Fellowship (JRF) awarded by the University Grants Commission (UGC), India. We further acknowledge DST-SERB for the Swarnajayanti fellowship (File No. SB/SJF/2020-21/19), MATRICS Grant No. MTR/2022/000454 from SERB, Government of India, National Supercomputing Mission for providing computing resources of ``PARAM Brahma'' at IISER Pune, which is implemented by C-DAC and supported by the Ministry of Electronics and Information Technology and Department of Science and Technology (DST), Government of India, and acknowledge National Mission on Interdisciplinary Cyber-Physical Systems of the Department of Science and Technology, Government of India, through the I-HUB Quantum Technology Foundation, Pune, India.

\end{acknowledgments}


\appendix

\section{Two Non-Interacting Qubits - Krylov Basis and Lanczos Coefficients}\label{appendix:two-non-interacting-qubits}

Here, we compute the Krylov basis and the Lanczos coefficients for a pair of non-interacting qubits discussed in Section \ref{sec:two_ni_qubits}, governed by the Hamiltonian, $\hat{H} = \hat{H}_1 + \hat{H}_2$, where $\hat{H}_i = (\omega_i/2) \hat{\sigma}_z^{(i)}$ with energy eigenstates $|\pm \rangle_i$ and eigenvalues $E_{i,\pm} = \pm \omega_i/2$. Consider an initial product state: $|\psi_0\rangle = |K_0^{(1)} K_0^{(2)} \rangle$, where the single-qubit states in Bloch sphere representation are $|K_0^{(i)} \rangle = \cos (\theta_i/2) |+\rangle_i + \sin (\theta_i/2) e^{i\phi_i} |-\rangle_i$. The other Krylov basis state for qubit $i$ is given by $|K_1^{(i)} \rangle = \sin (\theta_i/2) |+\rangle_i - \cos (\theta_i/2) e^{i\phi_i} |-\rangle_i$. The Lanczos coefficients for each qubit are given by $a_0^{(i)} = - a_1^{(i)} = (\omega_i \cos \theta_i)/2$ and $b_1^{(i)} = (\omega_i \sin \theta_i)/2$. We define $a_{01}^{(i)} = a_0^{(i)} - a_1^{(i)} = \omega_i \cos \theta_i$. Following the Lanczos algorithm, we then obtain the Krylov basis states for a pair of non-interacting qubits as, 

\begin{align}
    |K_0 \rangle &= |K_0^{(1)}K_0^{(2)} \rangle \\
    |K_1 \rangle &= \dfrac{b_1^{(1)}}{b_1} |K_1^{(1)} K_0^{(2)} \rangle + \dfrac{b_1^{(2)}}{b_1} |K_0^{(1)} K_1^{(2)} \rangle \\
    |K_2\rangle &= \dfrac{a_{01}^{(1)} - a_{01}^{(2)}}{\sqrt{\left( a_{01}^{(1)} - a_{01}^{(2)}\right)^2 + 4 b_1^2 }} \left[ - \dfrac{b_1^{(2)}}{b_1} |K_1^{(1)} K_0^{(2)} \rangle + \dfrac{b_1^{(1)}}{b_1} |K_0^{(1)} K_1^{(2)} \rangle \right] \nonumber \\
    &\hspace{1cm} + \dfrac{2 b_1}{\sqrt{\left( a_{01}^{(1)} - a_{01}^{(2)}\right)^2 + 4 b_1^2}} |K_1^{(1)} K_1^{(2)} \rangle \\
    |K_3\rangle &= \dfrac{2 b_1}{\sqrt{\left( a_{01}^{(1)} - a_{01}^{(2)}\right)^2 + 4 b_1^2}} \left[ \dfrac{b_1^{(2)}}{b_1} |K_1^{(1)} K_0^{(2)} \rangle - \dfrac{b_1^{(1)}}{b_1} |K_0^{(1)} K_1^{(2)} \rangle \right] \nonumber \\
    &\hspace{1cm} + \dfrac{a_{01}^{(1)} - a_{01}^{(2)}}{\sqrt{\left( a_{01}^{(1)} - a_{01}^{(2)}\right)^2 + 4 b_1^2}} |K_1^{(1)} K_1^{(2)} \rangle
\end{align}
where

\begin{align}
    b_1 &= \sqrt{\left(b_1^{(1)} \right)^2 + \left(b_1^{(2)} \right)^2} = \dfrac{\sqrt{\omega_1^2 \sin^2 \theta_1 + \omega_2^2 \sin^2 \theta_2}}{2} \\
    b_2 &= \dfrac{b_1^{(1)} b_1^{(2)} \left[ \left(a_{01}^{(1)} - a_{01}^{(2)} \right)^2  + 4 b_1^2 \right]^{1/2} }{b_1^2} {\notag} \\
    &= \dfrac{\omega_1 \omega_2 \sin \theta_1 \sin \theta_2}{\omega_1^2 \sin^2 \theta_1 + \omega_2^2 \sin^2 \theta_2} {\notag} \\
    &\hspace{0.5cm} \times \sqrt{(\omega_1 \cos \theta_1 - \omega_2 \cos \theta_2)^2 + \omega_1^2 \sin^2 \theta_1 + \omega_2^2 \sin^2 \theta_2} \\
    b_3 &= b_1 \left| \dfrac{\left(a_{01}^{(1)}\right)^2 - \left(a_{01}^{(2)}\right)^2 + 4 \left(b_1^{(1)}\right)^2 - 4 \left(b_1^{(2)}\right)^2}{\left(a_{01}^{(1)} - a_{01}^{(2)}\right)^2 + 4 (b_1^{(1)})^2 + 4 (b_1^{(2)})^2} \right| {\notag} \\
    &= \dfrac{\left|\omega_1^2 - \omega_2^2 \right| \sqrt{\omega_1^2 \sin^2 \theta_1 + \omega_2^2 \sin^2 \theta_2^2}}{(\omega_1 \cos \theta_1 - \omega_2 \cos \theta_2)^2 + \omega_1^2 \sin^2 \theta_1 + \omega_2^2 \sin^2 \theta_2}
\end{align}
and

\begin{align}
    a_0 &= a_0^{(1)} + a_0^{(2)} = \dfrac{\omega_1 \cos \theta_1 + \omega_2 \cos \theta_2}{2} \\
    a_1 &= \dfrac{\left(b_1^{(1)} \right)^2 \left(a_1^{(1)} + a_0^{(2)}\right) + \left(b_1^{(2)} \right)^2\left(a_0^{(1)} + a_1^{(2)} \right)}{b_1^2} {\notag} \\
    &= - \left(\dfrac{\omega_1^2 \sin^2 \theta_1 - \omega_2^2 \sin^2 \theta_2}{\omega_1^2 \sin^2 \theta_1 + \omega_2^2 \sin^2 \theta_2} \right) \left(\dfrac{\omega_1 \cos \theta_1 - \omega_2 \cos \theta_2}{2} \right)
\end{align}

\begin{align}
    a_2 &= \dfrac{ \left( a_{01}^{(1)} - a_{01}^{(2)}\right)^2 \left[ \left( a_0^{(1)} + a_1^{(2)} \right) \left( b_1^{(1)} \right)^2 + \left( a_0^{(2)} + a_1^{(1)} \right) \left( b_1^{(2)} \right)^2 \right] }{b_1^2 \left[ \left(a_{01}^{(1)} - a_{01}^{(2)} \right)^2 + 4 b_1^2 \right]} {\notag} \\
    &\hspace{0.5cm} + \dfrac{4 \left( a_{0}^{(1)} - a_0^{(2)} \right) \left[ \left(b_1^{(1)} \right)^4 - \left(b_1^{(2)} \right)^4 \right]}{b_1^2 \left[ \left(a_{01}^{(1)} - a_{01}^{(2)} \right)^2 + 4 b_1^2 \right]} \nonumber \\
    &\hspace{0.5cm} + \dfrac{8 \left( a_1^{(2)} \left( b_1^{(1)} \right)^4 + a_1^{(1)} \left( b_1^{(2)} \right)^4 \right) + 8 \left(a_1^{(1)} + a_1^{(2)} \right) \left(b_1^{(1)}\right)^2 \left(b_1^{(2)} \right)^2}{b_1^2 \left[ \left(a_{01}^{(1)} - a_{01}^{(2)} \right)^2 + 4 b_1^2 \right]} \nonumber \\
    &=\dfrac{(\omega_1 \cos \theta_1 - \omega_2 \cos \theta_2) (\omega_1^2 \sin^2 \theta_1 - \omega_2^2 \sin^2 \theta_2) }{2 (\omega_1^2 \sin^2 \theta_1 + \omega_2^2 \sin^2 \theta_2)} \nonumber \\
    & \hspace{0.5cm} - \dfrac{\omega_1 \omega_2 (\omega_1 \sin^2 \theta_1 \cos \theta_2 + \omega_2 \sin^2 \theta_2 \cos \theta_1)}{(\omega_1 \cos \theta_1 - \omega_2 \cos \theta_2)^2 + \omega_1^2 \sin^2 \theta_1 + \omega_2^2 \sin^2 \theta_2} \\
    a_3 &= \dfrac{4 a_1 b_1^2 + \left(a_1^{(1)} + a_1^{(2)}\right) \left(a_{01}^{(1)} - a_{01}^{(2)} \right)^2}{\left[ \left( a_{01}^{(1)} - a_{01}^{(2)} \right)^2 + 4 b_1^2 \right]} {\notag} \\
    &= \dfrac{\left( \omega_1 \cos \theta_1 - \omega_2 \cos \theta_2 \right) (\omega_2^2 - \omega_1^2)}{2 \left[ \left(\omega_1 \cos \theta_1 - \omega_2 \cos \theta_2  \right)^2 + \omega_1^2 \sin^2 \theta_1 + \omega_2^2 \sin^2 \theta_2 \right]}
\end{align}
are the Lanczos coefficients. Using the projections of $|\psi_0^{(i)} \rangle$ on $|K_0^{(i)}\rangle$ and $|K_1^{(i)}\rangle$, the Krylov complexity can be obtained as in Eq. (\ref{eq:non-interacting-qubits-gen-krylov}). Note that when the two qubits are governed by identical Hamiltonians, i.e. $\omega_1 = \omega_2$, then $b_3 = a_3 = 0$ and the Krylov basis terminates at $|K_2\rangle$. As a consistency check, note that when one of the qubits, say qubit 2 without any loss of generality, is initially in an eigenstate of its Hamiltonian (i.e. when $\sin \theta_2 = 0$), $b_2 = 0$ and our Krylov basis only consists of $|K_0\rangle$ and $|K_1\rangle$ given by the Krylov basis states of qubit 1. 

Finally, we note for completeness that in terms of the Lanczos coefficients, the Hamiltonian in the Krylov basis for a general two-qubit system (i.e. not necessarily non-interacting), after discarding terms proportional to the identity matrix, can be written as:

\begin{equation}
    \hat{H} = \begin{pmatrix}
        \dfrac{a_{01} + a_{02} + a_{03}}{4} & b_1 & 0 & 0 \\
        b_1 & \dfrac{a_{10} + a_{12} + a_{13}}{4} & b_2 & 0 \\
        0 & b_2 & \dfrac{a_{20} + a_{21} + a_{23}}{4} & b_3 \\
        0 & 0 & b_3 & \dfrac{a_{30} + a_{31} + a_{32}}{4}
    \end{pmatrix}
\end{equation}
where we set $a_{ij} = a_i - a_j$. This Hamiltonian can also be written in terms of tensor products of Pauli matrices and the identity matrix for two qubits as $\hat{H} = \sum\limits_{i,j} \alpha_{ij} \hat{\sigma}_i^{(1)} \otimes \hat{\sigma}_j^{(2)}$, with $i = 0, x, y, z$ and $\hat{\sigma}_0^{(i)} = \mathcal{I}^{(i)}$. The non-zero coefficients $\alpha_{ij}$ are given by:
\begin{align}
    \alpha_{IX} &= \dfrac{b_1 + b_3}{2} \\
    \alpha_{ZY} &= \dfrac{b_1 - b_3}{2} \\
    \alpha_{XX} &= \alpha_{YY} = \dfrac{b_2}{2} \\
%
    \alpha_{IZ} &= \dfrac{(a_0 - a_1) + (a_2 - a_3)}{4} \\
    \alpha_{ZI} &= \dfrac{(a_0 - a_3) + (a_1 - a_2)}{4} \\
    \alpha_{ZZ} &= \dfrac{(a_0 - a_1) - (a_2 - a_3)}{4}.
\end{align}

\bibliography{apssamp}

\providecommand{\noopsort}[1]{}\providecommand{\singleletter}[1]{#1}%
\begin{thebibliography}{89}%
\makeatletter
\providecommand \@ifxundefined [1]{%
 \@ifx{#1\undefined}
}%
\providecommand \@ifnum [1]{%
 \ifnum #1\expandafter \@firstoftwo
 \else \expandafter \@secondoftwo
 \fi
}%
\providecommand \@ifx [1]{%
 \ifx #1\expandafter \@firstoftwo
 \else \expandafter \@secondoftwo
 \fi
}%
\providecommand \natexlab [1]{#1}%
\providecommand \enquote  [1]{``#1''}%
\providecommand \bibnamefont  [1]{#1}%
\providecommand \bibfnamefont [1]{#1}%
\providecommand \citenamefont [1]{#1}%
\providecommand \href@noop [0]{\@secondoftwo}%
\providecommand \href [0]{\begingroup \@sanitize@url \@href}%
\providecommand \@href[1]{\@@startlink{#1}\@@href}%
\providecommand \@@href[1]{\endgroup#1\@@endlink}%
\providecommand \@sanitize@url [0]{\catcode `\\12\catcode `\$12\catcode `\&12\catcode `\#12\catcode `\^12\catcode `\_12\catcode `\%12\relax}%
\providecommand \@@startlink[1]{}%
\providecommand \@@endlink[0]{}%
\providecommand \url  [0]{\begingroup\@sanitize@url \@url }%
\providecommand \@url [1]{\endgroup\@href {#1}{\urlprefix }}%
\providecommand \urlprefix  [0]{URL }%
\providecommand \Eprint [0]{\href }%
\providecommand \doibase [0]{https://doi.org/}%
\providecommand \selectlanguage [0]{\@gobble}%
\providecommand \bibinfo  [0]{\@secondoftwo}%
\providecommand \bibfield  [0]{\@secondoftwo}%
\providecommand \translation [1]{[#1]}%
\providecommand \BibitemOpen [0]{}%
\providecommand \bibitemStop [0]{}%
\providecommand \bibitemNoStop [0]{.\EOS\space}%
\providecommand \EOS [0]{\spacefactor3000\relax}%
\providecommand \BibitemShut  [1]{\csname bibitem#1\endcsname}%
\let\auto@bib@innerbib\@empty
\bibitem [{\citenamefont {Watrous}(2009)}]{Complexity-Watrous2009}%
  \BibitemOpen
  \bibfield  {author} {\bibinfo {author} {\bibfnamefont {J.}~\bibnamefont {Watrous}},\ }\bibfield  {title} {\bibinfo {title} {Quantum computational complexity},\ }\href {https://link.springer.com/referenceworkentry/10.1007/978-0-387-30440-3_428} {\bibfield  {journal} {\bibinfo  {journal} {Encyclopedia of Complexity and Systems Science}\ ,\ \bibinfo {pages} {7174}} (\bibinfo {year} {2009})}\BibitemShut {NoStop}%
\bibitem [{\citenamefont {Osborne}(2012)}]{osborne_Hamiltonian_complexity_2012}%
  \BibitemOpen
  \bibfield  {author} {\bibinfo {author} {\bibfnamefont {T.~J.}\ \bibnamefont {Osborne}},\ }\bibfield  {title} {\bibinfo {title} {Hamiltonian complexity},\ }\href {https://doi.org/10.1088/0034-4885/75/2/022001} {\bibfield  {journal} {\bibinfo  {journal} {Rep. Prog. Phys.}\ }\textbf {\bibinfo {volume} {75}},\ \bibinfo {pages} {022001} (\bibinfo {year} {2012})}\BibitemShut {NoStop}%
\bibitem [{\citenamefont {Aaronson}(2016)}]{aaronson2016_complexity_quantumstates_transformations}%
  \BibitemOpen
  \bibfield  {author} {\bibinfo {author} {\bibfnamefont {S.}~\bibnamefont {Aaronson}},\ }\href {https://arxiv.org/abs/1607.05256} {\bibinfo {title} {The complexity of quantum states and transformations: From quantum money to black holes}} (\bibinfo {year} {2016}),\ \Eprint {https://arxiv.org/abs/1607.05256} {arXiv:1607.05256 [quant-ph]} \BibitemShut {NoStop}%
\bibitem [{\citenamefont {Nielsen}\ \emph {et~al.}(2006{\natexlab{a}})\citenamefont {Nielsen}, \citenamefont {Dowling}, \citenamefont {Gu},\ and\ \citenamefont {Doherty}}]{Nielson2006}%
  \BibitemOpen
  \bibfield  {author} {\bibinfo {author} {\bibfnamefont {M.~A.}\ \bibnamefont {Nielsen}}, \bibinfo {author} {\bibfnamefont {M.~R.}\ \bibnamefont {Dowling}}, \bibinfo {author} {\bibfnamefont {M.}~\bibnamefont {Gu}},\ and\ \bibinfo {author} {\bibfnamefont {A.~C.}\ \bibnamefont {Doherty}},\ }\bibfield  {title} {\bibinfo {title} {Quantum computation as geometry},\ }\href {https://doi.org/10.1126/science.1121541} {\bibfield  {journal} {\bibinfo  {journal} {Science}\ }\textbf {\bibinfo {volume} {311}},\ \bibinfo {pages} {1133} (\bibinfo {year} {2006}{\natexlab{a}})}\BibitemShut {NoStop}%
\bibitem [{\citenamefont {Dowling}\ and\ \citenamefont {Nielsen}(2008)}]{Nielson-Dowling-2008}%
  \BibitemOpen
  \bibfield  {author} {\bibinfo {author} {\bibfnamefont {M.~R.}\ \bibnamefont {Dowling}}\ and\ \bibinfo {author} {\bibfnamefont {M.~A.}\ \bibnamefont {Nielsen}},\ }\bibfield  {title} {\bibinfo {title} {The geometry of quantum computation},\ }\href {https://dl.acm.org/doi/10.5555/2016985.2016986} {\bibfield  {journal} {\bibinfo  {journal} {Quantum Info. Comput.}\ }\textbf {\bibinfo {volume} {8}},\ \bibinfo {pages} {861–899} (\bibinfo {year} {2008})}\BibitemShut {NoStop}%
\bibitem [{\citenamefont {Nielsen}(2006)}]{Nielson-Quantum-Computing-Journal-2006}%
  \BibitemOpen
  \bibfield  {author} {\bibinfo {author} {\bibfnamefont {M.~A.}\ \bibnamefont {Nielsen}},\ }\bibfield  {title} {\bibinfo {title} {A geometric approach to quantum circuit lower bounds},\ }\href {https://dl.acm.org/doi/10.5555/2011686.2011688} {\bibfield  {journal} {\bibinfo  {journal} {Quantum Info. Comput.}\ }\textbf {\bibinfo {volume} {6}},\ \bibinfo {pages} {213–262} (\bibinfo {year} {2006})}\BibitemShut {NoStop}%
\bibitem [{\citenamefont {Nielsen}\ \emph {et~al.}(2006{\natexlab{b}})\citenamefont {Nielsen}, \citenamefont {Dowling}, \citenamefont {Gu},\ and\ \citenamefont {Doherty}}]{Nielsen_Optimal_control_Geometry_QuantumComputing_PRA2006}%
  \BibitemOpen
  \bibfield  {author} {\bibinfo {author} {\bibfnamefont {M.~A.}\ \bibnamefont {Nielsen}}, \bibinfo {author} {\bibfnamefont {M.~R.}\ \bibnamefont {Dowling}}, \bibinfo {author} {\bibfnamefont {M.}~\bibnamefont {Gu}},\ and\ \bibinfo {author} {\bibfnamefont {A.~C.}\ \bibnamefont {Doherty}},\ }\bibfield  {title} {\bibinfo {title} {Optimal control, geometry, and quantum computing},\ }\href {https://doi.org/10.1103/PhysRevA.73.062323} {\bibfield  {journal} {\bibinfo  {journal} {Phys. Rev. A}\ }\textbf {\bibinfo {volume} {73}},\ \bibinfo {pages} {062323} (\bibinfo {year} {2006}{\natexlab{b}})}\BibitemShut {NoStop}%
\bibitem [{\citenamefont {Nielsen}\ and\ \citenamefont {Chuang}(2010)}]{Nielsen_Chuang_Book_2010}%
  \BibitemOpen
  \bibfield  {author} {\bibinfo {author} {\bibfnamefont {M.~A.}\ \bibnamefont {Nielsen}}\ and\ \bibinfo {author} {\bibfnamefont {I.~L.}\ \bibnamefont {Chuang}},\ }\href {https://www.cambridge.org/highereducation/books/quantum-computation-and-quantum-information/01E10196D0A682A6AEFFEA52D53BE9AE#overview} {\emph {\bibinfo {title} {Quantum Computation and Quantum Information: 10th Anniversary Edition}}}\ (\bibinfo  {publisher} {Cambridge University Press},\ \bibinfo {year} {2010})\BibitemShut {NoStop}%
\bibitem [{\citenamefont {Chapman}\ and\ \citenamefont {Policastro}(2022)}]{Complexity_Black_Hole_Review_Chapman2022}%
  \BibitemOpen
  \bibfield  {author} {\bibinfo {author} {\bibfnamefont {S.}~\bibnamefont {Chapman}}\ and\ \bibinfo {author} {\bibfnamefont {G.}~\bibnamefont {Policastro}},\ }\bibfield  {title} {\bibinfo {title} {Quantum computational complexity from quantum information to black holes and back},\ }\href {https://doi.org/10.1140/epjc/s10052-022-10037-1} {\bibfield  {journal} {\bibinfo  {journal} {The European Physical Journal C}\ }\textbf {\bibinfo {volume} {82}},\ \bibinfo {pages} {128} (\bibinfo {year} {2022})}\BibitemShut {NoStop}%
\bibitem [{\citenamefont {Balasubramanian}\ \emph {et~al.}(2020)\citenamefont {Balasubramanian}, \citenamefont {DeCross}, \citenamefont {Kar},\ and\ \citenamefont {Parrikar}}]{Quantum_Complexity_Time_Evolution_Chaos_VijayBala}%
  \BibitemOpen
  \bibfield  {author} {\bibinfo {author} {\bibfnamefont {V.}~\bibnamefont {Balasubramanian}}, \bibinfo {author} {\bibfnamefont {M.}~\bibnamefont {DeCross}}, \bibinfo {author} {\bibfnamefont {A.}~\bibnamefont {Kar}},\ and\ \bibinfo {author} {\bibfnamefont {O.}~\bibnamefont {Parrikar}},\ }\bibfield  {title} {\bibinfo {title} {Quantum complexity of time evolution with chaotic hamiltonians},\ }\href {https://doi.org/10.1007/JHEP01(2020)134} {\bibfield  {journal} {\bibinfo  {journal} {Journal of High Energy Physics}\ }\textbf {\bibinfo {volume} {2020}},\ \bibinfo {pages} {134} (\bibinfo {year} {2020})}\BibitemShut {NoStop}%
\bibitem [{\citenamefont {Balasubramanian}\ \emph {et~al.}(2021)\citenamefont {Balasubramanian}, \citenamefont {DeCross}, \citenamefont {Kar}, \citenamefont {Li},\ and\ \citenamefont {Parrikar}}]{Complexity_Growth_VijayBala2021}%
  \BibitemOpen
  \bibfield  {author} {\bibinfo {author} {\bibfnamefont {V.}~\bibnamefont {Balasubramanian}}, \bibinfo {author} {\bibfnamefont {M.}~\bibnamefont {DeCross}}, \bibinfo {author} {\bibfnamefont {A.}~\bibnamefont {Kar}}, \bibinfo {author} {\bibfnamefont {Y.~C.}\ \bibnamefont {Li}},\ and\ \bibinfo {author} {\bibfnamefont {O.}~\bibnamefont {Parrikar}},\ }\bibfield  {title} {\bibinfo {title} {Complexity growth in integrable and chaotic models},\ }\href {https://doi.org/10.1007/JHEP07(2021)011} {\bibfield  {journal} {\bibinfo  {journal} {Journal of High Energy Physics}\ }\textbf {\bibinfo {volume} {2021}},\ \bibinfo {pages} {11} (\bibinfo {year} {2021})}\BibitemShut {NoStop}%
\bibitem [{\citenamefont {Bueno}\ \emph {et~al.}(2021)\citenamefont {Bueno}, \citenamefont {Mag{\'a}n},\ and\ \citenamefont {Shahbazi}}]{Complexity_Measures_Magan_JHEP2021}%
  \BibitemOpen
  \bibfield  {author} {\bibinfo {author} {\bibfnamefont {P.}~\bibnamefont {Bueno}}, \bibinfo {author} {\bibfnamefont {J.~M.}\ \bibnamefont {Mag{\'a}n}},\ and\ \bibinfo {author} {\bibfnamefont {C.~S.}\ \bibnamefont {Shahbazi}},\ }\bibfield  {title} {\bibinfo {title} {Complexity measures in qft and constrained geometric actions},\ }\href {https://doi.org/10.1007/JHEP09(2021)200} {\bibfield  {journal} {\bibinfo  {journal} {Journal of High Energy Physics}\ }\textbf {\bibinfo {volume} {2021}},\ \bibinfo {pages} {200} (\bibinfo {year} {2021})}\BibitemShut {NoStop}%
\bibitem [{\citenamefont {Brand\~ao}\ \emph {et~al.}(2021)\citenamefont {Brand\~ao}, \citenamefont {Chemissany}, \citenamefont {Hunter-Jones}, \citenamefont {Kueng},\ and\ \citenamefont {Preskill}}]{Preskill_Models_of_Quantum_Complexity_Growth_PRXQuantum2021}%
  \BibitemOpen
  \bibfield  {author} {\bibinfo {author} {\bibfnamefont {F.~G. S.~L.}\ \bibnamefont {Brand\~ao}}, \bibinfo {author} {\bibfnamefont {W.}~\bibnamefont {Chemissany}}, \bibinfo {author} {\bibfnamefont {N.}~\bibnamefont {Hunter-Jones}}, \bibinfo {author} {\bibfnamefont {R.}~\bibnamefont {Kueng}},\ and\ \bibinfo {author} {\bibfnamefont {J.}~\bibnamefont {Preskill}},\ }\bibfield  {title} {\bibinfo {title} {Models of quantum complexity growth},\ }\href {https://doi.org/10.1103/PRXQuantum.2.030316} {\bibfield  {journal} {\bibinfo  {journal} {PRX Quantum}\ }\textbf {\bibinfo {volume} {2}},\ \bibinfo {pages} {030316} (\bibinfo {year} {2021})}\BibitemShut {NoStop}%
\bibitem [{\citenamefont {Bulchandani}\ and\ \citenamefont {Sondhi}(2021)}]{Smooth_Quantum_Complexity_Sondhi_JHEP2021}%
  \BibitemOpen
  \bibfield  {author} {\bibinfo {author} {\bibfnamefont {V.~B.}\ \bibnamefont {Bulchandani}}\ and\ \bibinfo {author} {\bibfnamefont {S.~L.}\ \bibnamefont {Sondhi}},\ }\bibfield  {title} {\bibinfo {title} {How smooth is quantum complexity?},\ }\href {https://doi.org/10.1007/JHEP10(2021)230} {\bibfield  {journal} {\bibinfo  {journal} {Journal of High Energy Physics}\ }\textbf {\bibinfo {volume} {2021}},\ \bibinfo {pages} {230} (\bibinfo {year} {2021})}\BibitemShut {NoStop}%
\bibitem [{\citenamefont {Brown}(2023)}]{ABrown_Quantum_Complexity_LowerBound_Nature2023}%
  \BibitemOpen
  \bibfield  {author} {\bibinfo {author} {\bibfnamefont {A.~R.}\ \bibnamefont {Brown}},\ }\bibfield  {title} {\bibinfo {title} {A quantum complexity lower bound from differential geometry},\ }\href {https://www.nature.com/articles/s41567-022-01884-6} {\bibfield  {journal} {\bibinfo  {journal} {Nat. Phys.}\ }\textbf {\bibinfo {volume} {19}},\ \bibinfo {pages} {401} (\bibinfo {year} {2023})}\BibitemShut {NoStop}%
\bibitem [{\citenamefont {Mag{\'a}n}(2018)}]{Black_Holes_Complexity_Chaos_JMagan_JHEP2018}%
  \BibitemOpen
  \bibfield  {author} {\bibinfo {author} {\bibfnamefont {J.~M.}\ \bibnamefont {Mag{\'a}n}},\ }\bibfield  {title} {\bibinfo {title} {Black holes, complexity and quantum chaos},\ }\href {https://doi.org/10.1007/JHEP09(2018)043} {\bibfield  {journal} {\bibinfo  {journal} {Journal of High Energy Physics}\ }\textbf {\bibinfo {volume} {2018}},\ \bibinfo {pages} {43} (\bibinfo {year} {2018})}\BibitemShut {NoStop}%
\bibitem [{\citenamefont {Auzzi}\ \emph {et~al.}(2021)\citenamefont {Auzzi}, \citenamefont {Baiguera}, \citenamefont {De~Luca}, \citenamefont {Legramandi}, \citenamefont {Nardelli},\ and\ \citenamefont {Zenoni}}]{Geometry_of_Quantum_Complexity_Nielsen_Auzzi_PRD2021}%
  \BibitemOpen
  \bibfield  {author} {\bibinfo {author} {\bibfnamefont {R.}~\bibnamefont {Auzzi}}, \bibinfo {author} {\bibfnamefont {S.}~\bibnamefont {Baiguera}}, \bibinfo {author} {\bibfnamefont {G.~B.}\ \bibnamefont {De~Luca}}, \bibinfo {author} {\bibfnamefont {A.}~\bibnamefont {Legramandi}}, \bibinfo {author} {\bibfnamefont {G.}~\bibnamefont {Nardelli}},\ and\ \bibinfo {author} {\bibfnamefont {N.}~\bibnamefont {Zenoni}},\ }\bibfield  {title} {\bibinfo {title} {Geometry of quantum complexity},\ }\href {https://doi.org/10.1103/PhysRevD.103.106021} {\bibfield  {journal} {\bibinfo  {journal} {Phys. Rev. D}\ }\textbf {\bibinfo {volume} {103}},\ \bibinfo {pages} {106021} (\bibinfo {year} {2021})}\BibitemShut {NoStop}%
\bibitem [{\citenamefont {Susskind}(2016)}]{susskind2014_computational_complexity_blackhole}%
  \BibitemOpen
  \bibfield  {author} {\bibinfo {author} {\bibfnamefont {L.}~\bibnamefont {Susskind}},\ }\bibfield  {title} {\bibinfo {title} {Computational complexity and black hole horizons},\ }\href {https://doi.org/10.1002/prop.201500092} {\bibfield  {journal} {\bibinfo  {journal} {Fortschr. Phys.}\ }\textbf {\bibinfo {volume} {64}},\ \bibinfo {pages} {24} (\bibinfo {year} {2016})}\BibitemShut {NoStop}%
\bibitem [{\citenamefont {Stanford}\ and\ \citenamefont {Susskind}(2014)}]{Susskind_Complexity_ShockWaveGeom_PhysRevD2014}%
  \BibitemOpen
  \bibfield  {author} {\bibinfo {author} {\bibfnamefont {D.}~\bibnamefont {Stanford}}\ and\ \bibinfo {author} {\bibfnamefont {L.}~\bibnamefont {Susskind}},\ }\bibfield  {title} {\bibinfo {title} {Complexity and shock wave geometries},\ }\href {https://doi.org/10.1103/PhysRevD.90.126007} {\bibfield  {journal} {\bibinfo  {journal} {Phys. Rev. D}\ }\textbf {\bibinfo {volume} {90}},\ \bibinfo {pages} {126007} (\bibinfo {year} {2014})}\BibitemShut {NoStop}%
\bibitem [{\citenamefont {Susskind}\ and\ \citenamefont {Zhao}(2014)}]{susskind2014_switchback_complexity}%
  \BibitemOpen
  \bibfield  {author} {\bibinfo {author} {\bibfnamefont {L.}~\bibnamefont {Susskind}}\ and\ \bibinfo {author} {\bibfnamefont {Y.}~\bibnamefont {Zhao}},\ }\href {https://arxiv.org/abs/1408.2823} {\bibinfo {title} {Switchbacks and the bridge to nowhere}} (\bibinfo {year} {2014}),\ \Eprint {https://arxiv.org/abs/1408.2823} {arXiv:1408.2823 [hep-th]} \BibitemShut {NoStop}%
\bibitem [{\citenamefont {Alishahiha}(2015)}]{Alishahiha_Holographic_Complexity_PRD2015}%
  \BibitemOpen
  \bibfield  {author} {\bibinfo {author} {\bibfnamefont {M.}~\bibnamefont {Alishahiha}},\ }\bibfield  {title} {\bibinfo {title} {Holographic complexity},\ }\href {https://doi.org/10.1103/PhysRevD.92.126009} {\bibfield  {journal} {\bibinfo  {journal} {Phys. Rev. D}\ }\textbf {\bibinfo {volume} {92}},\ \bibinfo {pages} {126009} (\bibinfo {year} {2015})}\BibitemShut {NoStop}%
\bibitem [{\citenamefont {Brown}\ \emph {et~al.}(2016{\natexlab{a}})\citenamefont {Brown}, \citenamefont {Roberts}, \citenamefont {Susskind}, \citenamefont {Swingle},\ and\ \citenamefont {Zhao}}]{Susskind_Holographic_Complexity_PRL2016}%
  \BibitemOpen
  \bibfield  {author} {\bibinfo {author} {\bibfnamefont {A.~R.}\ \bibnamefont {Brown}}, \bibinfo {author} {\bibfnamefont {D.~A.}\ \bibnamefont {Roberts}}, \bibinfo {author} {\bibfnamefont {L.}~\bibnamefont {Susskind}}, \bibinfo {author} {\bibfnamefont {B.}~\bibnamefont {Swingle}},\ and\ \bibinfo {author} {\bibfnamefont {Y.}~\bibnamefont {Zhao}},\ }\bibfield  {title} {\bibinfo {title} {Holographic complexity equals bulk action?},\ }\href {https://doi.org/10.1103/PhysRevLett.116.191301} {\bibfield  {journal} {\bibinfo  {journal} {Phys. Rev. Lett.}\ }\textbf {\bibinfo {volume} {116}},\ \bibinfo {pages} {191301} (\bibinfo {year} {2016}{\natexlab{a}})}\BibitemShut {NoStop}%
\bibitem [{\citenamefont {Brown}\ \emph {et~al.}(2016{\natexlab{b}})\citenamefont {Brown}, \citenamefont {Roberts}, \citenamefont {Susskind}, \citenamefont {Swingle},\ and\ \citenamefont {Zhao}}]{susskind_complexity_action_blackholes_PRD2016}%
  \BibitemOpen
  \bibfield  {author} {\bibinfo {author} {\bibfnamefont {A.~R.}\ \bibnamefont {Brown}}, \bibinfo {author} {\bibfnamefont {D.~A.}\ \bibnamefont {Roberts}}, \bibinfo {author} {\bibfnamefont {L.}~\bibnamefont {Susskind}}, \bibinfo {author} {\bibfnamefont {B.}~\bibnamefont {Swingle}},\ and\ \bibinfo {author} {\bibfnamefont {Y.}~\bibnamefont {Zhao}},\ }\bibfield  {title} {\bibinfo {title} {Complexity, action, and black holes},\ }\href {https://doi.org/10.1103/PhysRevD.93.086006} {\bibfield  {journal} {\bibinfo  {journal} {Phys. Rev. D}\ }\textbf {\bibinfo {volume} {93}},\ \bibinfo {pages} {086006} (\bibinfo {year} {2016}{\natexlab{b}})}\BibitemShut {NoStop}%
\bibitem [{\citenamefont {Chemissany}\ and\ \citenamefont {Osborne}(2016)}]{Holographic_fluc_and_principle_of_minimal_complexity_Chemissany_JHEP2016}%
  \BibitemOpen
  \bibfield  {author} {\bibinfo {author} {\bibfnamefont {W.}~\bibnamefont {Chemissany}}\ and\ \bibinfo {author} {\bibfnamefont {T.~J.}\ \bibnamefont {Osborne}},\ }\bibfield  {title} {\bibinfo {title} {Holographic fluctuations and the principle of minimal complexity},\ }\href {https://doi.org/10.1007/JHEP12(2016)055} {\bibfield  {journal} {\bibinfo  {journal} {Journal of High Energy Physics}\ }\textbf {\bibinfo {volume} {2016}},\ \bibinfo {pages} {55} (\bibinfo {year} {2016})}\BibitemShut {NoStop}%
\bibitem [{\citenamefont {Brown}\ \emph {et~al.}(2017)\citenamefont {Brown}, \citenamefont {Susskind},\ and\ \citenamefont {Zhao}}]{Sussking_Quantum_Complexity_Neg_Curvature_PRD2017}%
  \BibitemOpen
  \bibfield  {author} {\bibinfo {author} {\bibfnamefont {A.~R.}\ \bibnamefont {Brown}}, \bibinfo {author} {\bibfnamefont {L.}~\bibnamefont {Susskind}},\ and\ \bibinfo {author} {\bibfnamefont {Y.}~\bibnamefont {Zhao}},\ }\bibfield  {title} {\bibinfo {title} {Quantum complexity and negative curvature},\ }\href {https://doi.org/10.1103/PhysRevD.95.045010} {\bibfield  {journal} {\bibinfo  {journal} {Phys. Rev. D}\ }\textbf {\bibinfo {volume} {95}},\ \bibinfo {pages} {045010} (\bibinfo {year} {2017})}\BibitemShut {NoStop}%
\bibitem [{\citenamefont {Bouland}\ \emph {et~al.}(2019)\citenamefont {Bouland}, \citenamefont {Fefferman},\ and\ \citenamefont {Vazirani}}]{bouland2019-computational-pseudorandomness-wormhole-growth}%
  \BibitemOpen
  \bibfield  {author} {\bibinfo {author} {\bibfnamefont {A.}~\bibnamefont {Bouland}}, \bibinfo {author} {\bibfnamefont {B.}~\bibnamefont {Fefferman}},\ and\ \bibinfo {author} {\bibfnamefont {U.}~\bibnamefont {Vazirani}},\ }\href {https://arxiv.org/abs/1910.14646} {\bibinfo {title} {Computational pseudorandomness, the wormhole growth paradox, and constraints on the ads/cft duality}} (\bibinfo {year} {2019}),\ \Eprint {https://arxiv.org/abs/1910.14646} {arXiv:1910.14646 [quant-ph]} \BibitemShut {NoStop}%
\bibitem [{\citenamefont {Chen}\ \emph {et~al.}(2022)\citenamefont {Chen}, \citenamefont {Czech},\ and\ \citenamefont {Wang}}]{Holographic_Duality_Review_Chen2022}%
  \BibitemOpen
  \bibfield  {author} {\bibinfo {author} {\bibfnamefont {B.}~\bibnamefont {Chen}}, \bibinfo {author} {\bibfnamefont {B.}~\bibnamefont {Czech}},\ and\ \bibinfo {author} {\bibfnamefont {Z.-Z.}\ \bibnamefont {Wang}},\ }\bibfield  {title} {\bibinfo {title} {Quantum information in holographic duality},\ }\href {https://doi.org/10.1088/1361-6633/ac51b5} {\bibfield  {journal} {\bibinfo  {journal} {Reports on Progress in Physics}\ }\textbf {\bibinfo {volume} {85}},\ \bibinfo {pages} {046001} (\bibinfo {year} {2022})}\BibitemShut {NoStop}%
\bibitem [{\citenamefont {Balasubramanian}\ \emph {et~al.}(2022)\citenamefont {Balasubramanian}, \citenamefont {Caputa}, \citenamefont {Magan},\ and\ \citenamefont {Wu}}]{Caputa2022-State-Complexity-Original}%
  \BibitemOpen
  \bibfield  {author} {\bibinfo {author} {\bibfnamefont {V.}~\bibnamefont {Balasubramanian}}, \bibinfo {author} {\bibfnamefont {P.}~\bibnamefont {Caputa}}, \bibinfo {author} {\bibfnamefont {J.~M.}\ \bibnamefont {Magan}},\ and\ \bibinfo {author} {\bibfnamefont {Q.}~\bibnamefont {Wu}},\ }\bibfield  {title} {\bibinfo {title} {Quantum chaos and the complexity of spread of states},\ }\href {https://doi.org/10.1103/PhysRevD.106.046007} {\bibfield  {journal} {\bibinfo  {journal} {Phys. Rev. D}\ }\textbf {\bibinfo {volume} {106}},\ \bibinfo {pages} {046007} (\bibinfo {year} {2022})}\BibitemShut {NoStop}%
\bibitem [{\citenamefont {Parker}\ \emph {et~al.}(2019)\citenamefont {Parker}, \citenamefont {Cao}, \citenamefont {Avdoshkin}, \citenamefont {Scaffidi},\ and\ \citenamefont {Altman}}]{Universal-Growth-Hypothesis-2019}%
  \BibitemOpen
  \bibfield  {author} {\bibinfo {author} {\bibfnamefont {D.~E.}\ \bibnamefont {Parker}}, \bibinfo {author} {\bibfnamefont {X.}~\bibnamefont {Cao}}, \bibinfo {author} {\bibfnamefont {A.}~\bibnamefont {Avdoshkin}}, \bibinfo {author} {\bibfnamefont {T.}~\bibnamefont {Scaffidi}},\ and\ \bibinfo {author} {\bibfnamefont {E.}~\bibnamefont {Altman}},\ }\bibfield  {title} {\bibinfo {title} {A universal operator growth hypothesis},\ }\href {https://doi.org/10.1103/PhysRevX.9.041017} {\bibfield  {journal} {\bibinfo  {journal} {Phys. Rev. X}\ }\textbf {\bibinfo {volume} {9}},\ \bibinfo {pages} {041017} (\bibinfo {year} {2019})}\BibitemShut {NoStop}%
\bibitem [{\citenamefont {Nandy}\ \emph {et~al.}(2024{\natexlab{a}})\citenamefont {Nandy}, \citenamefont {Matsoukas-Roubeas}, \citenamefont {Martínez-Azcona}, \citenamefont {Dymarsky},\ and\ \citenamefont {del Campo}}]{Krylov-Review2024}%
  \BibitemOpen
  \bibfield  {author} {\bibinfo {author} {\bibfnamefont {P.}~\bibnamefont {Nandy}}, \bibinfo {author} {\bibfnamefont {A.~S.}\ \bibnamefont {Matsoukas-Roubeas}}, \bibinfo {author} {\bibfnamefont {P.}~\bibnamefont {Martínez-Azcona}}, \bibinfo {author} {\bibfnamefont {A.}~\bibnamefont {Dymarsky}},\ and\ \bibinfo {author} {\bibfnamefont {A.}~\bibnamefont {del Campo}},\ }\href {https://arxiv.org/abs/2405.09628} {\bibinfo {title} {Quantum dynamics in krylov space: Methods and applications}} (\bibinfo {year} {2024}{\natexlab{a}}),\ \Eprint {https://arxiv.org/abs/2405.09628} {arXiv:2405.09628 [quant-ph]} \BibitemShut {NoStop}%
\bibitem [{\citenamefont {Sánchez-Garrido}(2024)}]{sanchezgarrido_thesis_2024krylovcomplexity}%
  \BibitemOpen
  \bibfield  {author} {\bibinfo {author} {\bibfnamefont {A.}~\bibnamefont {Sánchez-Garrido}},\ }\href {https://arxiv.org/abs/2407.03866} {\bibinfo {title} {On krylov complexity}} (\bibinfo {year} {2024}),\ \Eprint {https://arxiv.org/abs/2407.03866} {arXiv:2407.03866 [hep-th]} \BibitemShut {NoStop}%
\bibitem [{\citenamefont {Lanczos}(1950)}]{Lanczos1950}%
  \BibitemOpen
  \bibfield  {author} {\bibinfo {author} {\bibfnamefont {C.}~\bibnamefont {Lanczos}},\ }\bibfield  {title} {\bibinfo {title} {{An iteration method for the solution of the eigenvalue problem of linear differential and integral operators}},\ }\href {https://doi.org/10.6028/jres.045.026} {\bibfield  {journal} {\bibinfo  {journal} {J. Res. Natl. Bur. Stand. B}\ }\textbf {\bibinfo {volume} {45}},\ \bibinfo {pages} {255} (\bibinfo {year} {1950})}\BibitemShut {NoStop}%
\bibitem [{\citenamefont {Viswanath}\ and\ \citenamefont {Mueller}(1994)}]{Vishwanath_Mueller_Book_Recursion_Method}%
  \BibitemOpen
  \bibfield  {author} {\bibinfo {author} {\bibfnamefont {V.~S.}\ \bibnamefont {Viswanath}}\ and\ \bibinfo {author} {\bibfnamefont {G.}~\bibnamefont {Mueller}},\ }\href {http://inis.iaea.org/search/search.aspx?orig_q=RN:26012244} {\emph {\bibinfo {title} {The recursion method Application to many-body dynamics}}}\ (\bibinfo  {publisher} {Springer},\ \bibinfo {address} {Germany},\ \bibinfo {year} {1994})\BibitemShut {NoStop}%
\bibitem [{\citenamefont {Hashimoto}\ \emph {et~al.}(2023)\citenamefont {Hashimoto}, \citenamefont {Murata}, \citenamefont {Tanahashi},\ and\ \citenamefont {Watanabe}}]{Hashimoto2023-Billiards}%
  \BibitemOpen
  \bibfield  {author} {\bibinfo {author} {\bibfnamefont {K.}~\bibnamefont {Hashimoto}}, \bibinfo {author} {\bibfnamefont {K.}~\bibnamefont {Murata}}, \bibinfo {author} {\bibfnamefont {N.}~\bibnamefont {Tanahashi}},\ and\ \bibinfo {author} {\bibfnamefont {R.}~\bibnamefont {Watanabe}},\ }\bibfield  {title} {\bibinfo {title} {Krylov complexity and chaos in quantum mechanics},\ }\href {https://doi.org/10.1007/JHEP11(2023)040} {\bibfield  {journal} {\bibinfo  {journal} {Journal of High Energy Physics}\ }\textbf {\bibinfo {volume} {2023}},\ \bibinfo {pages} {40} (\bibinfo {year} {2023})}\BibitemShut {NoStop}%
\bibitem [{\citenamefont {Erdmenger}\ \emph {et~al.}(2023)\citenamefont {Erdmenger}, \citenamefont {Jian},\ and\ \citenamefont {Xian}}]{Xian2023-Universal-chaotic-dynamics}%
  \BibitemOpen
  \bibfield  {author} {\bibinfo {author} {\bibfnamefont {J.}~\bibnamefont {Erdmenger}}, \bibinfo {author} {\bibfnamefont {S.-K.}\ \bibnamefont {Jian}},\ and\ \bibinfo {author} {\bibfnamefont {Z.-Y.}\ \bibnamefont {Xian}},\ }\bibfield  {title} {\bibinfo {title} {Universal chaotic dynamics from krylov space},\ }\href {https://doi.org/10.1007/JHEP08(2023)176} {\bibfield  {journal} {\bibinfo  {journal} {Journal of High Energy Physics}\ }\textbf {\bibinfo {volume} {2023}},\ \bibinfo {pages} {176} (\bibinfo {year} {2023})}\BibitemShut {NoStop}%
\bibitem [{\citenamefont {Scialchi}\ \emph {et~al.}(2024)\citenamefont {Scialchi}, \citenamefont {Roncaglia},\ and\ \citenamefont {Wisniacki}}]{scialchi2023_integrability_chaos_krylov_arxiv}%
  \BibitemOpen
  \bibfield  {author} {\bibinfo {author} {\bibfnamefont {G.~F.}\ \bibnamefont {Scialchi}}, \bibinfo {author} {\bibfnamefont {A.~J.}\ \bibnamefont {Roncaglia}},\ and\ \bibinfo {author} {\bibfnamefont {D.~A.}\ \bibnamefont {Wisniacki}},\ }\bibfield  {title} {\bibinfo {title} {Integrability-to-chaos transition through the krylov approach for state evolution},\ }\href {https://doi.org/10.1103/PhysRevE.109.054209} {\bibfield  {journal} {\bibinfo  {journal} {Phys. Rev. E}\ }\textbf {\bibinfo {volume} {109}},\ \bibinfo {pages} {054209} (\bibinfo {year} {2024})}\BibitemShut {NoStop}%
\bibitem [{\citenamefont {Balasubramanian}\ \emph {et~al.}(2025)\citenamefont {Balasubramanian}, \citenamefont {Magan},\ and\ \citenamefont {Wu}}]{vijay_bala_dec2023_quantumchaos_integrability_arxiv}%
  \BibitemOpen
  \bibfield  {author} {\bibinfo {author} {\bibfnamefont {V.}~\bibnamefont {Balasubramanian}}, \bibinfo {author} {\bibfnamefont {J.~M.}\ \bibnamefont {Magan}},\ and\ \bibinfo {author} {\bibfnamefont {Q.}~\bibnamefont {Wu}},\ }\bibfield  {title} {\bibinfo {title} {Quantum chaos, integrability, and late times in the krylov basis},\ }\href {https://doi.org/10.1103/PhysRevE.111.014218} {\bibfield  {journal} {\bibinfo  {journal} {Phys. Rev. E}\ }\textbf {\bibinfo {volume} {111}},\ \bibinfo {pages} {014218} (\bibinfo {year} {2025})}\BibitemShut {NoStop}%
\bibitem [{\citenamefont {Bhattacharjee}\ \emph {et~al.}(2022)\citenamefont {Bhattacharjee}, \citenamefont {Sur},\ and\ \citenamefont {Nandy}}]{Bhattacharjee-PXP-PRB2022}%
  \BibitemOpen
  \bibfield  {author} {\bibinfo {author} {\bibfnamefont {B.}~\bibnamefont {Bhattacharjee}}, \bibinfo {author} {\bibfnamefont {S.}~\bibnamefont {Sur}},\ and\ \bibinfo {author} {\bibfnamefont {P.}~\bibnamefont {Nandy}},\ }\bibfield  {title} {\bibinfo {title} {Probing quantum scars and weak ergodicity breaking through quantum complexity},\ }\href {https://doi.org/10.1103/PhysRevB.106.205150} {\bibfield  {journal} {\bibinfo  {journal} {Phys. Rev. B}\ }\textbf {\bibinfo {volume} {106}},\ \bibinfo {pages} {205150} (\bibinfo {year} {2022})}\BibitemShut {NoStop}%
\bibitem [{\citenamefont {Nandy}\ \emph {et~al.}(2024{\natexlab{b}})\citenamefont {Nandy}, \citenamefont {Mukherjee}, \citenamefont {Bhattacharyya},\ and\ \citenamefont {Banerjee}}]{SNandy-PXP-Numerical-IOP2024}%
  \BibitemOpen
  \bibfield  {author} {\bibinfo {author} {\bibfnamefont {S.}~\bibnamefont {Nandy}}, \bibinfo {author} {\bibfnamefont {B.}~\bibnamefont {Mukherjee}}, \bibinfo {author} {\bibfnamefont {A.}~\bibnamefont {Bhattacharyya}},\ and\ \bibinfo {author} {\bibfnamefont {A.}~\bibnamefont {Banerjee}},\ }\bibfield  {title} {\bibinfo {title} {Quantum state complexity meets many-body scars},\ }\href {https://doi.org/10.1088/1361-648X/ad1a7b} {\bibfield  {journal} {\bibinfo  {journal} {Journal of Physics: Condensed Matter}\ }\textbf {\bibinfo {volume} {36}},\ \bibinfo {pages} {155601} (\bibinfo {year} {2024}{\natexlab{b}})}\BibitemShut {NoStop}%
\bibitem [{\citenamefont {Huh}\ \emph {et~al.}(2024)\citenamefont {Huh}, \citenamefont {Jeong},\ and\ \citenamefont {Pedraza}}]{Krylov_Complexity_Saddle-dominated_Scrambling_Chaos_Pedraza_JHEP2024}%
  \BibitemOpen
  \bibfield  {author} {\bibinfo {author} {\bibfnamefont {K.-B.}\ \bibnamefont {Huh}}, \bibinfo {author} {\bibfnamefont {H.-S.}\ \bibnamefont {Jeong}},\ and\ \bibinfo {author} {\bibfnamefont {J.~F.}\ \bibnamefont {Pedraza}},\ }\bibfield  {title} {\bibinfo {title} {Spread complexity in saddle-dominated scrambling},\ }\href {https://doi.org/10.1007/JHEP05(2024)137} {\bibfield  {journal} {\bibinfo  {journal} {Journal of High Energy Physics}\ }\textbf {\bibinfo {volume} {2024}},\ \bibinfo {pages} {137} (\bibinfo {year} {2024})}\BibitemShut {NoStop}%
\bibitem [{\citenamefont {Camargo}\ \emph {et~al.}(2024{\natexlab{a}})\citenamefont {Camargo}, \citenamefont {Huh}, \citenamefont {Jahnke}, \citenamefont {Jeong}, \citenamefont {Kim},\ and\ \citenamefont {Nishida}}]{Krylov_Chaos_Mixed_Field_Ising_Model_2024}%
  \BibitemOpen
  \bibfield  {author} {\bibinfo {author} {\bibfnamefont {H.~A.}\ \bibnamefont {Camargo}}, \bibinfo {author} {\bibfnamefont {K.-B.}\ \bibnamefont {Huh}}, \bibinfo {author} {\bibfnamefont {V.}~\bibnamefont {Jahnke}}, \bibinfo {author} {\bibfnamefont {H.-S.}\ \bibnamefont {Jeong}}, \bibinfo {author} {\bibfnamefont {K.-Y.}\ \bibnamefont {Kim}},\ and\ \bibinfo {author} {\bibfnamefont {M.}~\bibnamefont {Nishida}},\ }\bibfield  {title} {\bibinfo {title} {Spread and spectral complexity in quantum spin chains: from integrability to chaos},\ }\href {https://doi.org/10.1007/JHEP08(2024)241} {\bibfield  {journal} {\bibinfo  {journal} {Journal of High Energy Physics}\ }\textbf {\bibinfo {volume} {2024}},\ \bibinfo {pages} {241} (\bibinfo {year} {2024}{\natexlab{a}})}\BibitemShut {NoStop}%
\bibitem [{\citenamefont {Nizami}\ and\ \citenamefont {Shrestha}(2024)}]{nizami2024_Krylov_chaos_periodic_driving_arxiv}%
  \BibitemOpen
  \bibfield  {author} {\bibinfo {author} {\bibfnamefont {A.~A.}\ \bibnamefont {Nizami}}\ and\ \bibinfo {author} {\bibfnamefont {A.~W.}\ \bibnamefont {Shrestha}},\ }\bibfield  {title} {\bibinfo {title} {Spread complexity and quantum chaos for periodically driven spin chains},\ }\href {https://doi.org/10.1103/PhysRevE.110.034201} {\bibfield  {journal} {\bibinfo  {journal} {Phys. Rev. E}\ }\textbf {\bibinfo {volume} {110}},\ \bibinfo {pages} {034201} (\bibinfo {year} {2024})}\BibitemShut {NoStop}%
\bibitem [{\citenamefont {Camargo}\ \emph {et~al.}(2024{\natexlab{b}})\citenamefont {Camargo}, \citenamefont {Jahnke}, \citenamefont {Jeong}, \citenamefont {Kim},\ and\ \citenamefont {Nishida}}]{Nishida_Krylov_Billiard_MostlyOperator_PRD2024}%
  \BibitemOpen
  \bibfield  {author} {\bibinfo {author} {\bibfnamefont {H.~A.}\ \bibnamefont {Camargo}}, \bibinfo {author} {\bibfnamefont {V.}~\bibnamefont {Jahnke}}, \bibinfo {author} {\bibfnamefont {H.-S.}\ \bibnamefont {Jeong}}, \bibinfo {author} {\bibfnamefont {K.-Y.}\ \bibnamefont {Kim}},\ and\ \bibinfo {author} {\bibfnamefont {M.}~\bibnamefont {Nishida}},\ }\bibfield  {title} {\bibinfo {title} {Spectral and krylov complexity in billiard systems},\ }\href {https://doi.org/10.1103/PhysRevD.109.046017} {\bibfield  {journal} {\bibinfo  {journal} {Phys. Rev. D}\ }\textbf {\bibinfo {volume} {109}},\ \bibinfo {pages} {046017} (\bibinfo {year} {2024}{\natexlab{b}})}\BibitemShut {NoStop}%
\bibitem [{\citenamefont {Balasubramanian}\ \emph {et~al.}(2024)\citenamefont {Balasubramanian}, \citenamefont {Das}, \citenamefont {Erdmenger},\ and\ \citenamefont {Xian}}]{Krylov_triangular_billiards_balasubramanian_arxiv2024}%
  \BibitemOpen
  \bibfield  {author} {\bibinfo {author} {\bibfnamefont {V.}~\bibnamefont {Balasubramanian}}, \bibinfo {author} {\bibfnamefont {R.~N.}\ \bibnamefont {Das}}, \bibinfo {author} {\bibfnamefont {J.}~\bibnamefont {Erdmenger}},\ and\ \bibinfo {author} {\bibfnamefont {Z.-Y.}\ \bibnamefont {Xian}},\ }\href {https://arxiv.org/abs/2407.11114} {\bibinfo {title} {Chaos and integrability in triangular billiards}} (\bibinfo {year} {2024}),\ \Eprint {https://arxiv.org/abs/2407.11114} {arXiv:2407.11114 [hep-th]} \BibitemShut {NoStop}%
\bibitem [{\citenamefont {Bhattacharjee}\ and\ \citenamefont {Nandy}(2025)}]{krylov_fractality_ergodic_localized_regimes_transitions_generic_pnandy_arxiv2024}%
  \BibitemOpen
  \bibfield  {author} {\bibinfo {author} {\bibfnamefont {B.}~\bibnamefont {Bhattacharjee}}\ and\ \bibinfo {author} {\bibfnamefont {P.}~\bibnamefont {Nandy}},\ }\bibfield  {title} {\bibinfo {title} {Krylov fractality and complexity in generic random matrix ensembles},\ }\href {https://doi.org/10.1103/PhysRevB.111.L060202} {\bibfield  {journal} {\bibinfo  {journal} {Phys. Rev. B}\ }\textbf {\bibinfo {volume} {111}},\ \bibinfo {pages} {L060202} (\bibinfo {year} {2025})}\BibitemShut {NoStop}%
\bibitem [{\citenamefont {Baggioli}\ \emph {et~al.}(2024)\citenamefont {Baggioli}, \citenamefont {Huh}, \citenamefont {Jeong}, \citenamefont {Kim},\ and\ \citenamefont {Pedraza}}]{krylov_complexity_order_parameter_chaotic_integrable_baggioli_hyun_sik_jeong_arxiv2024}%
  \BibitemOpen
  \bibfield  {author} {\bibinfo {author} {\bibfnamefont {M.}~\bibnamefont {Baggioli}}, \bibinfo {author} {\bibfnamefont {K.-B.}\ \bibnamefont {Huh}}, \bibinfo {author} {\bibfnamefont {H.-S.}\ \bibnamefont {Jeong}}, \bibinfo {author} {\bibfnamefont {K.-Y.}\ \bibnamefont {Kim}},\ and\ \bibinfo {author} {\bibfnamefont {J.~F.}\ \bibnamefont {Pedraza}},\ }\href {https://arxiv.org/abs/2407.17054} {\bibinfo {title} {Krylov complexity as an order parameter for quantum chaotic-integrable transitions}} (\bibinfo {year} {2024}),\ \Eprint {https://arxiv.org/abs/2407.17054} {arXiv:2407.17054 [hep-th]} \BibitemShut {NoStop}%
\bibitem [{\citenamefont {Caputa}\ and\ \citenamefont {Liu}(2022)}]{Caputa2022-Topological-Phases-of-Matter}%
  \BibitemOpen
  \bibfield  {author} {\bibinfo {author} {\bibfnamefont {P.}~\bibnamefont {Caputa}}\ and\ \bibinfo {author} {\bibfnamefont {S.}~\bibnamefont {Liu}},\ }\bibfield  {title} {\bibinfo {title} {Quantum complexity and topological phases of matter},\ }\href {https://doi.org/10.1103/PhysRevB.106.195125} {\bibfield  {journal} {\bibinfo  {journal} {Phys. Rev. B}\ }\textbf {\bibinfo {volume} {106}},\ \bibinfo {pages} {195125} (\bibinfo {year} {2022})}\BibitemShut {NoStop}%
\bibitem [{\citenamefont {Caputa}\ \emph {et~al.}(2023)\citenamefont {Caputa}, \citenamefont {Gupta}, \citenamefont {Haque}, \citenamefont {Liu}, \citenamefont {Murugan},\ and\ \citenamefont {Van~Zyl}}]{Caputa2023-Kitaev}%
  \BibitemOpen
  \bibfield  {author} {\bibinfo {author} {\bibfnamefont {P.}~\bibnamefont {Caputa}}, \bibinfo {author} {\bibfnamefont {N.}~\bibnamefont {Gupta}}, \bibinfo {author} {\bibfnamefont {S.~S.}\ \bibnamefont {Haque}}, \bibinfo {author} {\bibfnamefont {S.}~\bibnamefont {Liu}}, \bibinfo {author} {\bibfnamefont {J.}~\bibnamefont {Murugan}},\ and\ \bibinfo {author} {\bibfnamefont {H.~J.~R.}\ \bibnamefont {Van~Zyl}},\ }\bibfield  {title} {\bibinfo {title} {Spread complexity and topological transitions in the kitaev chain},\ }\href {https://doi.org/10.1007/JHEP01(2023)120} {\bibfield  {journal} {\bibinfo  {journal} {Journal of High Energy Physics}\ }\textbf {\bibinfo {volume} {2023}},\ \bibinfo {pages} {120} (\bibinfo {year} {2023})}\BibitemShut {NoStop}%
\bibitem [{\citenamefont {Afrasiar}\ \emph {et~al.}(2023)\citenamefont {Afrasiar}, \citenamefont {Basak}, \citenamefont {Dey}, \citenamefont {Pal},\ and\ \citenamefont {Pal}}]{Krylov_Quenched_LMG_QPT_Afrasiar_JStat2023}%
  \BibitemOpen
  \bibfield  {author} {\bibinfo {author} {\bibfnamefont {M.}~\bibnamefont {Afrasiar}}, \bibinfo {author} {\bibfnamefont {J.~K.}\ \bibnamefont {Basak}}, \bibinfo {author} {\bibfnamefont {B.}~\bibnamefont {Dey}}, \bibinfo {author} {\bibfnamefont {K.}~\bibnamefont {Pal}},\ and\ \bibinfo {author} {\bibfnamefont {K.}~\bibnamefont {Pal}},\ }\bibfield  {title} {\bibinfo {title} {Time evolution of spread complexity in quenched lipkin--meshkov--glick model},\ }\href {https://doi.org/10.1088/1742-5468/ad0032} {\bibfield  {journal} {\bibinfo  {journal} {Journal of Statistical Mechanics: Theory and Experiment}\ }\textbf {\bibinfo {volume} {2023}},\ \bibinfo {pages} {103101} (\bibinfo {year} {2023})}\BibitemShut {NoStop}%
\bibitem [{\citenamefont {Gautam}\ \emph {et~al.}(2024{\natexlab{a}})\citenamefont {Gautam}, \citenamefont {Jaiswal},\ and\ \citenamefont {Gill}}]{Krylov_state_free_fermion_theories_QPT_Gautam_EurPhysJ_2024}%
  \BibitemOpen
  \bibfield  {author} {\bibinfo {author} {\bibfnamefont {M.}~\bibnamefont {Gautam}}, \bibinfo {author} {\bibfnamefont {N.}~\bibnamefont {Jaiswal}},\ and\ \bibinfo {author} {\bibfnamefont {A.}~\bibnamefont {Gill}},\ }\bibfield  {title} {\bibinfo {title} {Spread complexity in free fermion models},\ }\href {https://doi.org/10.1140/epjb/s10051-023-00636-6} {\bibfield  {journal} {\bibinfo  {journal} {The European Physical Journal B}\ }\textbf {\bibinfo {volume} {97}},\ \bibinfo {pages} {3} (\bibinfo {year} {2024}{\natexlab{a}})}\BibitemShut {NoStop}%
\bibitem [{\citenamefont {Bento}\ \emph {et~al.}(2024)\citenamefont {Bento}, \citenamefont {del Campo},\ and\ \citenamefont {C\'eleri}}]{Krylov-DynamicalPhaseTransition-LMGModel-PRB2024}%
  \BibitemOpen
  \bibfield  {author} {\bibinfo {author} {\bibfnamefont {P.~H.~S.}\ \bibnamefont {Bento}}, \bibinfo {author} {\bibfnamefont {A.}~\bibnamefont {del Campo}},\ and\ \bibinfo {author} {\bibfnamefont {L.~C.}\ \bibnamefont {C\'eleri}},\ }\bibfield  {title} {\bibinfo {title} {Krylov complexity and dynamical phase transition in the quenched lipkin-meshkov-glick model},\ }\href {https://doi.org/10.1103/PhysRevB.109.224304} {\bibfield  {journal} {\bibinfo  {journal} {Phys. Rev. B}\ }\textbf {\bibinfo {volume} {109}},\ \bibinfo {pages} {224304} (\bibinfo {year} {2024})}\BibitemShut {NoStop}%
\bibitem [{\citenamefont {Cohen}\ \emph {et~al.}(2024)\citenamefont {Cohen}, \citenamefont {Oz},\ and\ \citenamefont {Zhong}}]{Lanczos-Krylov-PowerLawRandomBandedModel-2024}%
  \BibitemOpen
  \bibfield  {author} {\bibinfo {author} {\bibfnamefont {K.}~\bibnamefont {Cohen}}, \bibinfo {author} {\bibfnamefont {Y.}~\bibnamefont {Oz}},\ and\ \bibinfo {author} {\bibfnamefont {D.-l.}\ \bibnamefont {Zhong}},\ }\bibfield  {title} {\bibinfo {title} {Complexity measure diagnostics of ergodic to many-body localization transition},\ }\href {https://doi.org/10.1103/PhysRevB.110.L180101} {\bibfield  {journal} {\bibinfo  {journal} {Phys. Rev. B}\ }\textbf {\bibinfo {volume} {110}},\ \bibinfo {pages} {L180101} (\bibinfo {year} {2024})}\BibitemShut {NoStop}%
\bibitem [{\citenamefont {M\"uck}(2024)}]{Krylov_Complexity_Black_Hole_States_Wolfgang_Muck_PRD2024}%
  \BibitemOpen
  \bibfield  {author} {\bibinfo {author} {\bibfnamefont {W.}~\bibnamefont {M\"uck}},\ }\bibfield  {title} {\bibinfo {title} {Black holes and marchenko-pastur distribution},\ }\href {https://doi.org/10.1103/PhysRevD.109.126001} {\bibfield  {journal} {\bibinfo  {journal} {Phys. Rev. D}\ }\textbf {\bibinfo {volume} {109}},\ \bibinfo {pages} {126001} (\bibinfo {year} {2024})}\BibitemShut {NoStop}%
\bibitem [{\citenamefont {Dixit}\ \emph {et~al.}(2024)\citenamefont {Dixit}, \citenamefont {Haque},\ and\ \citenamefont {Razzaque}}]{Krylov_Neutrino_Oscillations_Dixit2024_Arxiv}%
  \BibitemOpen
  \bibfield  {author} {\bibinfo {author} {\bibfnamefont {K.}~\bibnamefont {Dixit}}, \bibinfo {author} {\bibfnamefont {S.~S.}\ \bibnamefont {Haque}},\ and\ \bibinfo {author} {\bibfnamefont {S.}~\bibnamefont {Razzaque}},\ }\bibfield  {title} {\bibinfo {title} {Quantum spread complexity in neutrino oscillations},\ }\href {https://doi.org/10.1140/epjc/s10052-024-12620-0} {\bibfield  {journal} {\bibinfo  {journal} {The European Physical Journal C}\ }\textbf {\bibinfo {volume} {84}},\ \bibinfo {pages} {260} (\bibinfo {year} {2024})}\BibitemShut {NoStop}%
\bibitem [{\citenamefont {Caputa}\ \emph {et~al.}(2024{\natexlab{a}})\citenamefont {Caputa}, \citenamefont {Magan}, \citenamefont {Patramanis},\ and\ \citenamefont {Tonni}}]{Caputa_Modular_Hamiltonian_PRD2024}%
  \BibitemOpen
  \bibfield  {author} {\bibinfo {author} {\bibfnamefont {P.}~\bibnamefont {Caputa}}, \bibinfo {author} {\bibfnamefont {J.~M.}\ \bibnamefont {Magan}}, \bibinfo {author} {\bibfnamefont {D.}~\bibnamefont {Patramanis}},\ and\ \bibinfo {author} {\bibfnamefont {E.}~\bibnamefont {Tonni}},\ }\bibfield  {title} {\bibinfo {title} {Krylov complexity of modular hamiltonian evolution},\ }\href {https://doi.org/10.1103/PhysRevD.109.086004} {\bibfield  {journal} {\bibinfo  {journal} {Phys. Rev. D}\ }\textbf {\bibinfo {volume} {109}},\ \bibinfo {pages} {086004} (\bibinfo {year} {2024}{\natexlab{a}})}\BibitemShut {NoStop}%
\bibitem [{\citenamefont {Gautam}\ \emph {et~al.}(2024{\natexlab{b}})\citenamefont {Gautam}, \citenamefont {Pal}, \citenamefont {Pal}, \citenamefont {Gill}, \citenamefont {Jaiswal},\ and\ \citenamefont {Sarkar}}]{Krylov_Comp_Quenched_Interacting_Systems_Tapobrata_PRB2024}%
  \BibitemOpen
  \bibfield  {author} {\bibinfo {author} {\bibfnamefont {M.}~\bibnamefont {Gautam}}, \bibinfo {author} {\bibfnamefont {K.}~\bibnamefont {Pal}}, \bibinfo {author} {\bibfnamefont {K.}~\bibnamefont {Pal}}, \bibinfo {author} {\bibfnamefont {A.}~\bibnamefont {Gill}}, \bibinfo {author} {\bibfnamefont {N.}~\bibnamefont {Jaiswal}},\ and\ \bibinfo {author} {\bibfnamefont {T.}~\bibnamefont {Sarkar}},\ }\bibfield  {title} {\bibinfo {title} {Spread complexity evolution in quenched interacting quantum systems},\ }\href {https://doi.org/10.1103/PhysRevB.109.014312} {\bibfield  {journal} {\bibinfo  {journal} {Phys. Rev. B}\ }\textbf {\bibinfo {volume} {109}},\ \bibinfo {pages} {014312} (\bibinfo {year} {2024}{\natexlab{b}})}\BibitemShut {NoStop}%
\bibitem [{\citenamefont {Gill}\ \emph {et~al.}(2024)\citenamefont {Gill}, \citenamefont {Pal}, \citenamefont {Pal},\ and\ \citenamefont {Sarkar}}]{Krylov_Quench_TwoPoint_Measurement_Schemes_Tapobrata_PRB2024}%
  \BibitemOpen
  \bibfield  {author} {\bibinfo {author} {\bibfnamefont {A.}~\bibnamefont {Gill}}, \bibinfo {author} {\bibfnamefont {K.}~\bibnamefont {Pal}}, \bibinfo {author} {\bibfnamefont {K.}~\bibnamefont {Pal}},\ and\ \bibinfo {author} {\bibfnamefont {T.}~\bibnamefont {Sarkar}},\ }\bibfield  {title} {\bibinfo {title} {Complexity in two-point measurement schemes},\ }\href {https://doi.org/10.1103/PhysRevB.109.104303} {\bibfield  {journal} {\bibinfo  {journal} {Phys. Rev. B}\ }\textbf {\bibinfo {volume} {109}},\ \bibinfo {pages} {104303} (\bibinfo {year} {2024})}\BibitemShut {NoStop}%
\bibitem [{\citenamefont {Bhattacharya}\ \emph {et~al.}(2023)\citenamefont {Bhattacharya}, \citenamefont {Nandy}, \citenamefont {Nath},\ and\ \citenamefont {Sahu}}]{Krylov_NonHermitian_Appendix_PNandy_JHEP2023}%
  \BibitemOpen
  \bibfield  {author} {\bibinfo {author} {\bibfnamefont {A.}~\bibnamefont {Bhattacharya}}, \bibinfo {author} {\bibfnamefont {P.}~\bibnamefont {Nandy}}, \bibinfo {author} {\bibfnamefont {P.~P.}\ \bibnamefont {Nath}},\ and\ \bibinfo {author} {\bibfnamefont {H.}~\bibnamefont {Sahu}},\ }\bibfield  {title} {\bibinfo {title} {On krylov complexity in open systems: an approach via bi-lanczos algorithm},\ }\href {https://doi.org/10.1007/JHEP12(2023)066} {\bibfield  {journal} {\bibinfo  {journal} {Journal of High Energy Physics}\ }\textbf {\bibinfo {volume} {2023}},\ \bibinfo {pages} {66} (\bibinfo {year} {2023})}\BibitemShut {NoStop}%
\bibitem [{\citenamefont {Bhattacharya}\ \emph {et~al.}(2024{\natexlab{a}})\citenamefont {Bhattacharya}, \citenamefont {Das}, \citenamefont {Dey},\ and\ \citenamefont {Erdmenger}}]{Krylov-NonHermitianSkinEffect-IPR-2024}%
  \BibitemOpen
  \bibfield  {author} {\bibinfo {author} {\bibfnamefont {A.}~\bibnamefont {Bhattacharya}}, \bibinfo {author} {\bibfnamefont {R.~N.}\ \bibnamefont {Das}}, \bibinfo {author} {\bibfnamefont {B.}~\bibnamefont {Dey}},\ and\ \bibinfo {author} {\bibfnamefont {J.}~\bibnamefont {Erdmenger}},\ }\bibfield  {title} {\bibinfo {title} {Spread complexity and localization in $\mathcal{PT}$-symmetric systems},\ }\href {https://doi.org/10.1103/PhysRevB.110.064320} {\bibfield  {journal} {\bibinfo  {journal} {Phys. Rev. B}\ }\textbf {\bibinfo {volume} {110}},\ \bibinfo {pages} {064320} (\bibinfo {year} {2024}{\natexlab{a}})}\BibitemShut {NoStop}%
\bibitem [{\citenamefont {Nizami}\ and\ \citenamefont {Shrestha}(2023)}]{Krylov_Driven_Systems_Nizami_PRE2023}%
  \BibitemOpen
  \bibfield  {author} {\bibinfo {author} {\bibfnamefont {A.~A.}\ \bibnamefont {Nizami}}\ and\ \bibinfo {author} {\bibfnamefont {A.~W.}\ \bibnamefont {Shrestha}},\ }\bibfield  {title} {\bibinfo {title} {Krylov construction and complexity for driven quantum systems},\ }\href {https://doi.org/10.1103/PhysRevE.108.054222} {\bibfield  {journal} {\bibinfo  {journal} {Phys. Rev. E}\ }\textbf {\bibinfo {volume} {108}},\ \bibinfo {pages} {054222} (\bibinfo {year} {2023})}\BibitemShut {NoStop}%
\bibitem [{\citenamefont {Alishahiha}\ and\ \citenamefont {Vasli}(2025)}]{Thermalization_KrylovBasis_Alishahiha2024arxiv}%
  \BibitemOpen
  \bibfield  {author} {\bibinfo {author} {\bibfnamefont {M.}~\bibnamefont {Alishahiha}}\ and\ \bibinfo {author} {\bibfnamefont {M.~J.}\ \bibnamefont {Vasli}},\ }\bibfield  {title} {\bibinfo {title} {Thermalization in krylov basis},\ }\href {https://doi.org/10.1140/epjc/s10052-025-13757-2} {\bibfield  {journal} {\bibinfo  {journal} {The European Physical Journal C}\ }\textbf {\bibinfo {volume} {85}},\ \bibinfo {pages} {39} (\bibinfo {year} {2025})}\BibitemShut {NoStop}%
\bibitem [{\citenamefont {Bhattacharya}\ \emph {et~al.}(2024{\natexlab{b}})\citenamefont {Bhattacharya}, \citenamefont {Das}, \citenamefont {Dey},\ and\ \citenamefont {Erdmenger}}]{Krylov_Nonunitary_Zeno_Effect_Erdmenger_JHEP2024}%
  \BibitemOpen
  \bibfield  {author} {\bibinfo {author} {\bibfnamefont {A.}~\bibnamefont {Bhattacharya}}, \bibinfo {author} {\bibfnamefont {R.~N.}\ \bibnamefont {Das}}, \bibinfo {author} {\bibfnamefont {B.}~\bibnamefont {Dey}},\ and\ \bibinfo {author} {\bibfnamefont {J.}~\bibnamefont {Erdmenger}},\ }\bibfield  {title} {\bibinfo {title} {Spread complexity for measurement-induced non-unitary dynamics and zeno effect},\ }\href {https://doi.org/10.1007/JHEP03(2024)179} {\bibfield  {journal} {\bibinfo  {journal} {Journal of High Energy Physics}\ }\textbf {\bibinfo {volume} {2024}},\ \bibinfo {pages} {179} (\bibinfo {year} {2024}{\natexlab{b}})}\BibitemShut {NoStop}%
\bibitem [{\citenamefont {Zhou}\ and\ \citenamefont {Chen}(2024)}]{zhou_Krylov_2ModeBEC_Arxiv2024}%
  \BibitemOpen
  \bibfield  {author} {\bibinfo {author} {\bibfnamefont {B.}~\bibnamefont {Zhou}}\ and\ \bibinfo {author} {\bibfnamefont {S.}~\bibnamefont {Chen}},\ }\bibfield  {title} {\bibinfo {title} {Spread complexity and dynamical transition in multimode bose-einstein condensates},\ }\href {https://doi.org/10.1103/PhysRevB.110.064318} {\bibfield  {journal} {\bibinfo  {journal} {Phys. Rev. B}\ }\textbf {\bibinfo {volume} {110}},\ \bibinfo {pages} {064318} (\bibinfo {year} {2024})}\BibitemShut {NoStop}%
\bibitem [{\citenamefont {Jha}\ and\ \citenamefont {Roy}(2024)}]{krylov_syk_sparse_holographic_non-holographic_jha_arxiv2024}%
  \BibitemOpen
  \bibfield  {author} {\bibinfo {author} {\bibfnamefont {R.~G.}\ \bibnamefont {Jha}}\ and\ \bibinfo {author} {\bibfnamefont {R.}~\bibnamefont {Roy}},\ }\href {https://arxiv.org/abs/2407.20569} {\bibinfo {title} {Sparsity dependence of krylov state complexity in the syk model}} (\bibinfo {year} {2024}),\ \Eprint {https://arxiv.org/abs/2407.20569} {arXiv:2407.20569 [hep-th]} \BibitemShut {NoStop}%
\bibitem [{\citenamefont {Craps}\ \emph {et~al.}(2024)\citenamefont {Craps}, \citenamefont {Evnin},\ and\ \citenamefont {Pascuzzi}}]{Krylov-Nielsen-PRL2024}%
  \BibitemOpen
  \bibfield  {author} {\bibinfo {author} {\bibfnamefont {B.}~\bibnamefont {Craps}}, \bibinfo {author} {\bibfnamefont {O.}~\bibnamefont {Evnin}},\ and\ \bibinfo {author} {\bibfnamefont {G.}~\bibnamefont {Pascuzzi}},\ }\bibfield  {title} {\bibinfo {title} {A relation between krylov and nielsen complexity},\ }\href {https://doi.org/10.1103/PhysRevLett.132.160402} {\bibfield  {journal} {\bibinfo  {journal} {Phys. Rev. Lett.}\ }\textbf {\bibinfo {volume} {132}},\ \bibinfo {pages} {160402} (\bibinfo {year} {2024})}\BibitemShut {NoStop}%
\bibitem [{\citenamefont {Alishahiha}\ and\ \citenamefont {Banerjee}(2023)}]{Density-matrix-SciPost2023}%
  \BibitemOpen
  \bibfield  {author} {\bibinfo {author} {\bibfnamefont {M.}~\bibnamefont {Alishahiha}}\ and\ \bibinfo {author} {\bibfnamefont {S.}~\bibnamefont {Banerjee}},\ }\bibfield  {title} {\bibinfo {title} {{A universal approach to Krylov state and operator complexities}},\ }\href {https://doi.org/10.21468/SciPostPhys.15.3.080} {\bibfield  {journal} {\bibinfo  {journal} {SciPost Phys.}\ }\textbf {\bibinfo {volume} {15}},\ \bibinfo {pages} {080} (\bibinfo {year} {2023})}\BibitemShut {NoStop}%
\bibitem [{\citenamefont {Caputa}\ \emph {et~al.}(2024{\natexlab{b}})\citenamefont {Caputa}, \citenamefont {Jeong}, \citenamefont {Liu}, \citenamefont {Pedraza},\ and\ \citenamefont {Qu}}]{Caputa-Density-Matrix2024}%
  \BibitemOpen
  \bibfield  {author} {\bibinfo {author} {\bibfnamefont {P.}~\bibnamefont {Caputa}}, \bibinfo {author} {\bibfnamefont {H.-S.}\ \bibnamefont {Jeong}}, \bibinfo {author} {\bibfnamefont {S.}~\bibnamefont {Liu}}, \bibinfo {author} {\bibfnamefont {J.~F.}\ \bibnamefont {Pedraza}},\ and\ \bibinfo {author} {\bibfnamefont {L.-C.}\ \bibnamefont {Qu}},\ }\bibfield  {title} {\bibinfo {title} {Krylov complexity of density matrix operators},\ }\href {https://doi.org/10.1007/JHEP05(2024)337} {\bibfield  {journal} {\bibinfo  {journal} {Journal of High Energy Physics}\ }\textbf {\bibinfo {volume} {2024}},\ \bibinfo {pages} {337} (\bibinfo {year} {2024}{\natexlab{b}})}\BibitemShut {NoStop}%
\bibitem [{\citenamefont {Chattopadhyay}\ \emph {et~al.}(2023)\citenamefont {Chattopadhyay}, \citenamefont {Mitra},\ and\ \citenamefont {van Zyl}}]{Krylov_state_comp_dilaton_geometrical_vanZyl_PRD2023}%
  \BibitemOpen
  \bibfield  {author} {\bibinfo {author} {\bibfnamefont {A.}~\bibnamefont {Chattopadhyay}}, \bibinfo {author} {\bibfnamefont {A.}~\bibnamefont {Mitra}},\ and\ \bibinfo {author} {\bibfnamefont {H.~J.~R.}\ \bibnamefont {van Zyl}},\ }\bibfield  {title} {\bibinfo {title} {Spread complexity as classical dilaton solutions},\ }\href {https://doi.org/10.1103/PhysRevD.108.025013} {\bibfield  {journal} {\bibinfo  {journal} {Phys. Rev. D}\ }\textbf {\bibinfo {volume} {108}},\ \bibinfo {pages} {025013} (\bibinfo {year} {2023})}\BibitemShut {NoStop}%
\bibitem [{\citenamefont {Aguilar-Gutierrez}\ and\ \citenamefont {Rolph}(2024)}]{Aguilar_Gutierrez_Krylov_Not_Distance_PRD2024}%
  \BibitemOpen
  \bibfield  {author} {\bibinfo {author} {\bibfnamefont {S.~E.}\ \bibnamefont {Aguilar-Gutierrez}}\ and\ \bibinfo {author} {\bibfnamefont {A.}~\bibnamefont {Rolph}},\ }\bibfield  {title} {\bibinfo {title} {Krylov complexity is not a measure of distance between states or operators},\ }\href {https://doi.org/10.1103/PhysRevD.109.L081701} {\bibfield  {journal} {\bibinfo  {journal} {Phys. Rev. D}\ }\textbf {\bibinfo {volume} {109}},\ \bibinfo {pages} {L081701} (\bibinfo {year} {2024})}\BibitemShut {NoStop}%
\bibitem [{\citenamefont {Rabinovici}\ \emph {et~al.}(2023)\citenamefont {Rabinovici}, \citenamefont {S{\'a}nchez-Garrido}, \citenamefont {Shir},\ and\ \citenamefont {Sonner}}]{Wormhole-length-JHEP2023}%
  \BibitemOpen
  \bibfield  {author} {\bibinfo {author} {\bibfnamefont {E.}~\bibnamefont {Rabinovici}}, \bibinfo {author} {\bibfnamefont {A.}~\bibnamefont {S{\'a}nchez-Garrido}}, \bibinfo {author} {\bibfnamefont {R.}~\bibnamefont {Shir}},\ and\ \bibinfo {author} {\bibfnamefont {J.}~\bibnamefont {Sonner}},\ }\bibfield  {title} {\bibinfo {title} {A bulk manifestation of krylov complexity},\ }\href {https://doi.org/10.1007/JHEP08(2023)213} {\bibfield  {journal} {\bibinfo  {journal} {Journal of High Energy Physics}\ }\textbf {\bibinfo {volume} {2023}},\ \bibinfo {pages} {213} (\bibinfo {year} {2023})}\BibitemShut {NoStop}%
\bibitem [{\citenamefont {Gallagher}(1994)}]{Gallagher_1994}%
  \BibitemOpen
  \bibfield  {author} {\bibinfo {author} {\bibfnamefont {T.~F.}\ \bibnamefont {Gallagher}},\ }\href {https://www.cambridge.org/core/books/rydberg-atoms/B610BDE54694936F496F59F326C1A81B} {\emph {\bibinfo {title} {Rydberg Atoms}}},\ Cambridge Monographs on Atomic, Molecular and Chemical Physics\ (\bibinfo  {publisher} {Cambridge University Press},\ \bibinfo {year} {1994})\BibitemShut {NoStop}%
\bibitem [{\citenamefont {Saffman}\ \emph {et~al.}(2010)\citenamefont {Saffman}, \citenamefont {Walker},\ and\ \citenamefont {M\o{}lmer}}]{Quantum-Info-Rydberg-Review-Saffman2010}%
  \BibitemOpen
  \bibfield  {author} {\bibinfo {author} {\bibfnamefont {M.}~\bibnamefont {Saffman}}, \bibinfo {author} {\bibfnamefont {T.~G.}\ \bibnamefont {Walker}},\ and\ \bibinfo {author} {\bibfnamefont {K.}~\bibnamefont {M\o{}lmer}},\ }\bibfield  {title} {\bibinfo {title} {Quantum information with rydberg atoms},\ }\href {https://doi.org/10.1103/RevModPhys.82.2313} {\bibfield  {journal} {\bibinfo  {journal} {Rev. Mod. Phys.}\ }\textbf {\bibinfo {volume} {82}},\ \bibinfo {pages} {2313} (\bibinfo {year} {2010})}\BibitemShut {NoStop}%
\bibitem [{\citenamefont {Shao}\ \emph {et~al.}(2024)\citenamefont {Shao}, \citenamefont {Su}, \citenamefont {Li}, \citenamefont {Nath}, \citenamefont {Wu},\ and\ \citenamefont {Li}}]{Rydberg_Review_Rejish_2024}%
  \BibitemOpen
  \bibfield  {author} {\bibinfo {author} {\bibfnamefont {X.-Q.}\ \bibnamefont {Shao}}, \bibinfo {author} {\bibfnamefont {S.-L.}\ \bibnamefont {Su}}, \bibinfo {author} {\bibfnamefont {L.}~\bibnamefont {Li}}, \bibinfo {author} {\bibfnamefont {R.}~\bibnamefont {Nath}}, \bibinfo {author} {\bibfnamefont {J.-H.}\ \bibnamefont {Wu}},\ and\ \bibinfo {author} {\bibfnamefont {W.}~\bibnamefont {Li}},\ }\bibfield  {title} {\bibinfo {title} {Rydberg superatoms: An artificial quantum system for quantum information processing and quantum optics},\ }\href {https://doi.org/10.1063/5.0211071} {\bibfield  {journal} {\bibinfo  {journal} {Applied Physics Reviews}\ }\textbf {\bibinfo {volume} {11}},\ \bibinfo {pages} {031320} (\bibinfo {year} {2024})}\BibitemShut {NoStop}%
\bibitem [{\citenamefont {Browaeys}\ and\ \citenamefont {Lahaye}(2020)}]{Browaeys-Rydberg-Review-NatPhys2020}%
  \BibitemOpen
  \bibfield  {author} {\bibinfo {author} {\bibfnamefont {A.}~\bibnamefont {Browaeys}}\ and\ \bibinfo {author} {\bibfnamefont {T.}~\bibnamefont {Lahaye}},\ }\bibfield  {title} {\bibinfo {title} {Many-body phsyics with individually controlled rydberg atoms},\ }\href {https://www.nature.com/articles/s41567-019-0733-z} {\bibfield  {journal} {\bibinfo  {journal} {Nat. Phys.}\ }\textbf {\bibinfo {volume} {16}},\ \bibinfo {pages} {132} (\bibinfo {year} {2020})}\BibitemShut {NoStop}%
\bibitem [{\citenamefont {Bernien}\ \emph {et~al.}(2017)\citenamefont {Bernien}, \citenamefont {Schwartz}, \citenamefont {Keesling}, \citenamefont {Levine}, \citenamefont {Omran}, \citenamefont {Pichler}, \citenamefont {Choi}, \citenamefont {Zibrov}, \citenamefont {Endres}, \citenamefont {Greiner}, \citenamefont {Vuleti{\'c}},\ and\ \citenamefont {Lukin}}]{Lukin_51Qbits_Nature2017}%
  \BibitemOpen
  \bibfield  {author} {\bibinfo {author} {\bibfnamefont {H.}~\bibnamefont {Bernien}}, \bibinfo {author} {\bibfnamefont {S.}~\bibnamefont {Schwartz}}, \bibinfo {author} {\bibfnamefont {A.}~\bibnamefont {Keesling}}, \bibinfo {author} {\bibfnamefont {H.}~\bibnamefont {Levine}}, \bibinfo {author} {\bibfnamefont {A.}~\bibnamefont {Omran}}, \bibinfo {author} {\bibfnamefont {H.}~\bibnamefont {Pichler}}, \bibinfo {author} {\bibfnamefont {S.}~\bibnamefont {Choi}}, \bibinfo {author} {\bibfnamefont {A.~S.}\ \bibnamefont {Zibrov}}, \bibinfo {author} {\bibfnamefont {M.}~\bibnamefont {Endres}}, \bibinfo {author} {\bibfnamefont {M.}~\bibnamefont {Greiner}}, \bibinfo {author} {\bibfnamefont {V.}~\bibnamefont {Vuleti{\'c}}},\ and\ \bibinfo {author} {\bibfnamefont {M.~D.}\ \bibnamefont {Lukin}},\ }\bibfield  {title} {\bibinfo {title} {Probing many-body dynamics on a 51-atom quantum simulator},\ }\href {https://doi.org/10.1038/nature24622} {\bibfield  {journal} {\bibinfo  {journal} {Nature}\ }\textbf {\bibinfo {volume} {551}},\
  \bibinfo {pages} {579} (\bibinfo {year} {2017})}\BibitemShut {NoStop}%
\bibitem [{\citenamefont {Ebadi}\ \emph {et~al.}(2021)\citenamefont {Ebadi}, \citenamefont {Wang}, \citenamefont {Levine}, \citenamefont {Keesling}, \citenamefont {Semeghini}, \citenamefont {Omran}, \citenamefont {Bluvstein}, \citenamefont {Samajdar}, \citenamefont {Pichler}, \citenamefont {Ho}, \citenamefont {Choi}, \citenamefont {Sachdev}, \citenamefont {Greiner}, \citenamefont {Vuleti{\'c}},\ and\ \citenamefont {Lukin}}]{Lukin_256Qbits_Nature2021}%
  \BibitemOpen
  \bibfield  {author} {\bibinfo {author} {\bibfnamefont {S.}~\bibnamefont {Ebadi}}, \bibinfo {author} {\bibfnamefont {T.~T.}\ \bibnamefont {Wang}}, \bibinfo {author} {\bibfnamefont {H.}~\bibnamefont {Levine}}, \bibinfo {author} {\bibfnamefont {A.}~\bibnamefont {Keesling}}, \bibinfo {author} {\bibfnamefont {G.}~\bibnamefont {Semeghini}}, \bibinfo {author} {\bibfnamefont {A.}~\bibnamefont {Omran}}, \bibinfo {author} {\bibfnamefont {D.}~\bibnamefont {Bluvstein}}, \bibinfo {author} {\bibfnamefont {R.}~\bibnamefont {Samajdar}}, \bibinfo {author} {\bibfnamefont {H.}~\bibnamefont {Pichler}}, \bibinfo {author} {\bibfnamefont {W.~W.}\ \bibnamefont {Ho}}, \bibinfo {author} {\bibfnamefont {S.}~\bibnamefont {Choi}}, \bibinfo {author} {\bibfnamefont {S.}~\bibnamefont {Sachdev}}, \bibinfo {author} {\bibfnamefont {M.}~\bibnamefont {Greiner}}, \bibinfo {author} {\bibfnamefont {V.}~\bibnamefont {Vuleti{\'c}}},\ and\ \bibinfo {author} {\bibfnamefont {M.~D.}\ \bibnamefont {Lukin}},\ }\bibfield  {title} {\bibinfo {title}
  {Quantum phases of matter on a 256-atom programmable quantum simulator},\ }\href {https://doi.org/10.1038/s41586-021-03582-4} {\bibfield  {journal} {\bibinfo  {journal} {Nature}\ }\textbf {\bibinfo {volume} {595}},\ \bibinfo {pages} {227} (\bibinfo {year} {2021})}\BibitemShut {NoStop}%
\bibitem [{\citenamefont {Barb{\'o}n}\ \emph {et~al.}(2019)\citenamefont {Barb{\'o}n}, \citenamefont {Rabinovici}, \citenamefont {Shir},\ and\ \citenamefont {Sinha}}]{K-Shannon-Entropy-OperatorComplexity-JHEP2019}%
  \BibitemOpen
  \bibfield  {author} {\bibinfo {author} {\bibfnamefont {J.~L.~F.}\ \bibnamefont {Barb{\'o}n}}, \bibinfo {author} {\bibfnamefont {E.}~\bibnamefont {Rabinovici}}, \bibinfo {author} {\bibfnamefont {R.}~\bibnamefont {Shir}},\ and\ \bibinfo {author} {\bibfnamefont {R.}~\bibnamefont {Sinha}},\ }\bibfield  {title} {\bibinfo {title} {On the evolution of operator complexity beyond scrambling},\ }\href {https://doi.org/10.1007/JHEP10(2019)264} {\bibfield  {journal} {\bibinfo  {journal} {Journal of High Energy Physics}\ }\textbf {\bibinfo {volume} {2019}},\ \bibinfo {pages} {264} (\bibinfo {year} {2019})}\BibitemShut {NoStop}%
\bibitem [{\citenamefont {Chougale}\ \emph {et~al.}(2020)\citenamefont {Chougale}, \citenamefont {Talukdar}, \citenamefont {Ramos},\ and\ \citenamefont {Nath}}]{PhysRevA.102.022816}%
  \BibitemOpen
  \bibfield  {author} {\bibinfo {author} {\bibfnamefont {Y.}~\bibnamefont {Chougale}}, \bibinfo {author} {\bibfnamefont {J.}~\bibnamefont {Talukdar}}, \bibinfo {author} {\bibfnamefont {T.}~\bibnamefont {Ramos}},\ and\ \bibinfo {author} {\bibfnamefont {R.}~\bibnamefont {Nath}},\ }\bibfield  {title} {\bibinfo {title} {Dynamics of rydberg excitations and quantum correlations in an atomic array coupled to a photonic crystal waveguide},\ }\href {https://doi.org/10.1103/PhysRevA.102.022816} {\bibfield  {journal} {\bibinfo  {journal} {Phys. Rev. A}\ }\textbf {\bibinfo {volume} {102}},\ \bibinfo {pages} {022816} (\bibinfo {year} {2020})}\BibitemShut {NoStop}%
\bibitem [{\citenamefont {Lv}\ \emph {et~al.}(2024)\citenamefont {Lv}, \citenamefont {Zhang},\ and\ \citenamefont {Zhou}}]{building_krylov_complexity_from_circuit_complexity_chenwei_PRR2024}%
  \BibitemOpen
  \bibfield  {author} {\bibinfo {author} {\bibfnamefont {C.}~\bibnamefont {Lv}}, \bibinfo {author} {\bibfnamefont {R.}~\bibnamefont {Zhang}},\ and\ \bibinfo {author} {\bibfnamefont {Q.}~\bibnamefont {Zhou}},\ }\bibfield  {title} {\bibinfo {title} {Building krylov complexity from circuit complexity},\ }\href {https://doi.org/10.1103/PhysRevResearch.6.L042001} {\bibfield  {journal} {\bibinfo  {journal} {Phys. Rev. Res.}\ }\textbf {\bibinfo {volume} {6}},\ \bibinfo {pages} {L042001} (\bibinfo {year} {2024})}\BibitemShut {NoStop}%
\bibitem [{\citenamefont {Jaksch}\ \emph {et~al.}(2000)\citenamefont {Jaksch}, \citenamefont {Cirac}, \citenamefont {Zoller}, \citenamefont {Rolston}, \citenamefont {C\^ot\'e},\ and\ \citenamefont {Lukin}}]{Rydberg-Blockade-Jaksch-PRL2000}%
  \BibitemOpen
  \bibfield  {author} {\bibinfo {author} {\bibfnamefont {D.}~\bibnamefont {Jaksch}}, \bibinfo {author} {\bibfnamefont {J.~I.}\ \bibnamefont {Cirac}}, \bibinfo {author} {\bibfnamefont {P.}~\bibnamefont {Zoller}}, \bibinfo {author} {\bibfnamefont {S.~L.}\ \bibnamefont {Rolston}}, \bibinfo {author} {\bibfnamefont {R.}~\bibnamefont {C\^ot\'e}},\ and\ \bibinfo {author} {\bibfnamefont {M.~D.}\ \bibnamefont {Lukin}},\ }\bibfield  {title} {\bibinfo {title} {Fast quantum gates for neutral atoms},\ }\href {https://doi.org/10.1103/PhysRevLett.85.2208} {\bibfield  {journal} {\bibinfo  {journal} {Phys. Rev. Lett.}\ }\textbf {\bibinfo {volume} {85}},\ \bibinfo {pages} {2208} (\bibinfo {year} {2000})}\BibitemShut {NoStop}%
\bibitem [{\citenamefont {Lukin}\ \emph {et~al.}(2001)\citenamefont {Lukin}, \citenamefont {Fleischhauer}, \citenamefont {Cote}, \citenamefont {Duan}, \citenamefont {Jaksch}, \citenamefont {Cirac},\ and\ \citenamefont {Zoller}}]{Rydberg-Blockade-Lukin-PRL2001}%
  \BibitemOpen
  \bibfield  {author} {\bibinfo {author} {\bibfnamefont {M.~D.}\ \bibnamefont {Lukin}}, \bibinfo {author} {\bibfnamefont {M.}~\bibnamefont {Fleischhauer}}, \bibinfo {author} {\bibfnamefont {R.}~\bibnamefont {Cote}}, \bibinfo {author} {\bibfnamefont {L.~M.}\ \bibnamefont {Duan}}, \bibinfo {author} {\bibfnamefont {D.}~\bibnamefont {Jaksch}}, \bibinfo {author} {\bibfnamefont {J.~I.}\ \bibnamefont {Cirac}},\ and\ \bibinfo {author} {\bibfnamefont {P.}~\bibnamefont {Zoller}},\ }\bibfield  {title} {\bibinfo {title} {Dipole blockade and quantum information processing in mesoscopic atomic ensembles},\ }\href {https://doi.org/10.1103/PhysRevLett.87.037901} {\bibfield  {journal} {\bibinfo  {journal} {Phys. Rev. Lett.}\ }\textbf {\bibinfo {volume} {87}},\ \bibinfo {pages} {037901} (\bibinfo {year} {2001})}\BibitemShut {NoStop}%
\bibitem [{\citenamefont {Heidemann}\ \emph {et~al.}(2007)\citenamefont {Heidemann}, \citenamefont {Krohn}, \citenamefont {Bendkowsky}, \citenamefont {Butscher}, \citenamefont {L\"ow}, \citenamefont {Santos},\ and\ \citenamefont {Pfau}}]{Coherent-Collective-Excitation-Rydberg-Blockade-PRL2007}%
  \BibitemOpen
  \bibfield  {author} {\bibinfo {author} {\bibfnamefont {R.}~\bibnamefont {Heidemann}}, \bibinfo {author} {\bibfnamefont {U.}~\bibnamefont {Krohn}}, \bibinfo {author} {\bibfnamefont {V.}~\bibnamefont {Bendkowsky}}, \bibinfo {author} {\bibfnamefont {B.}~\bibnamefont {Butscher}}, \bibinfo {author} {\bibfnamefont {R.}~\bibnamefont {L\"ow}}, \bibinfo {author} {\bibfnamefont {L.}~\bibnamefont {Santos}},\ and\ \bibinfo {author} {\bibfnamefont {T.}~\bibnamefont {Pfau}},\ }\bibfield  {title} {\bibinfo {title} {Evidence for coherent collective rydberg excitation in the strong blockade regime},\ }\href {https://doi.org/10.1103/PhysRevLett.99.163601} {\bibfield  {journal} {\bibinfo  {journal} {Phys. Rev. Lett.}\ }\textbf {\bibinfo {volume} {99}},\ \bibinfo {pages} {163601} (\bibinfo {year} {2007})}\BibitemShut {NoStop}%
\bibitem [{\citenamefont {Urban}\ \emph {et~al.}(2009)\citenamefont {Urban}, \citenamefont {Johnson}, \citenamefont {Henage}, \citenamefont {Isenhower}, \citenamefont {Yavuz}, \citenamefont {Walker},\ and\ \citenamefont {Saffman}}]{Rydberg-Blockade-Exp-NatPhys2009}%
  \BibitemOpen
  \bibfield  {author} {\bibinfo {author} {\bibfnamefont {E.}~\bibnamefont {Urban}}, \bibinfo {author} {\bibfnamefont {T.~A.}\ \bibnamefont {Johnson}}, \bibinfo {author} {\bibfnamefont {T.}~\bibnamefont {Henage}}, \bibinfo {author} {\bibfnamefont {L.}~\bibnamefont {Isenhower}}, \bibinfo {author} {\bibfnamefont {D.~D.}\ \bibnamefont {Yavuz}}, \bibinfo {author} {\bibfnamefont {T.~G.}\ \bibnamefont {Walker}},\ and\ \bibinfo {author} {\bibfnamefont {M.}~\bibnamefont {Saffman}},\ }\bibfield  {title} {\bibinfo {title} {Observation of rydberg blockade between two atoms},\ }\href {https://www.nature.com/articles/nphys1178} {\bibfield  {journal} {\bibinfo  {journal} {Nat. Phys.}\ }\textbf {\bibinfo {volume} {5}},\ \bibinfo {pages} {110} (\bibinfo {year} {2009})}\BibitemShut {NoStop}%
\bibitem [{\citenamefont {Ga$\ddot{\text{e}}$tan}\ \emph {et~al.}(2009)\citenamefont {Ga$\ddot{\text{e}}$tan}, \citenamefont {Miroshnychenko}, \citenamefont {Wilk}, \citenamefont {Chotia}, \citenamefont {Vitteau}, \citenamefont {Comparat}, \citenamefont {Pillet}, \citenamefont {Browaeys},\ and\ \citenamefont {Grangier}}]{Collective-Excitation-NatPhys2009}%
  \BibitemOpen
  \bibfield  {author} {\bibinfo {author} {\bibfnamefont {A.}~\bibnamefont {Ga$\ddot{\text{e}}$tan}}, \bibinfo {author} {\bibfnamefont {Y.}~\bibnamefont {Miroshnychenko}}, \bibinfo {author} {\bibfnamefont {T.}~\bibnamefont {Wilk}}, \bibinfo {author} {\bibfnamefont {A.}~\bibnamefont {Chotia}}, \bibinfo {author} {\bibfnamefont {M.}~\bibnamefont {Vitteau}}, \bibinfo {author} {\bibfnamefont {D.}~\bibnamefont {Comparat}}, \bibinfo {author} {\bibfnamefont {P.}~\bibnamefont {Pillet}}, \bibinfo {author} {\bibfnamefont {A.}~\bibnamefont {Browaeys}},\ and\ \bibinfo {author} {\bibfnamefont {P.}~\bibnamefont {Grangier}},\ }\bibfield  {title} {\bibinfo {title} {Observation of collective excitation of two individual atoms in the rydberg blockade regime},\ }\href {https://www.nature.com/articles/nphys1183} {\bibfield  {journal} {\bibinfo  {journal} {Nat. Phys.}\ }\textbf {\bibinfo {volume} {5}},\ \bibinfo {pages} {115} (\bibinfo {year} {2009})}\BibitemShut {NoStop}%
\bibitem [{\citenamefont {Wilk}\ \emph {et~al.}(2010)\citenamefont {Wilk}, \citenamefont {Ga\"etan}, \citenamefont {Evellin}, \citenamefont {Wolters}, \citenamefont {Miroshnychenko}, \citenamefont {Grangier},\ and\ \citenamefont {Browaeys}}]{Entanglement-Rydberg-Blockade-PRL2010}%
  \BibitemOpen
  \bibfield  {author} {\bibinfo {author} {\bibfnamefont {T.}~\bibnamefont {Wilk}}, \bibinfo {author} {\bibfnamefont {A.}~\bibnamefont {Ga\"etan}}, \bibinfo {author} {\bibfnamefont {C.}~\bibnamefont {Evellin}}, \bibinfo {author} {\bibfnamefont {J.}~\bibnamefont {Wolters}}, \bibinfo {author} {\bibfnamefont {Y.}~\bibnamefont {Miroshnychenko}}, \bibinfo {author} {\bibfnamefont {P.}~\bibnamefont {Grangier}},\ and\ \bibinfo {author} {\bibfnamefont {A.}~\bibnamefont {Browaeys}},\ }\bibfield  {title} {\bibinfo {title} {Entanglement of two individual neutral atoms using rydberg blockade},\ }\href {https://doi.org/10.1103/PhysRevLett.104.010502} {\bibfield  {journal} {\bibinfo  {journal} {Phys. Rev. Lett.}\ }\textbf {\bibinfo {volume} {104}},\ \bibinfo {pages} {010502} (\bibinfo {year} {2010})}\BibitemShut {NoStop}%
\bibitem [{\citenamefont {Dudin}\ \emph {et~al.}(2012)\citenamefont {Dudin}, \citenamefont {Li}, \citenamefont {Bariani},\ and\ \citenamefont {Kuzmich}}]{Many-Body-Rabi-Oscillations-Exp-NatPhys2012}%
  \BibitemOpen
  \bibfield  {author} {\bibinfo {author} {\bibfnamefont {Y.~O.}\ \bibnamefont {Dudin}}, \bibinfo {author} {\bibfnamefont {L.}~\bibnamefont {Li}}, \bibinfo {author} {\bibfnamefont {F.}~\bibnamefont {Bariani}},\ and\ \bibinfo {author} {\bibfnamefont {A.}~\bibnamefont {Kuzmich}},\ }\bibfield  {title} {\bibinfo {title} {Observation of coherent many-body rabi oscillations},\ }\href {https://www.nature.com/articles/nphys2413} {\bibfield  {journal} {\bibinfo  {journal} {Nat. Phys.}\ }\textbf {\bibinfo {volume} {8}},\ \bibinfo {pages} {790} (\bibinfo {year} {2012})}\BibitemShut {NoStop}%
\bibitem [{\citenamefont {Srivastava}\ \emph {et~al.}(2019)\citenamefont {Srivastava}, \citenamefont {Niranjan},\ and\ \citenamefont {Nath}}]{Rydberg-Biased-Freezing-IOP2019}%
  \BibitemOpen
  \bibfield  {author} {\bibinfo {author} {\bibfnamefont {V.}~\bibnamefont {Srivastava}}, \bibinfo {author} {\bibfnamefont {A.}~\bibnamefont {Niranjan}},\ and\ \bibinfo {author} {\bibfnamefont {R.}~\bibnamefont {Nath}},\ }\bibfield  {title} {\bibinfo {title} {Dynamics and quantum correlations in two independently driven rydberg atoms with distinct laser fields},\ }\href {https://doi.org/10.1088/1361-6455/ab32a2} {\bibfield  {journal} {\bibinfo  {journal} {Journal of Physics B: Atomic, Molecular and Optical Physics}\ }\textbf {\bibinfo {volume} {52}},\ \bibinfo {pages} {184001} (\bibinfo {year} {2019})}\BibitemShut {NoStop}%
\bibitem [{\citenamefont {Krithika}\ \emph {et~al.}(2021)\citenamefont {Krithika}, \citenamefont {Pal}, \citenamefont {Nath},\ and\ \citenamefont {Mahesh}}]{Rydberg-Biased-Freezing-Expt-PRR2021}%
  \BibitemOpen
  \bibfield  {author} {\bibinfo {author} {\bibfnamefont {V.~R.}\ \bibnamefont {Krithika}}, \bibinfo {author} {\bibfnamefont {S.}~\bibnamefont {Pal}}, \bibinfo {author} {\bibfnamefont {R.}~\bibnamefont {Nath}},\ and\ \bibinfo {author} {\bibfnamefont {T.~S.}\ \bibnamefont {Mahesh}},\ }\bibfield  {title} {\bibinfo {title} {Observation of interaction induced blockade and local spin freezing in a nmr quantum simulator},\ }\href {https://doi.org/10.1103/PhysRevResearch.3.033035} {\bibfield  {journal} {\bibinfo  {journal} {Phys. Rev. Res.}\ }\textbf {\bibinfo {volume} {3}},\ \bibinfo {pages} {033035} (\bibinfo {year} {2021})}\BibitemShut {NoStop}%
\bibitem [{\citenamefont {Bravyi}\ \emph {et~al.}(2011)\citenamefont {Bravyi}, \citenamefont {DiVincenzo},\ and\ \citenamefont {Loss}}]{Schrieffer-Wolff-Other-Effective-Bravyi2011}%
  \BibitemOpen
  \bibfield  {author} {\bibinfo {author} {\bibfnamefont {S.}~\bibnamefont {Bravyi}}, \bibinfo {author} {\bibfnamefont {D.~P.}\ \bibnamefont {DiVincenzo}},\ and\ \bibinfo {author} {\bibfnamefont {D.}~\bibnamefont {Loss}},\ }\bibfield  {title} {\bibinfo {title} {Schrieffer--wolff transformation for quantum many-body systems},\ }\href {https://www.sciencedirect.com/science/article/pii/S0003491611001059} {\bibfield  {journal} {\bibinfo  {journal} {Annals of Physics}\ }\textbf {\bibinfo {volume} {326}},\ \bibinfo {pages} {2793} (\bibinfo {year} {2011})}\BibitemShut {NoStop}%
\end{thebibliography}%

\end{document}